\documentclass[twoside,twocolumn,9pt]{article}
\usepackage{extsizes}
\usepackage[super,sort&compress,comma]{natbib} 
\usepackage[version=3]{mhchem}
\usepackage[left=1.5cm, right=1.5cm, top=1.785cm, bottom=2.0cm]{geometry}
\usepackage{balance}
\usepackage{mathptmx}
\usepackage{sectsty}
\usepackage{graphicx} 
\usepackage{lastpage}
\usepackage[format=plain,justification=justified,singlelinecheck=false,font={stretch=1.125,small,sf},labelfont=bf,labelsep=space]{caption}
\usepackage{float}
\usepackage{fancyhdr}
\usepackage{fnpos}
\usepackage[english]{babel}
\addto{\captionsenglish}{
}
\usepackage{array}
\usepackage{droidsans}
\usepackage{charter}
\usepackage[T1]{fontenc}
\usepackage[usenames,dvipsnames]{xcolor}
\usepackage{setspace}
\usepackage[compact]{titlesec}
\usepackage{hyperref}
\hypersetup{hidelinks}

\usepackage{epstopdf}

\usepackage{layouts}
\usepackage{csquotes}
\usepackage{siunitx}
\usepackage{todonotes}
\renewcommand{\d}{\mathrm{d}}

\definecolor{cream}{RGB}{222,217,201}

\begin{document}

\pagestyle{fancy}
\thispagestyle{plain}
\fancypagestyle{plain}{
\renewcommand{\headrulewidth}{0pt}
}

\makeFNbottom
\makeatletter
\renewcommand\LARGE{\@setfontsize\LARGE{15pt}{17}}
\renewcommand\Large{\@setfontsize\Large{12pt}{14}}
\renewcommand\large{\@setfontsize\large{10pt}{12}}
\renewcommand\footnotesize{\@setfontsize\footnotesize{7pt}{10}}
\makeatother

\renewcommand{\thefootnote}{\fnsymbol{footnote}}
\renewcommand\footnoterule{\vspace*{1pt}\color{cream}\hrule width 3.5in height 0.4pt \color{black}\vspace*{5pt}}
\setcounter{secnumdepth}{5}

\makeatletter
\renewcommand\@biblabel[1]{#1}
\renewcommand\@makefntext[1]{\noindent\makebox[0pt][r]{\@thefnmark\,}#1}
\makeatother
\renewcommand{\figurename}{\small{Fig.}~}
\sectionfont{\sffamily\Large}
\subsectionfont{\normalsize}
\subsubsectionfont{\bf}
\setstretch{1.125} \setlength{\skip\footins}{0.8cm}
\setlength{\footnotesep}{0.25cm}
\setlength{\jot}{10pt}
\titlespacing*{\section}{0pt}{4pt}{4pt}
\titlespacing*{\subsection}{0pt}{15pt}{1pt}

\fancyfoot{}
\fancyfoot[LO,RE]{\vspace{-7.1pt}\includegraphics[height=9pt]{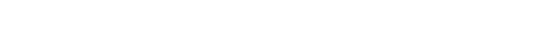}}
\fancyfoot[CO]{\vspace{-7.1pt}\hspace{13.2cm}\includegraphics{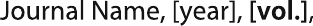}}
\fancyfoot[CE]{\vspace{-7.2pt}\hspace{-14.2cm}\includegraphics{head_foot/RF}}
\fancyfoot[RO]{\footnotesize{\sffamily{1--\pageref{LastPage}~\textbar\hspace{2pt}\thepage}}}
\fancyfoot[LE]{\footnotesize{\sffamily{\thepage~\textbar\hspace{3.45cm} 1--\pageref{LastPage}}}}
\fancyhead{}
\renewcommand{\headrulewidth}{0pt}
\renewcommand{\footrulewidth}{0pt}
\setlength{\arrayrulewidth}{1pt}
\setlength{\columnsep}{6.5mm}
\setlength\bibsep{1pt}

\makeatletter
\newlength{\figrulesep}
\setlength{\figrulesep}{0.5\textfloatsep}

\newcommand{\topfigrule}{\vspace*{-1pt}\noindent{\color{cream}\rule[-\figrulesep]{\columnwidth}{1.5pt}} }

\newcommand{\botfigrule}{\vspace*{-2pt}\noindent{\color{cream}\rule[\figrulesep]{\columnwidth}{1.5pt}} }

\newcommand{\dblfigrule}{\vspace*{-1pt}\noindent{\color{cream}\rule[-\figrulesep]{\textwidth}{1.5pt}} }

\makeatother

\twocolumn[
{\includegraphics[height=30pt]{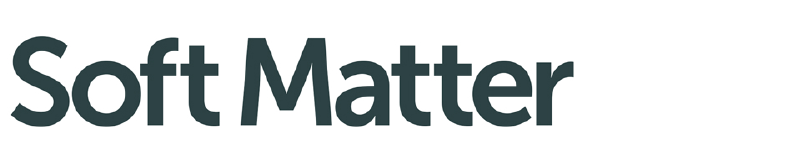}\hfill\raisebox{0pt}[0pt][0pt]{\includegraphics[height=55pt]{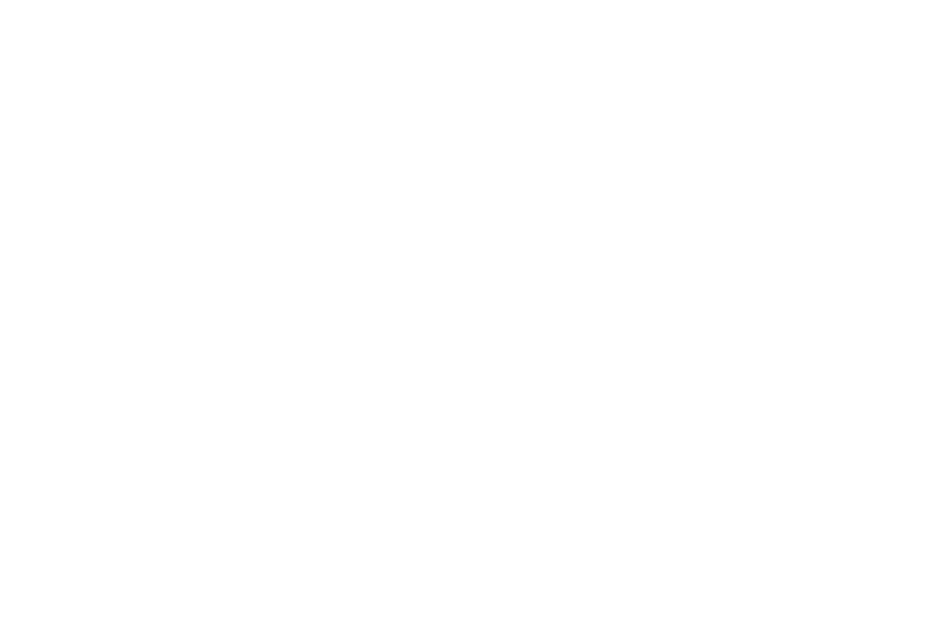}}\\[1ex]
			\includegraphics[width=18.5cm]{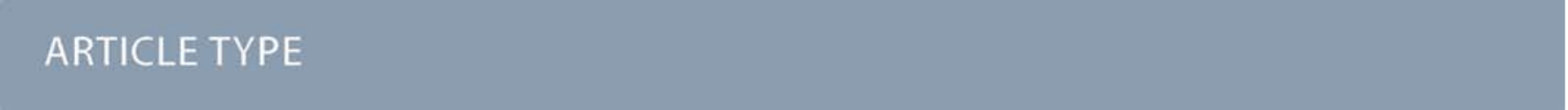}}\par
	\vspace{1em}
	\sffamily
	\begin{tabular}{m{4.5cm} p{13.5cm}}
		\includegraphics{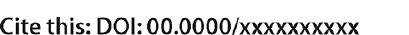}   & \noindent\LARGE{\textbf{Stick-slip Dynamics in the Forced Wetting of Polymer Brushes}}\\
		\vspace{0.3cm}                    & \vspace{0.3cm}\\
		                                  & \noindent\large{Daniel Greve,\textit{$^{a,\ast}$} Simon Hartmann,\textit{$^{a,b,\dag}$} 
		                                  and Uwe Thiele\textit{$^{a,b,\ddag}$}}\\
		\includegraphics{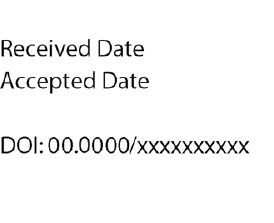} & \noindent\normalsize{We study the static and dynamic wetting of adaptive substrates using a mesoscopic hydrodynamic model for a liquid droplet on a solid substrate covered by a polymer brush. First, we show that on the macroscale Young's law still holds for the equilibrium contact angle and that on the mesoscale a Neumann-type law governs the shape of the wetting ridge. Following an analytic and numeric assessment of the static profiles of droplet and wetting ridge, we examine the dynamics of the wetting ridge for a liquid meniscus that is advanced at constant speed. In other words, we consider an inverse Landau-Levich case where a brush-covered plate is introduced into (and not drawn from) a liquid bath. We find a characteristic stick-slip motion that emerges when the dynamic contact angle of the stationary moving meniscus decreases with increasing velocity, and relate the onset of slip to Gibbs' inequality and to a cross-over in relevant time scales.}
	\end{tabular}
\vspace{0.6cm}
]

\renewcommand*\rmdefault{bch}\normalfont\upshape
\rmfamily
\section*{}
\vspace{-1cm}

\footnotetext{\textit{$^{a}$~Institut f\"ur Theoretische Physik, Westf\"alische Wilhelms-Universit\"at M\"unster, Wilhelm-Klemm-Str.\ 9, 48149 M\"unster, Germany.}}
\footnotetext{\textit{$^{b}$~Center for Nonlinear Science (CeNoS), Westf{\"a}lische Wilhelms-Universit\"at M\"unster, Corrensstr.\ 2, 48149 M\"unster, Germany}}

\footnotetext{$\ast$~E-mail:~daniel.greve@wwu.de, ORCID:~0000-0001-7767-3375}
\footnotetext{$\dag$~E-mail:~s.hartmann@wwu.de, ORCID:~0000-0002-3127-136X}
\footnotetext{$\ddag$~E-mail:~u.thiele@uni-muenster.de, ORCID:~0000-0001-7989-9271}

\section{Introduction}\label{sec:intro}

Many hydrodynamic processes of practical importance involve the motion of three-phase contact lines. In consequence, for many years phenomena involving static and dynamic contact lines have been of much interdisciplinary interest. This includes shapes of static and dynamic drops and bubbles at solid substrates and liquid bridges between solids as well as wetting and dewetting processes.\cite{Genn1985rmp,TeDS1988rpap,StVe2009jpm,BEIM2009rmp,CrMa2009rmp,SnAn2013arfm,EWGT2016prf}
An important class of problems that has gained much attention in recent years is the wetting of soft adaptive substrates,\cite{BiRR2018arfm,BBSV2018l,AnSn2020arfm} i.e., substrates with an intrinsic dynamics that responds to the dynamics of the liquids on top of them.

To comprehend the intricacies of dynamic wetting processes on substrates like hydrogels and polymer brushes, first, a profound understanding of the statics of wetting is essential. For a sessile liquid drop on a substrate, two limiting cases are often considered: on the one hand, for a rigid solid substrate, the equilibrium contact angle $\theta_\mathrm{Y}$ at the three-phase contact line is given by a macroscopic horizontal force balance, namely, the Young law.\cite{Youn1805ptrs}
If, on the other hand, the substrate is liquid, also the vertical force balance has to be taken into account as the liquid-gas interface tension exerts a traction force that deforms the substrate. The two conditions form the Neumann law.\cite{DeGennesBrochard2004,MDSA2012prl} It determines the two independent angles between the three involved interfaces and is invariant under a rigid rotation of the three-phase contact region.

For other non-rigid substrates as soft elastic or otherwise adaptive substrates, normally, the Neumann law applies at least in the close vicinity of the contact line.\cite{PAKZ2020prx,AnSn2020arfm} At larger distances, the bulk influence of substrate elasticity becomes relevant and results in features like the wetting ridge and viscoelastic breaking.\cite{AnSn2020arfm}

Here, we are specifically concerned with liquid droplets and menisci on a rigid solid substrate covered by a polymer brush that can elastically deform, and also absorbs liquid due to mass transfer and imbibition processes.\cite{MaMu2005jpm,LeMu2011jcp,BBSV2018l,MeSB2019m,EDSD2021m} As a result, brush swelling and deformation interact with contact line motion, e.g., for spreading and sliding drops.\cite{MMHM2011jpm,ThHa2020epjt,WHNK2020l,LSSK2021l} In contrast to soft solid substrates, where, depending on substrate softness, large elastic deformations can be found even on the scale of macroscopic droplets,\cite{HeST2021sm} the comparatively small length (nanometer to micrometer) of the grafted polymer molecules restricts brush deformations to mesoscopic scales. We will show that, in consequence, the wetting behaviour of polymer brushes corresponds to an intermediate case, where characteristics of both, the macroscopic Young law and Neumann law are encountered.

Due to the involved small scales of the deformations, many of the features known from soft solid substrates, e.g., the occurrence of a wetting ridge at the contact line, have, to our knowledge, not yet been assessed in experiments involving three-phase contact lines on polymer brushes. However, they may be responsible for intricate observed macroscopic behaviour, namely, the emergence of stick-slip dynamics, i.e., stick-slip motion of the contact line. This occurs, e.g., in forced wetting experiments with expanding or moving droplets on polymer brushes.\cite{WMYT2010scc,SHNF2021acis,HLSW2022l}
The length scale-bridging relation of stick-slip motion and the formation of wetting ridges and associated pinning effects is already more widely investigated for contact line motion on soft solid substrates.\cite{KDNR2013sm, KBRD2014sm, KDGP2015nc, PBDJ2017sm, GASK2018prl, AnSn2020arfm,MoAK2022el} However, even there, the bifurcations underlying transitions between stationary and stick-slip contact line motion are not yet well understood. 

Note that related stick-slip phenomena occur well beyond systems involving soft and adaptive substrates. They are described in a wide range of (de)wetting, deposition and coating processes,\cite{HaLi2012acie,Thie2014acis} where they can occur for both, advancing and receding contact lines. Detailed investigations exist for droplets spreading or sliding on rigid substrates with imposed regular wettability or topography patterns.\cite{CuFe2001el,ThKn2006njp,ZhMi2009l,BKHT2011pre,SaKa2013jfm,VFFP2013prl,VSFP2014l,SBAV2014pre} Rough rigid substrates often reveal a stochastic stick-slip motion due to a contact line pinning induced by a random topography.\cite{ScWo1998prl,CuFe2004jcis,TYYA2006l,SaKP2010prl} Furthermore, self-organised stick-slip motion is of key importance to deposition processes like in the evaporative dewetting or dip coating of a solution or suspension,\cite{BoDG2010l,HaLi2012acie,Thie2014acis,Lars2014aj,JEZT2018l} and the Langmuir-Blodgett transfer of a surfactant layer from a bath onto a moving plate.\cite{SpCR1994el,LKGF2012s} This is also closely related to the fine structure of coffee rings.\cite{Deeg2000pre} Beyond various technical applications,\cite{HaLi2012acie} similar time-periodic behaviour may as well be found in biology as cells can show a stick-slip motility.\cite{ZGLA2017jpdp,RMGV2020prr} Detailed nonlinear analyses of the rich dynamical behaviour for the examples of evaporative dewetting,\cite{FrAT2012sm} Langmuir-Blodgett\cite{LKGF2012s,KoTh2014n} and dip-coating transfer\cite{TWGT2019prf} are available. They show that the onset of stick-slip motion is often related to time-periodic states that appear at Hopf bifurcations (at large contact line speeds) and global bifurcations (at low contact line speeds). The emergence of the entire branch of such states has recently also been investigated.\cite{MiMT2021prf} Here, we aim at a detailed understanding of stick-slip motion of advancing contact lines on adaptive brush-covered substrates.

Our work is structured as follows. In section~\ref{sec:model} we present a mesoscopic hydrodynamic model that allows us to study the coupled dynamics of the profiles of the brush-liquid and the liquid-gas interfaces under full consideration of absorption, swelling and imbibition processes.
In the subsequent section~\ref{sec:equilibrium} we consider steady sessile droplets on the swollen polymer brush and analytically obtain laws characterizing the macroscopic contact angle and the mesoscopic wetting ridge. Using numerical simulations we then analyze in section~\ref{sec:forced_wetting} the case of forced dynamic wetting, i.e., the inverse Landau-Levich case when a brush-covered plate is slowly pushed into a liquid bath. There, a particular interest lies on the dynamic behaviour of the wetting ridge close to the three-phase contact line. Then, section~\ref{sec:stickslip} focuses on the conditions for the emergence of stick-slip motion. Finally, section~\ref{sec:conclusions} provides a conclusion and outlook.

\section{Mesoscopic hydrodynamic model}\label{sec:model}
We develop a mesoscopic gradient dynamics model that allows for studies of situations involving the dynamics of a three-phase contact line on an adaptive substrate formed by a rigid smooth solid covered by a polymer brush. In particular, we extend a recently presented model by~\citet{ThHa2020epjt} to also capture the dependence of wettability on brush state. The additional incorporation of driving forces then allows us to study forced wetting.

\begin{figure}
	\centering
	\includegraphics{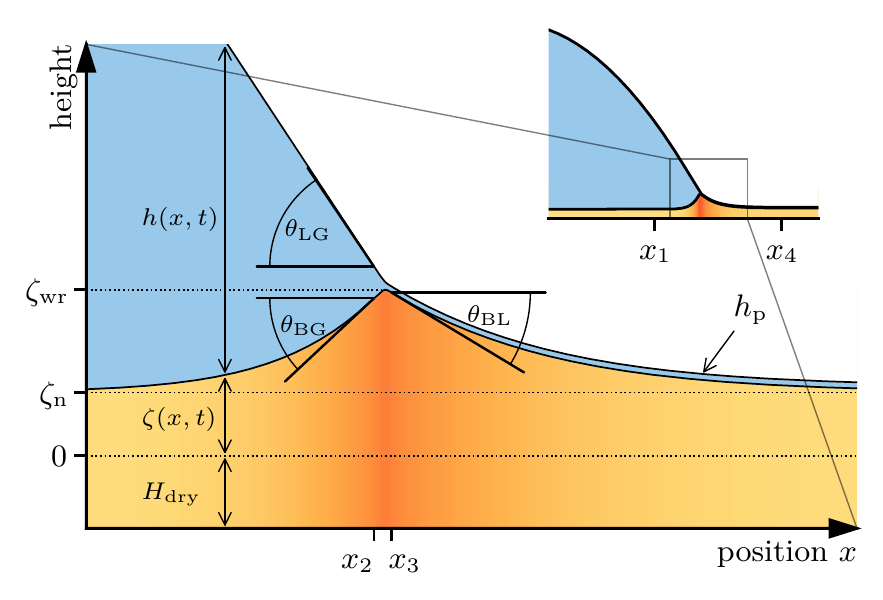}
	\caption{Shown is a sketch of the considered geometry close to a static three-phase contact line for the case of a liquid meniscus on a polymer brush. Such a meniscus may, e.g., be seen as the contact line region of a sessile liquid drop (inset). Indicated are the definitions of height profiles $h$ and $\zeta$, the equilibrium (Neumann) angles $\theta_\mathrm{LG}$, $\theta_\mathrm{BL}$ and $\theta_\mathrm{BG}$ defined at the tip of the wetting ridge. The horizontal dotted lines mark the dry brush thickness $H_\mathrm{dry}$, as well as the equilibrium brush height $\zeta_\mathrm{n}$ and the height of the wetting ridge $\zeta_\mathrm{wr}$, both above $H_\mathrm{dry}$. At positions $x_1$ and $x_4$ far away from the three-phase contact line region (inset) the brush is practically flat. Positions $x_2$ and $x_3$ mark the points of maximal and minimal slope of the brush-liquid interface, i.e., its inflection points. Far from the meniscus the liquid thickness approaches a mesoscopic adsorption layer height $h_\mathrm{p}$.}
	\label{fig:Winkel}
\end{figure}

In particular, we consider a liquid drop, layer or meniscus of height profile $h(\mathbf x, t)$ on a polymer brush of height $H(\mathrm x, t) = H_\mathrm{dry} + \zeta(\mathrm x, t)$. Here, $\mathbf x=(x,y)^T$ are the substrate coordinates, $H_\mathrm{dry}$ denotes the reference height of a completely dry brush and $\zeta(\mathrm x, t)$ is the local increase in brush thickness due to swelling, i.e., it corresponds to an effective height of the liquid contained within the brush (or effective liquid volume per area). The geometry in the case of a liquid meniscus is sketched in Fig.~\ref{fig:Winkel}.

The coupled dynamics of $h$ and $\zeta$ is then described within a gradient dynamics framework\cite{Thie2018csa} -- a common formulation of evolution equations for one-layer thin liquid films and shallow drops.\cite{Mitl1993jcis,Thie2010jpcm} This approach has been expanded to two-field systems as, for example, two-layer liquid films,\cite{PBMT2005jcp,BCJP2013epje} liquid drops/films covered by an insoluble surfactant,\cite{ThAP2016prf} drops on viscoelastic substrates,\cite{HeST2021sm} films of liquid mixtures,\cite{ThTL2013prl} and drops of volatile liquids in a vapor-filled gap.\cite{HDJT2022apa} It also forms a central building block for models for biofilms\cite{TrJT2018sm} and drops of active liquids.\cite{StJT2022sm}

To arrive at a gradient dynamics description for a meniscus on a brush-covered substrate we assume that the region of interest is sufficiently small to be able to neglect inertia, i.e., the dynamics is mainly driven by an underlying  free energy functional $\mathcal{F}[h, \zeta]$ that depends on the drop profile and the brush state.  Allowing for various occurring transport and transfer processes, the dynamical equations for the profiles $h$ and $\zeta$ have the form
\begin{equation}
	\begin{aligned}
		\partial_t h     &= \nabla \cdot \left[\frac{h^3}{3\eta}\,\nabla\frac{\delta\mathcal{F}}{\delta h}\right] \ \, - \ M\left[\frac{\delta\mathcal{F}}{\delta h}-\frac{\delta\mathcal{F}}{\delta\zeta}\right] \ + \ U\partial_x h\\
		\partial_t \zeta &=\nabla \cdot \left[D\zeta\,\nabla\frac{\delta\mathcal{F}}{\delta \zeta}\right] \ - \ M\left[\frac{\delta\mathcal{F}}{\delta \zeta}-\frac{\delta\mathcal{F}}{\delta h}\right] \ + \ U\partial_x \zeta.\label{eq:gradient_model}
	\end{aligned}
\end{equation}
where $U$ is an imposed velocity of the substrate that is drawn out ($U>0$) or pushed into ($U<0$) the bath/drop. For $U=0$, Eqs.~\eqref{eq:gradient_model} have the same form as the ones in \citet{ThHa2020epjt}. The first term on the r.h.s.\ of the first equation describes advective transport within the liquid layer (with the dynamic viscosity $\eta$), driven by the pressure gradient $\nabla (\delta\mathcal{F} / \delta h)$, while the second term describes the loss/gain of liquid via transfer to/from the brush (with rate constant $M$). This absorption process is driven by the pressure difference between liquid and brush. Without transfer, the equation becomes the standard mesoscopic thin-film (lubrication, long-wave) equation for a nonvolatile liquid on a rigid solid substrate.\cite{Genn1985rmp,OrDB1997rmp,StVe2009jpm,BEIM2009rmp,CrMa2009rmp,Thie2010jpcm} The first term on the r.h.s.\ of the second equation describes diffusive transport of liquid within the brush (diffusive imbibition with the diffusion constant $D$) driven by the gradient in chemical potential\footnote{Note that conceptionally it is a chemical potential that drives diffusion, however, here it is literally a pressure as our considered field $\zeta$ is a height and not a particle number.} $\nabla (\delta\mathcal{F} / \delta \zeta)$ while the second term describes the loss [gain] of liquid via transfer to [from] the meniscus. As we assume a nonvolatile liquid, the nonconserved terms in the two equations exactly compensate, i.e., $h+\zeta$ follows a continuity equation.

Assuming a long-wave setting, i.e., small interface slopes,\cite{OrDB1997rmp,CrMa2009rmp} $h^3/3\eta$ and $D\zeta$ are the viscous mobility in the liquid and the diffusive mobility within the brush, respectively. Note that there is no dynamics cross-coupling between drop and brush as we neglect advective transport of liquid within the brush. The driving term has the form of a Galilean transformation from a reference frame $\Sigma$, in which brush and liquid meniscus are at rest, into a frame $\Sigma'$ moving with velocity $U$. In consequence, as long as boundary effects are excluded, in the new reference frame $\Sigma'$ both the liquid film/drop and the substrate are translated horizontally at a constant speed $-U$ without any change in drop or brush profile.
However, incorporating boundary conditions within the moving frame $\Sigma'$, e.g., imposing the film inclination on one boundary of the finite (co-moving) domain, exhibits a driving force onto the film.
This geometry effectively models a liquid drop being pushed over the substrate with a sharp blade (in the reference frame $\Sigma'$ of the blade) or the substrate being pushed into a liquid bath (in the reference frame $\Sigma'$ of the bath), i.e. the dipping-phase of a dip-coating process.
Note that such a permanent external driving force is normally not captured by a free energy functional $\mathcal{F}$. In other words, it represents a nonvariational influence that persistently keeps the system out of equilibrium making time-periodic behaviour possible.

The considered free energy functional is
\begin{equation}
	\mathcal{F}[h,\zeta] = \int \left[ f_\mathrm{cap}(h,\zeta) + \xi_\zeta f_\mathrm{wet}(h, \zeta) + g_\mathrm{brush}(\zeta) \right]\mathrm{d}^2x.
	\label{eq:free_energy}
\end{equation}
It incorporates contributions due to capillarity of the brush-liquid and the liquid-gas interfaces
\begin{equation}
	f_\mathrm{cap}(h,\zeta)=\gamma_\mathrm{bl}(\zeta)\,\xi_\zeta + \gamma\,\xi_{h+\zeta},
\end{equation}
 wettability $\xi_\zeta f_\mathrm{wet}(h, \zeta)$, and the local free energy of the (dry or swollen) brush $g_\mathrm{brush}(\zeta)$. Here, $\gamma$ and $\gamma_\mathrm{bl}(\zeta)$ denote the interface energies of the liquid-gas and brush-liquid interfaces, respectively. In general, both, wetting energy $f_\mathrm{wet}(h, \zeta)$ and brush-liquid interface energy $\gamma_\mathrm{bl}(\zeta)$ will depend on the brush state encoded in $\zeta$. This dependency is necessary to ensure that at very high liquid concentration within the brush, the brush-liquid interface energy tends to zero while the brush-air interface energy (as encoded in $f_\mathrm{wet}$, cf.~\citet{Genn1985rmp,BEIM2009rmp,TSTJ2018l}) tends to $\gamma$. Note that this generalizes the approach in earlier work \cite{ThHa2020epjt} where such $\zeta$-dependencies were not considered.
Further note that the metric factors of the two interfaces,
\begin{equation}
	\xi_\zeta=\sqrt{1+(\nabla\zeta)^2}\quad
	\text{and}\quad
	\xi_{h+\zeta}=\sqrt{1+[\nabla(h+\zeta)]^2},
\end{equation}
enter the interface energies as well as the wetting energy. This ensures consistency with macroscopic relations (see below).

\begin{figure}
	\centering
	\includegraphics{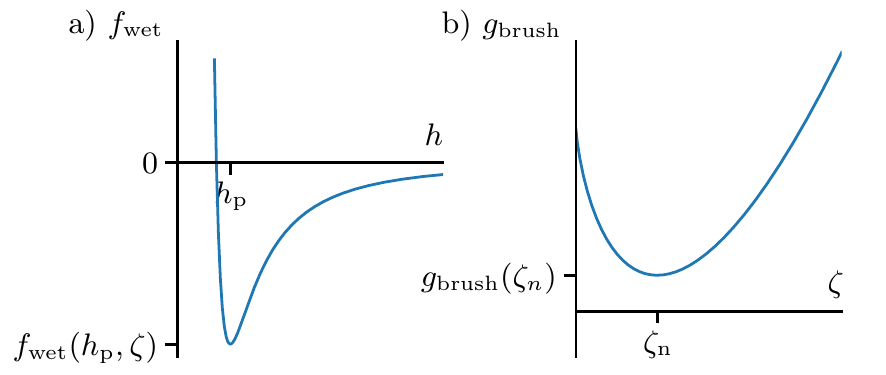}
	\caption{Displayed is the typical qualitative behaviour of (a) the wetting energy $f_{\mathrm{wet}}(h,\zeta)$ as a function of $h$ at fixed $\zeta$ [Eq.~\eqref{eq:wetting_potential}], and (b) the brush energy $g_{\mathrm{brush}}(\zeta)$ as a function of $\zeta$ [Eq.~\eqref{eq:brush-en-area}].}
	\label{fig:energies}
\end{figure}

The employed wetting energy we base on a simple standard form for partially wetting liquids that includes long-range attractive and short-range repulsive contributions, both as power laws\cite{Pism2001pre,Thie2010jpcm} (see \autoref{fig:energies}~(a))
\begin{equation}
	f_\mathrm{wet}(h, \zeta) = A(\zeta)\,\left[\frac{h_\mathrm{p}^3}{5h^5}-\frac{1}{2h^2}\right].\label{eq:wetting_potential}
\end{equation}
Here, we employ a brush state-dependent Hamaker constant $A$ effectively letting the equilibrium contact angle adjust to the swelling state of the brush. For a simple ansatz that fulfils the aforementioned limiting cases, we assume that both, the Hamaker constant $A(\zeta)$ and brush-liquid interface energy $\gamma_\mathrm{bl}(\zeta)$, depend linearly on the polymer concentration in the brush $c$, i.e.,
\begin{equation}
	A(\zeta) = A_0\,c(\zeta) \quad \text{and} \quad \gamma_\mathrm{bl}(\zeta) = \gamma_\mathrm{bl,0}\,c(\zeta),\label{eq:brush_dependencies}
\end{equation}
where the constants $A_0$ and $\gamma_\mathrm{bl,0}$ are the reference values obtained for a dry brush ($\zeta=0$). This in particular ensures that
\begin{equation}
    f_\mathrm{wet}(h, \zeta\to\infty)=0 \quad \text{and} \quad \gamma_\mathrm{bl}(\zeta\to\infty) = 0,
\end{equation}
i.e. the brush-liquid interface turns into a liquid-liquid interface. The implications for the macroscopic brush-air interface are discussed below in section~\ref{sec:consistency}.
The volume fraction of polymer within the brush layer is defined as
\begin{equation}
	c(\zeta) = \frac{H(\zeta=0)}{H(\zeta)} = \frac{H_\mathrm{dry}}{H_\mathrm{dry} + \zeta},
\end{equation}
i.e., it corresponds to the inverse of the local swelling ratio.
Finally, we employ the brush energy\cite{ThHa2020epjt}
\begin{equation}
	g_\mathrm{brush}(\zeta) = \frac{H_\mathrm{dry} k_B T}{\ell_K^3}\left[\frac{\sigma^2}{2 c^2} +
		(1/c - 1) \,\log\left(1-c\right) \right],
	\label{eq:brush-en-area}
\end{equation}
where $T$ is the temperature, $\ell_K$ is the Kuhn length (or the length of a unit cell in the lattice model) and $\sigma$ is the relative grafting density, i.e., the number of grafted chains per unit area. Note that in a simple Alexander-de Gennes geometry of the brush\cite{Genn1991crasi,Alex1977jp,Somm2017m} the grafting density relates to the collapsed brush height via the degree of polymerization $N$ as $H_\mathrm{dry} = \sigma N \ell_K$.

Within the brush energy \eqref{eq:brush-en-area}, the first term accounts for the elastic energy due to the stretching of polymers, whereas the second one represents an entropic contribution as obtained from a Flory-Huggins lattice model for polymer-solvent mixtures.\cite{Flor1953} Independently of all parameters, the brush energy is convex and therefore has one global minimum, see \autoref{fig:energies}~(b). Note that our approach assumes that the liquid content of the brush is vertically homogeneous, i.e., the distribution of liquid in the brush does not depend on the height coordinate.

In summary, compared to~\citet{ThHa2020epjt}, we have generalised the theoretical description in four ways: First, we incorporate dependencies of the brush-liquid interface energy and of the wetting energy on the brush state. Second, for consistency with macroscopic laws (considered in section~\ref{sec:consistency}), the wetting energy is scaled with the metric factor of the brush-liquid interface. Third, we incorporate a permanent driving force to be able to investigate dip-coating processes. Fourth, we improve the energy functional by using the exact metric factors resulting in the exact curvature in the time-evolution equations.\cite{GaRa1988ces,Snoe2006pf} It has been shown that such improved representations of the underlying energy functional are more important for a correct description of the physical behaviour than the details of the dynamics even though the resulting model is not asymptotically exact\cite{BoTH2018jfm} (also see section~3 of the review by~\citet{Thie2018csa}). Naturally, for small inclinations $\nabla \zeta$ and $\nabla h$, the metric factors can be Taylor-expanded up to second order to arrive at a standard long-wave approximation of both, mobilities and energies \cite{OrDB1997rmp,CrMa2009rmp} as used by~\citet{ThHa2020epjt}.

\section{Equilibrium states}\label{sec:equilibrium}
\subsection{Grand potential and mechanical analogue}
We start by analysing the variational case ($U=0$), where the gradient dynamics structure of Eqs.~\eqref{eq:gradient_model} implies a continuous decrease of the free energy, $\mathrm{d} \mathcal{F} / \mathrm{d}t \leq 0$ (see Appendix~\ref{sec:dissipation_derivation}). With other words, $\mathcal{F}$ is a Lyapunov functional and the system always approaches a steady state corresponding to a minimum of $\mathcal{F}$ under the constraint of an imposed total liquid volume. Then, all equilibria correspond to minima of the grand potential
\begin{equation}
	\mathcal{G}[h,\zeta] = \int \left[ f_\mathrm{cap} + \xi_\zeta f_\mathrm{wet} + g_\mathrm{brush} - P(h+\zeta) \right]\,\mathrm{d}^2x\label{eq:F_P}
\end{equation}
where $P$ is a Lagrange multiplier to ensure volume conservation, and equilibria fulfil the corresponding Euler-Lagrange equations
$\delta\mathcal{F}/\delta h=P$ and $\delta\mathcal{F}/\delta \zeta=P$.

As the integrand of $\mathcal{G}$ only depends on the fields $h$ and $\zeta$ and their first spatial derivatives, in the case of a one-dimensional substrate (1D),\footnote{With other words, we consider two-dimensional (2D) droplets, see inset of \autoref{fig:Winkel}, that may be seen as cross sections of liquid ridges in 3D that are translation-invariant in the transverse direction. Also in a dip coating geometry a 2D slice of a transversely translation-invariant configuration is considered.} there exists a strong analogue to classical Lagrangian and Hamiltonian mechanics: The functional $\mathcal{G}$ then corresponds to an action and its integrand to a Lagrangian $\mathcal{L}$, i.e., $\mathcal{G}=\int\mathcal{L}\,{\d}x$.
The fields $h(x)+\zeta(x)$ and $\zeta(x)$ are generalised coordinates, and the spatial coordinate $x$ becomes time. In consequence, the determination of steady interface profiles corresponds to solving the coupled generalised Newton equations
\begin{align}
	\frac{\delta \mathcal{G}}{\delta h} =     &-\gamma\frac{\partial_{xx}(h+\zeta)}{\xi_{h+\zeta}^{3}} + \xi_\zeta\partial_h f_\mathrm{wet}-P=0\label{eq:dFdh}\\
	\frac{\delta \mathcal{G}}{\delta \zeta} = &-\gamma\frac{\partial_{xx}(h+\zeta)}{\xi_{h+\zeta}^{3}} - \partial_x \cdot \left[ (\gamma_\mathrm{bl} + f_\mathrm{wet})\frac{\partial_x\zeta}{\xi_\zeta}\right]\nonumber\\
	&+ \xi_\zeta \partial_\zeta (\gamma_\mathrm{bl} + f_\mathrm{wet}) + \partial_\zeta g_\mathrm{brush}-P=0.\label{eq:dFdz}
\end{align}

Further employing the analogy, we define the appropriate generalised momenta $p_{h+\zeta}$ and $p_\zeta$
\begin{align}
	p_{h+\zeta} &:= \frac{\partial \mathcal{L}}{\partial [\partial_{x} (h+\zeta)]} = \gamma \frac{\partial_{x} (h+\zeta)}{\xi_{h+\zeta}}\label{eq:ph}\\
	p_\zeta     &:= \frac{\partial \mathcal{L}}{\partial (\partial_{x} \zeta)} = (\gamma_\mathrm{bl}+f_\mathrm{wet})\frac{\partial_{x} \zeta}{\xi_\zeta}\label{eq:pzeta}
\end{align}
and determine the first integral
\begin{equation}
	\begin{split}
		\mathcal{H}&=p_{h+\zeta}\partial_{x} (h+\zeta) + p_\zeta \partial_{x}\zeta-\mathcal{L}\\&=-\frac{\gamma}{\xi_{h+\zeta}} -\frac{\gamma_\mathrm{bl} + f_\mathrm{wet}}{\xi_{\zeta}}-g_\mathrm{brush}+P(h+\zeta),\label{eq:Hamiltonian}
	\end{split}
\end{equation}
that corresponds to the Hamiltonian. In consequence, $E=-\mathcal{H}$ is a constant energy density, which we can use to describe the balance of horizontal forces across the contact line. 
\subsection{Limiting cases and consistency condition}\label{sec:consistency}
\subsubsection{Contact angles}

Next, we consider a 2D slice through a straight liquid meniscus as depicted in \autoref{fig:Winkel}. Then, the angles between local tangents to the brush-liquid and liquid-gas interface and the horizontal, $\theta_\mathrm{bl}$ and $\theta_\mathrm{lg}$, respectively, are directly related to the metric factors via
\begin{equation}\label{eq:thet-metric}
	\begin{aligned}
		\cos\theta_\mathrm{lg} &= \frac{1}{\xi_{h+\zeta}},                    &
		\sin\theta_\mathrm{lg} &= \frac{\partial_x (h+\zeta)}{\xi_{h+\zeta}},\\
		\cos\theta_\mathrm{bl} &= \frac{1}{\xi_{\zeta}},\quad\text{and}       &
		\sin\theta_\mathrm{bl} &= \frac{\partial_x \zeta}{\xi_{\zeta}}.
	\end{aligned}
\end{equation}
As there is no trivial definition of equilibrium contact angles in the mesocopic picture,\cite{TMTT2013jcp} here, we introduce them as extremal values of $\theta_{\mathrm{lg}}$ and $\theta_{\mathrm{bl}}$, i.e., as local steepest slopes. Namely, considering the geometry in the main panel of \autoref{fig:Winkel} we use
\begin{equation}\label{eq:thet-metric2}
	\theta_{\mathrm{LG}}=\min(\theta_{\mathrm{lg}}), \quad\theta_{\mathrm{BL}}=\max(\theta_{\mathrm{bl}})\quad\text{and}\quad\theta_{\mathrm{BG}}=\min(\theta_{\mathrm{bl}}).
\end{equation}
Note that due to the identification with the metric factors, inclination and contact angles are signed values in the range $[-\pi/2,\pi/2]$
where a positive [negative] angle corresponds to a positive [negative] slope of the corresponding interface profile.
\subsubsection{Global Young law}\label{sec:young}
As the wetting energy~\eqref{eq:wetting_potential} and its derivatives approach zero at large film height $h$, for a macroscopic droplet far away from the contact line region (e.g., at position $x_1$ in Fig.~\ref{fig:Winkel}) Eq.~\eqref{eq:dFdh} becomes
\begin{equation}
	P = -\gamma \kappa \quad \text{with the curvature}\quad \kappa = \frac{\partial_{xx} (h+\zeta)}{\xi_{h+\zeta}^3}.
\end{equation}
Therefore, macroscopically,  $P$ corresponds to the Laplace pressure and in the studied 2D slice the liquid-gas interface approaches the shape of a circurlar arc. Next, we consider a very large drop, i.e., $\kappa\rightarrow 0$, at equilibrium on a brush. Aiming at a global picture that ignores the details of the configuration in the contact line region, we consider positions $x_1$ and $x_4$ far from the contact line (Fig.~\ref{fig:Winkel}). There, the brush approaches a flat state, i.e., all spatial derivatives of the brush profile approach zero.
In consequence, within the drop (at~$x_1$) Eqs.~\eqref{eq:dFdh} and \eqref{eq:dFdz} simplify to
\begin{equation}
	P = 0
	\quad \text{and} \quad
	\partial_\zeta (\gamma_\mathrm{bl} + g_\mathrm{brush}) = 0,
	\label{eq:zeta_n-x1}
\end{equation}
as the Derjaguin (or disjoining) pressure $- \partial_h f_\mathrm{wet}$ vanishes within the droplet due to large $h$ (as discussed above). The brush assumes a height $\zeta_\mathrm{d}$ given by the minimum of $\gamma_\mathrm{bl} + g_\mathrm{brush}$ with respect to $\zeta$ (Fig.~\ref{fig:energies}~(b)). In contrast, far away from the drop (at $x_4$), the Derjaguin pressure matters and Eqs.~\eqref{eq:dFdh} and \eqref{eq:dFdz} become
\begin{equation}
	\partial_h f_{\mathrm{wet}} = 0
	\quad \text{and} \quad
	\partial_\zeta (\gamma_\mathrm{bl} + f_\mathrm{wet} + g_\mathrm{brush}) = 0,
	\label{eq:zeta_n-x4}
\end{equation}
i.e., the minimum of the wetting energy w.r.t.\ $h$ gives the adsorption layer height  $h_p$ (Fig.~\ref{fig:energies}~(a)) while the minimum of $\gamma_\mathrm{bl} + f_\mathrm{wet} + g_\mathrm{brush}$  w.r.t.\ $\zeta$ gives the brush height $\zeta_\mathrm{p}$ far away from the drop.

Using this result and Eq.~\eqref{eq:Hamiltonian} we evaluate the asymptotic energy density $E=-\mathcal{H}$ inside and outside of the drop
\begin{align}
	E(x_1) &= \gamma\cos{\theta_{LG}}+\gamma_\mathrm{bl}(\zeta_\mathrm{d})+g_\mathrm{brush}({\zeta_\mathrm{d}})\label{eq:E(x1)}\\
	E(x_4) &= \gamma+\gamma_\mathrm{bl}(\zeta_\mathrm{p})+f_\mathrm{wet}(h_\mathrm{p}, \zeta_\mathrm{p})+g_\mathrm{brush}({\zeta_\mathrm{p}})\label{eq:E(x4)}.
\end{align}
We set the two energies equal, identify the occurring contact angle $\theta_\mathrm{LG}$ as a macroscopic angle $\theta_\mathrm{Y}$ and obtain an equivalent to Young's law\cite{Youn1805ptrs}\begin{equation}
	\begin{aligned}
		\gamma\cos\theta_\mathrm{Y} = &\ \gamma+f_\mathrm{wet}(h_\mathrm{p}, \zeta_\mathrm{p})\\
		&+ \gamma_\mathrm{bl}(\zeta_\mathrm{p}) + g_\mathrm{brush}(\zeta_\mathrm{p}) -  \gamma_\mathrm{bl}(\zeta_\mathrm{d}) - g_\mathrm{brush}(\zeta_\mathrm{d})\label{eq:young}
	\end{aligned}
\end{equation}
expressed in mesoscale quantities.
Note that the last four addends are solely due to the adaptive character of the brush and are not present in the classical result known from rigid non-adaptive substrates. However, as per the bulk equilibrium condition \eqref{eq:zeta_n-x1} the correction to Young's law is of second order in the height difference $\zeta_\mathrm{d}-\zeta_\mathrm{p}$ and therefore potentially small.

A similar consideration based solely on macroscopic quantities, see Appendix~\ref{sec:macroscopic}, gives an identically amended macroscopic Young's law
\begin{equation}
	\begin{aligned}
		\gamma\cos\theta_\mathrm{Y} = &\ \gamma_\mathrm{bg}(\zeta_p) - \gamma_\mathrm{bl}(\zeta_\mathrm{p})\\
		&+ \gamma_\mathrm{bl}(\zeta_\mathrm{p}) + g_\mathrm{brush}(\zeta_\mathrm{p}) - \gamma_\mathrm{bl}(\zeta_\mathrm{d}) - g_\mathrm{brush}(\zeta_\mathrm{d}).\label{eq:young_macro}
	\end{aligned}
\end{equation}
Note that here we added the terms $\gamma_\mathrm{bl}(\zeta_p)-\gamma_\mathrm{bl}(\zeta_p)=0$ to resemble the form of the mesoscopic Young law, namely Eq.~\eqref{eq:young}.
Comparing both versions of Young's law, we identify a consistency relation between the macroscopic and mesoscopic global picture
\begin{equation}
	f_\mathrm{wet}(h_\mathrm{p}, \zeta_p)=\gamma_\mathrm{bg}(\zeta_\mathrm{p})-\gamma_\mathrm{bl}(\zeta_\mathrm{p})-\gamma.\label{eq:consistency}
\end{equation}
The relation determines how the dependencies of interface tensions and wetting potential on the brush state $\zeta$ are related.
Together with our assumptions on the brush state-dependency of the mesoscopic wetting potential $f_\mathrm{wet}(h, \zeta)$ and the brush-liquid interface energy $\gamma_\mathrm{bl}(\zeta)$ in Eq.~\eqref{eq:brush_dependencies} this implies that the macroscopic brush-gas interface energy scales as
\begin{equation}
    \gamma_\mathrm{bg}(\zeta) = (\gamma_\mathrm{bg,0} - \gamma) c(\zeta) + \gamma,
\end{equation}
i.e., it interpolates between the dry brush-gas interface energy $\gamma_\mathrm{bg,0}$, and the energy of a liquid-gas interface in case of a fully swollen brush, as expected.

\subsubsection{Local Neumann law}\label{sec:neumann}
Having established the global picture for a large drop on a brush within a mesoscale and a macroscale description, we next retain the mesoscopic view and consider the local picture of the contact line region. Equating the brush and film pressure (Eqs.~\eqref{eq:dFdh} and \eqref{eq:dFdz}), we obtain for the curvature of the brush-liquid interface
\begin{equation}
	\partial_x \cdot (\gamma_\mathrm{bl} + f_\mathrm{wet})\frac{\partial_x\zeta}{\xi_\zeta^{3}} = \xi_\zeta \partial_\zeta (\gamma_\mathrm{bl} + f_\mathrm{wet}) + \partial_\zeta g_\mathrm{brush}-\xi_\zeta\partial_h  f_\mathrm{wet}.\label{eq:KrBrush}
\end{equation}
Furthermore, Eq.~\eqref{eq:dFdh} indicates that the Derjaguin pressure term $-\xi_\zeta \partial_h f_\mathrm{wet}$ has to balance the (strong) curvature of the liquid profile in the contact line region. Thus, generally, the r.h.s.\ of Eq.~\eqref{eq:KrBrush} is non-zero, i.e., the curvature of the brush profile can not vanish. We expect that the brush forms some type of wetting ridge as known from elastic substrates.\cite{AnSn2020arfm}

To investigate the ridge we consider the inflection points $x_2$ and $x_3$ of the brush profile as indicated in Fig~\ref{fig:Winkel}. Close to three-phase contact, the Derjaguin pressure will dominate Eq.~\eqref{eq:KrBrush} over a small length scale of the order of the height $h_p$. Hence, the distance between $x_2$ and $x_3$ will scale with $h_p/\theta_{LG}$. Outside this small region, the wetting ridge will decay to the equilibrium brush height $\zeta_\mathrm{p}$ (or $\zeta_\mathrm{d}$) and its shape is governed by the differential equation~\eqref{eq:KrBrush}.

To further advance, we next focus on situations where the wetting ridge is large compared to the adsorption layer height and assume, motivated by the previous argument, that the brush heights at $x_2$ and $x_3$ well approximate the peak height $\zeta_\mathrm{wr}$. Also assuming that the inclination angle of the liquid-gas interface at $x_2$ equals $\theta_{LG}$, we again use Eq.~\eqref{eq:Hamiltonian} to evaluate the energy $E=-\mathcal{H}$, this time at $x_2$ and $x_3$. We obtain
\begin{align}
	E(x_2) &=\gamma\cos{\theta_\mathrm{LG}}+\gamma_\mathrm{bl}\cos{\theta_\mathrm{BL}}+g_\mathrm{brush}({\zeta_\mathrm{wr}})\label{eq:E(x2)}\\
	E(x_3) &=[\gamma+\gamma_\mathrm{bl}+f_\mathrm{wet}(h_\mathrm{p},\zeta_\mathrm{wr})]\cos\theta_\mathrm{BG}+g_\mathrm{brush}({\zeta_\mathrm{wr}}).\label{eq:E(x3)}
\end{align}
Equating the two expressions yields the horizontal component of the Neumann law, namely,
\begin{equation}
	\gamma\cos\theta_\mathrm{LG}+\gamma_\mathrm{bl}\cos{\theta_\mathrm{BL}}=[\gamma+\gamma_\mathrm{bl}+f_\mathrm{wet}(h_\mathrm{p},\zeta_\mathrm{wr})]\cos\theta_\mathrm{BG}.	\label{eq:Neumann_hor}
\end{equation}

To obtain the vertical component of the law, we consider conservation of the total generalised momentum $p_\zeta+p_{h+\zeta}$ (Eqs.~\eqref{eq:ph} and~\eqref{eq:pzeta}) across the contact line region, i.e., from  $x_2$ to $x_3$.\footnote{In the mechanical analogue the approach of the brush-liquid and liquid-gas interfaces and subsequent fusion at the contact line into the brush-gas interface corresponds to a completely inelastic collision of two particles. The wetting energy takes the role of an internal energy like a spring that snaps in on close approach.}
Also using Eqs.~\eqref{eq:thet-metric} and~\eqref{eq:thet-metric2}, this directly gives
\begin{equation}
	\gamma\sin\theta_\mathrm{LG}+\gamma_\mathrm{bl}\sin\theta_\mathrm{BL}=[\gamma+\gamma_\mathrm{bl}+f_\mathrm{wet}(h_p, \zeta_\mathrm{wr})]\sin\theta_\mathrm{BG}. \label{eq:Neumann_ver}
\end{equation}
Note that the brush energy does not enter as we have used $\zeta(x_2)\approx\zeta(x_3)\approx\zeta_\mathrm{wr}$.

The obtained mesoscopic Neumann law, Eqs.~\eqref{eq:Neumann_hor} and~\eqref{eq:Neumann_ver}, corresponds to the usual macroscopic form when taking the consistency condition~\eqref{eq:consistency} into account, see Appendix~\ref{sec:macroscopic}. We emphasise that the full agreement critically depends on the above discussed scaling of the wetting energy with the metric factor of the brush-liquid interface.

We therefore conclude that in the limit of a wetting ridge that is large compared to the adsorption layer height $h_\mathrm{p}$, its shape is governed by the Neumann law expressed in mesoscale quantities. This is in line with similar findings for elastic substrates, where~\citet{PAKZ2020prx} have shown that the classic macroscopic Neumann law applies at the wetting ridge as substrate elasticity is negligible when approaching the contact line sufficiently closely.\cite{AnSn2020arfm}

\subsubsection{Height of the wetting ridge}
In a similar way one can equate the energy $E$ underneath the liquid far away from the wetting ridge \eqref{eq:E(x1)} and close to the peak of the wetting ridge \eqref{eq:E(x2)}. As a result we obtain the difference in brush energy at the peak and for the flat brush
\begin{equation}
	g_\mathrm{brush}({\zeta_\mathrm{wr}})-g_\mathrm{brush}({\zeta_\mathrm{d}})=\gamma_\mathrm{bl}(\zeta_\mathrm{d})-\gamma_\mathrm{bl}(\zeta_\mathrm{wr})\cos{\theta_\mathrm{BL}}.\label{eq:ridge_height_def}
\end{equation}
This is a transcendent equation for the wetting ridge height $\zeta_\mathrm{wr}$ at the contact line of large drops at equilibrium. Next, we derive an explicit expression for $\Delta \zeta = \zeta_\mathrm{wr} - \zeta_\mathrm{d}$ by expanding Eq.~\eqref{eq:ridge_height_def} up to the second order in small values of $\Delta \zeta$ and $\theta_\mathrm{BL}$.
\begin{equation}
	\begin{aligned}
		&\partial_\zeta g_\mathrm{brush}(\zeta_\mathrm{d}) \Delta \zeta + \partial_{\zeta\zeta} g_\mathrm{brush}(\zeta_\mathrm{d}) \frac{(\Delta \zeta)^2}{2}\\&= -\partial_\zeta \gamma_\mathrm{bl}(\zeta_\mathrm{d})\Delta \zeta - \partial_{\zeta\zeta} \gamma_\mathrm{bl}(\zeta_\mathrm{d})\frac{(\Delta \zeta)^2}{2} + \gamma_\mathrm{bl}(\zeta_\mathrm{d})\frac{\theta_\mathrm{BL}^2}{2}.
	\end{aligned}
\end{equation}
The first order contributions cancel due to Eq.~\eqref{eq:zeta_n-x1}. Hence, the height of the wetting ridge is approximately given by
\begin{equation}
	\Delta \zeta \approx \theta_\mathrm{BL} \sqrt{\frac{\gamma_\mathrm{bl}(\zeta_\mathrm{d})}{\partial_{\zeta\zeta} [g_\mathrm{brush}(\zeta_\mathrm{d}) +  \gamma_\mathrm{bl}(\zeta_\mathrm{d})]}}.\label{eq:ridge_height}
\end{equation}
The angle $\theta_\mathrm{BL}$ may be expressed via the Neumann relations given in the previous section.

Note that at no point of the above derivations the brush energy $g_\mathrm{brush}$ has to be specified, i.e., all results are generic for rather general adaptive substrates. For purely elastic substrates the square root in Eq.~\eqref{eq:ridge_height} corresponds to the elastocapillary length $\ell_\mathrm{ec}$.\cite{HeST2021sm}

\subsection{Approaching the limiting cases}\label{sec:NumerikGG}
Next we consider how the derived relations for large steady drops are approached when the volume of drops of finite size is increased. This is done numerically employing direct time simulations and pseudo-arclength path continuation using the C++ finite element library \textit{oomph-lib}.\cite{HeHa2006} In all calculations a nondimensional version of the dynamic model~\eqref{eq:gradient_model} is used. Note that we present results for both variants of the presented mesoscopic model, the full-curvature formulation and the long-wave approximation mentioned at the end of section~\ref{sec:model}. Details of the nondimensionalisation and the long-wave approximation are given in Appendix~\ref{sec:nondim_and_longwave}. For the sake of readability from here on we use the dimensionless formulation without the tildes.

As the boundary conditions (BC) for the numerical analysis of the steady states we employ homogeneous Neumann conditions
\begin{equation}
	\partial_x h = \partial_x \zeta = \partial_x \frac{\delta \mathcal{F}}{\delta h} = \partial_x \frac{\delta \mathcal{F}}{\delta \zeta} = 0
\end{equation}
at both boundaries of the 1D domain. This ensures, in particular, that for $U = 0$ there is no liquid flux through the domain boundaries, i.e., the total volume is conserved. For the forced wetting case $U > 0$ we need to amend the BC to retain the global balance of liquid volume as discussed in Appendix~\ref{sec:numerics}. In order to reach the limit of very large droplets $P\to 0$, we exploit that it is sufficient to only simulate the vicinity of the contact line region. We therefore replace the homogeneous Neumann condition $\partial_x h(x=0) = 0$ with an imposed interface slope $\partial_x h(x=0) = -\theta_\mathrm{0}$ at the boundary associated with the drop. When we increase $\theta_0$ letting it approach the equilibrium angle $|\theta_Y|$, the curvature of the bulk part of the drop decreases and approaches zero, i.e., $P\to 0$, and an infinitely large drop is approached.

\begin{figure}[!htb]
	\centering
	\includegraphics{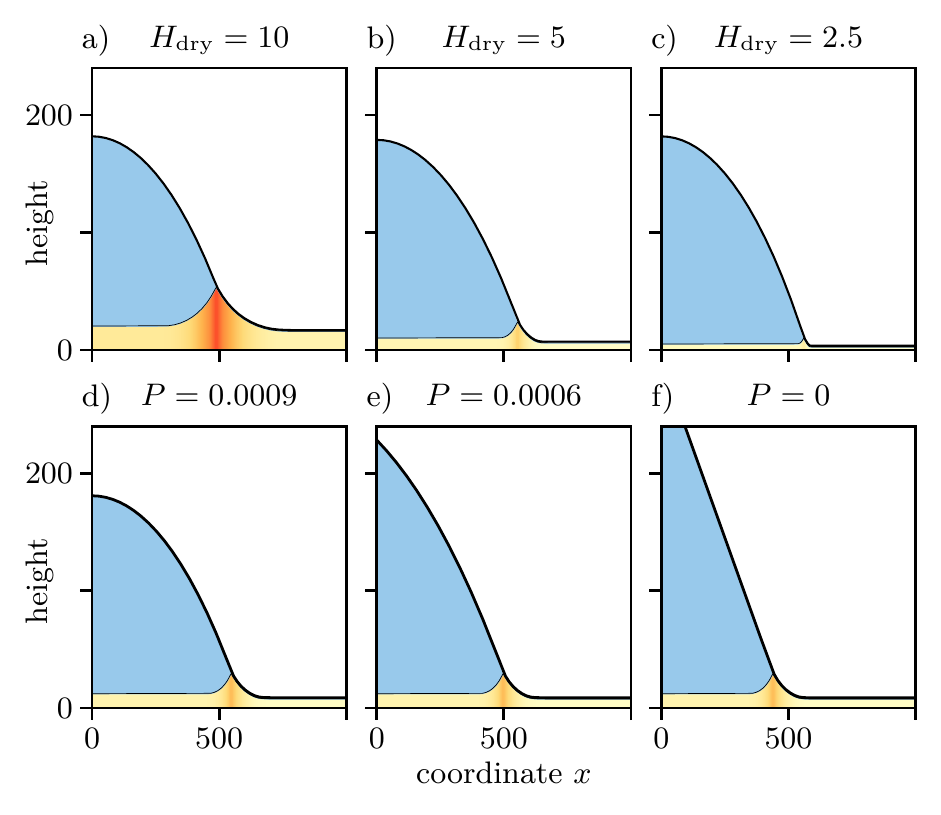}
	\caption{Steady state brush profiles for different values of (a-c) the dry brush height $H_\mathrm{dry}$  and (d-f) the drop size, demonstrating that the ridge scales with brush height but not drop size. The drop size is adjusted via the slope of the drop profile at $x=0$ that controls the Laplace pressure $P$. In particular, $P=0$ corresponds to an infinitely large drop. Note the small but significant difference $\zeta_\mathrm{d}-\zeta_\mathrm{p}$ between the brush heights inside and outside the drop. The remaining parameters are $T=0.05$, $\sigma=0.3$, and $\gamma_\mathrm{bl}=0.4$. The Laplace pressure in panels (a-c) is $P=\num{9e-4}$ and the dry brush height in panels (d-f) is $H_\mathrm{dry}=6$. The results are obtained from finite element simulations of the full model, Eqs.~\eqref{eq:full_model1}--\eqref{eq:full_model3}. A quantitative characterisation of the wetting ridge is provided in Figs.~\ref{fig:statics_variation_H_dry} and~\ref{fig:statics_variation_P}.}
	\label{fig:steady_profiles}
\end{figure}

In Fig.~\ref{fig:steady_profiles} we depict equilibrium states obtained in simulations for different dry brush heights and drop volumes. In this way we obtain an overview of possible shapes of the wetting ridge. Panels (a)-(c) only differ in the dry height of the brush $ H_\mathrm{dry} = \sigma  \ell$, whereas panels (d)-(f) have identical brushes with $H_\mathrm{dry}=6$ while drop size varies from (d) small to (f) the limit of an infinitely large drop. The dry brush height $H_\mathrm{dry}=\sigma N \ell_K$ is varied by controlling the polymer chain length $N \ell_K$ at fixed grafting density $\sigma=0.3$.
Careful inspection of Fig.~\ref{fig:steady_profiles} shows that there is always a significant difference $\zeta_\mathrm{d}-\zeta_\mathrm{p}$ of the brush swelling inside and outside the drop. Numerically, we find that in all panels the brush inside the drop is swollen to approximately $\SI{300}{\percent}$ of its dry height. Outside the drop, the brush is swollen only to approximately \SI{200}{\percent}. Here, we have used $1/c$ to calculate the swelling ratio.

\begin{figure}[!tb]
    \centering
    \includegraphics{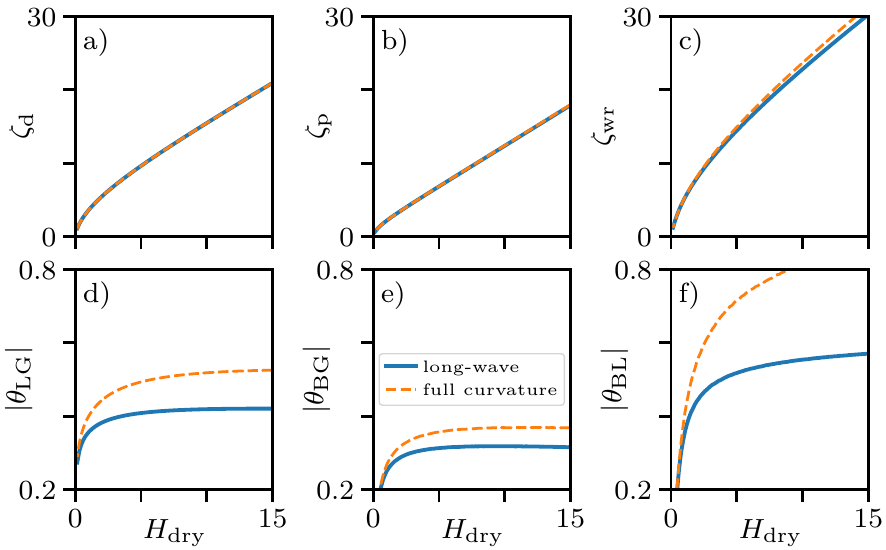}
    \caption{Shown are geometric measures of the equilibrium configuration in the contact line region in dependence of the dry brush height $H_\mathrm{dry}$. Given are the brush heights  (a) below the drop $\zeta_\mathrm{d}$, (b) outside the drop $\zeta_\mathrm{d}$, (c) the wetting ridge height $\zeta_\mathrm{wr}$, and (d-f) the three Neumann angles $\theta_\mathrm{LG}$, $\theta_\mathrm{BG}$, and $\theta_\mathrm{BL}$. For the three angles we compare the full-curvature (dashed lines) and long-wave (solid lines) results. All other parameters are as in Fig.~\ref{fig:steady_profiles}~(a-c).
    }
    \label{fig:statics_variation_H_dry}
\end{figure}

This is further quantified in Fig.~\ref{fig:statics_variation_H_dry} that provides various geometric measures of the equilibrium configuration in the contact line region in dependence of the dry brush height $H_\mathrm{dry}$. There we find that the equilibrium (swollen) brush heights inside ($\zeta_\mathrm{d}$) and outside ($\zeta_\mathrm{p}$) the drop, and the wetting ridge height $\zeta_\mathrm{wr}$ all increase approximately linearly with $H_\mathrm{dry}$ while the Neumann angles strongly [weakly] increase at small [large] $H_\mathrm{dry}$.

\begin{figure}[!tb]
    \centering
    \includegraphics{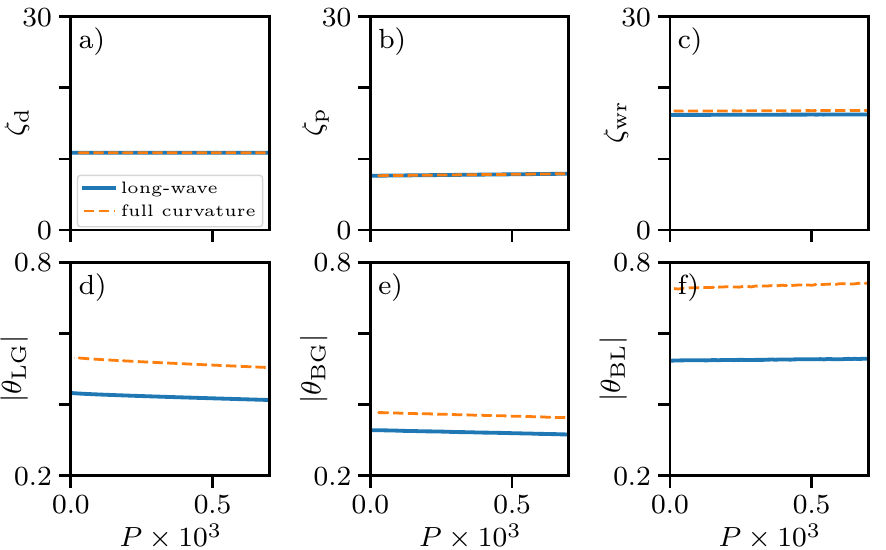}
    \caption{Shown are geometric measures of the equilibrium configuration in the contact line region in dependence of the drop volume as controlled by the Laplace pressure $P$. The limit $P=0$ corresponds to an infinitely large droplet. The shown quantities in panels (a-f) and linestyles are as in Fig.~\ref{fig:statics_variation_H_dry}. All other parameters are as in Fig.~\ref{fig:steady_profiles}~(d-f).}
    \label{fig:statics_variation_P}
\end{figure}

In contrast to the brush height, the drop size has only a very small influence on the swelling state and the wetting ridge. Namely, in Fig.~\ref{fig:steady_profiles}~(d-f) one discerns nearly no change in $\zeta_\mathrm{d,p,wr}$ upon varying the Laplace pressure $P$. This visual impression is quantitatively supported  by Fig.~\ref{fig:statics_variation_P}, where we display the geometric measures as obtained from simulated equilibrium drops at different $P$.

Note that the shape and size of the wetting ridge are also impacted by some of the parameters not discussed here. As Eq.~\eqref{eq:ridge_height} already indicates, the ridge height is governed by an interplay of the interface energy and the brush forces. Thus, the values of the interface energies as well as the scale of the brush energy, namely, the dimensionless temperature parameter $T$, are equally relevant. While the above considerations are entirely static, the situation may become more intricate when the system is taken outside of equilibrium. This will be addressed in the subsequent sections.

Besides demonstrating the influence of drop size and dry brush height on the wetting ridge, Figs.~\ref{fig:statics_variation_H_dry} and~\ref{fig:statics_variation_P} also compare the results obtained with the full-curvature and long-wave variants of the model (cf.~Section~\ref{sec:model}). In general, for the chosen values of the interface energies the differences are relatively small.  The Neumann angles tend to have slightly lower values for the long-wave model. Differences are largest for thin brushes. The brush swelling is practically identical in the two cases. Moreover, Fig~\ref{fig:statics_variation_P} suggests that the effect is rather independent of the drop size.
In consequence, we only use the model in long-wave approximation for all remaining numerical calculations.

\begin{figure}
    \centering
    \includegraphics[width=0.49\textwidth]{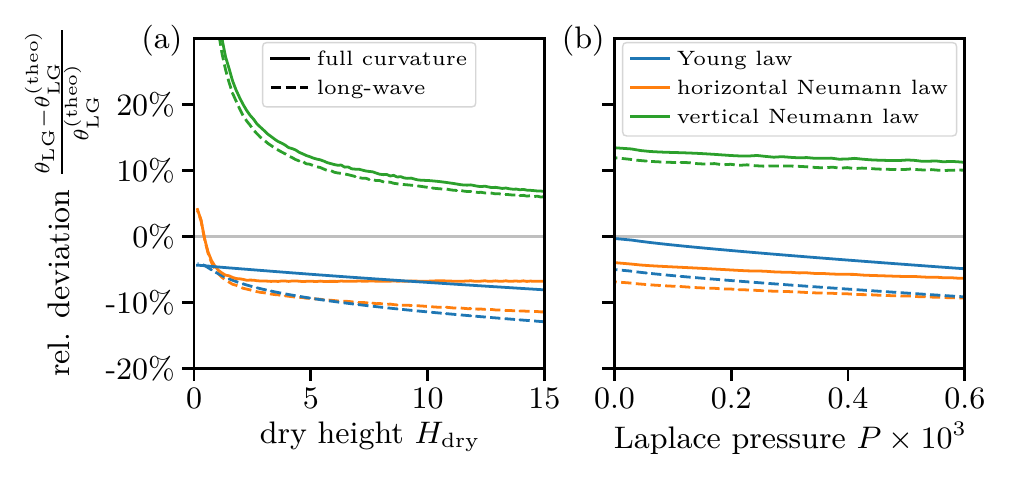}
    \caption{Relative deviation $(\theta_\mathrm{LG}-\theta_\mathrm{LG}^\mathrm{(theo)})/(\theta_\mathrm{LG}^\mathrm{(theo)})$ between the measured liquid-gas contact angle $\theta_\mathrm{LG}$ and the predicted contact angle $\theta_\mathrm{LG}^\mathrm{(theo)}$ using either the Young or the Neumann laws, namely, Eqs.~\eqref{eq:young},~\eqref{eq:Neumann_hor},~\&~\eqref{eq:Neumann_ver},. The result quantifies how well the relations are fulfilled in dependence of (a) the dry brush height $H_\mathrm{dry}$ and (b) the Laplace pressure $P$ (encoding the drop volume). Solid lines correspond to calculations based on the full-curvature version of the model and the long-wave results are shown as dashed lines. The simulation data corresponds to the one shown in Figs.~\ref{fig:statics_variation_H_dry}~\&~\ref{fig:statics_variation_P}. In particular panel (a) uses $P=\num{9e-4}$ and panel (b) uses $H_\mathrm{dry}=6$.}
    \label{fig:Neumann_deviation}
\end{figure}

Finally, we assess how well the Young and Neumann laws are fulfilled by the simulation results by comparing the measured value of the liquid-gas contact angle $\theta_\mathrm{LG}$ to the predicted value $\theta_\mathrm{LG}^\mathrm{(theo)}$ that we obtain from either the Young or the Neumann laws, Eqs.~\eqref{eq:young},~\eqref{eq:Neumann_hor},~\&~\eqref{eq:Neumann_ver}, and the measured values of the brush angles. The relative deviation is shown in Fig.~\ref{fig:Neumann_deviation} for both the full-curvature and the long-wave version of the model. There, the results are shown in dependence of the parameters varied in Figs.~\ref{fig:statics_variation_H_dry}~\&~\ref{fig:statics_variation_P}, namely, $H_\mathrm{dry}$ and $P$.

We find that in the full-curvature formulation the Young law is perfectly fulfilled in the limit of large droplets ($P=0$), whereas the long-wave approximation results in a slightly lower contact angle. For decreasing drop size (increasing Laplace pressure) the results for the Young law increasingly deviate. For the Neumann law a deviation is already notable in the full-curvature case for $P=0$: there, the horizontal [vertical] Neumann balance deviates by about \SI{4}{\percent} [more than \SI{10}{\percent}].

We suggest that the latter deviations are due to the mesoscopic modelling as they are connected to the intricacies hidden in the analytic derivation of the mesoscopic Neumann law. On the one hand, we have assumed that the wetting potential can be neglected at the inner inflection point $x_2$ of the wetting ridge, i.e., that it does not appear in Eq.~\eqref{eq:E(x2)}. Indeed the local film height is not sufficiently large, resulting in a contribution to the observed deviation in the horizontal Neumann balance. On the other hand, the conservation of the total generalised momentum $p_\zeta+p_{h+\zeta}$ is also only exact in the limit $h_p/\zeta_{\mathrm{wr}}\rightarrow 0$.\footnote{In the mechanical analogue the inelastic collision takes place in an external force field (corresponding to $\partial_\zeta g_\mathrm{brush}$), such that the total momentum is only approximately conserved during a relatively short collision time.} The finite height of the wetting ridge results in the observed deviation in the vertical Neumann balance. 
Note that as the convergence towards the limit $h_p/\zeta_{\mathrm{wr}}\rightarrow 0$ is very slow, and reducing the adsorption layer height $h_p$ increases the numerical effort significantly, here, we do not go beyond the choice of parameters employed above.

\section{Forced wetting}\label{sec:forced_wetting}
\subsection{General remarks}
Having analysed the case of drops and menisci at equilibrium, we next consider the case of forced wetting of a brush-covered substrate. Namely, we activate the additional advection terms in Eqs.~\eqref{eq:gradient_model} by considering substrate velocities $U>0$, i.e., pushing the substrate covered by a nearly dry brush into the liquid. In other words, we invert the typical Landau-Levich setting where a plate is drawn from a bath.

As this corresponds to a persistent energy influx, thermodynamically, we keep the system permanently out of equilibrium. This implies that occurring steady interface profiles  with $\partial_t h=\partial_t \zeta=0$ do not correspond to a quiescent state of the liquid but to an internal stationary flow profile. Neither do such steady profiles correspond to minima of the free energy $\mathcal{F}$, i.e., the discussion of equilibrium states in \autoref{sec:equilibrium} does only provide the behaviour in the static limiting case $U\rightarrow 0$ and can not be applied for $U\neq0$. Further, it is important to note that out of equilibrium other states are possible beside steady ones. In particular, we expect time-periodic stick-slip motion of the contact line to occur. First, however, we analyse the changes occurring in the steady profiles when the velocity $U$ is increased from zero.

As any analytic treatment becomes rather challenging, here, we entirely focus on numerical results. To facilitate the numerical computations and a clear discussion of the effect of the brush energy, from now on we only consider the dimensionless model in long-wave approximation. For simplification, in this section we neglect the dependency of capillarity and wettability on brush state, i.e., we use $\gamma_\mathrm{bl}$ and  $f_\mathrm{wet}$ that do not depend on $\zeta$, as the basic features of the stick-slip motion can be understood even without these brush-dependencies.
An exemplary comparison between dynamic results of the full model and the simplified version are provided and discussed in Appendix~\ref{app:full_dynamic_comparison}.
Also, from hereon the meniscus is modelled using the above mentioned inclined film boundary condition $\partial_x h(x=0) = -\theta_0$ in order to capture the scenario of a brush-covered plate being pushed into a large liquid reservoir rather than into a finite sized droplet.

\subsection{Small velocities -- Rotation of Neumann angles}\label{sec:rot_Neumann}
\begin{figure}[!htb]
	\centering
	\includegraphics[width=\hsize]{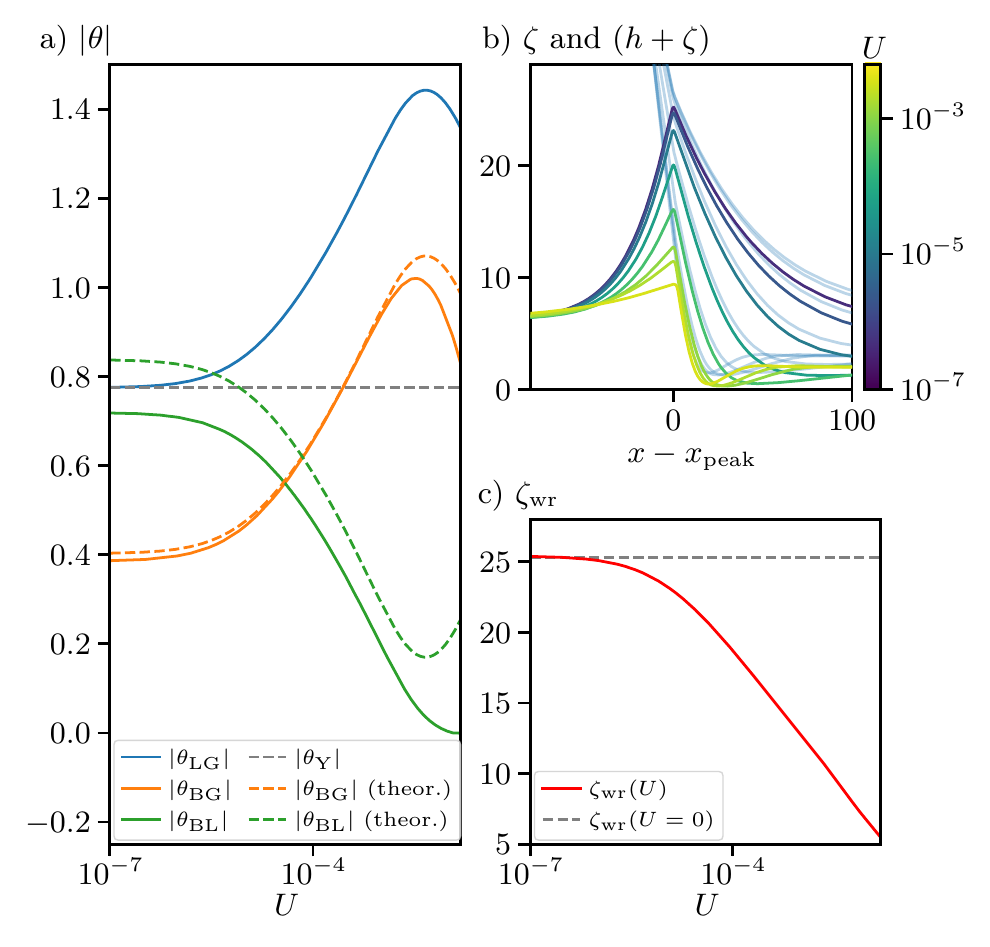}
	\caption{Characterization of the out-of-equilibrium configuration of the contact line region for (relatively small) plate velocities $U$. Shown are (a) the dynamic contact angles (as indicated in the legend) as a function of $U$. Numerical results (solid lines) are compared to  predictions obtained with the Neumann law based on the measured dynamic liquid-gas contact angle $\theta_{\mathrm{LG}}$ (dot-dashed line). The horizontal line gives the value of the equilibrium Young angle $\theta_{\mathrm{Y}}$. Panel~(b) presents examples of steady interface profiles at different velocities $U$ as indicated by the colour code of the brush-liquid interface. Gray lines indicate the liquid-air interfaces. Note that the profiles are shifted such that the peak positions of the wetting ridges coincide. Panel~(c) indicates the decrease of the ridge height $\zeta_\mathrm{wr}$ with increasing $U$. The long-wave approximation is used with parameters $T=0.02$, $\sigma=0.3$, $\gamma_\mathrm{bl}=0.3$, $\ell=20$, $M=0.1$ and $D=\SI{4e-3}{}$.}
	\label{fig:Braking}
\end{figure}

We start by considering stationary states that occur at relatively small advection velocities $U>0$. Fig.~\ref{fig:Braking} shows corresponding characteristics of the out-of-equilibrium configuration of the contact line region: As $U$ is increased from zero, all contact angles start to deviate from their equilibrium values (panel~(a)), also compare the typical profiles given in panel~(b). The wetting ridge height decreases with increasing $U$ (panel~(c)).

The solid lines in Fig.~\ref{fig:Braking}~(a) converge at very small values ($U\approx10^{-6}$) to the equilibrium values. Then, upon increasing $U$, the dynamic $|\theta_\mathrm{LG}|$ and $|\theta_\mathrm{BG}|$ increase while $\theta_\mathrm{BL}$ decreases. However, introducing the three dynamic angles into the Neumann law (Eqs.~\eqref{eq:Neumann_hor} and~\eqref{eq:Neumann_ver}) we find that it holds reasonably well at least up to $U\approx10^{-3}$. To show this we use the numerically determined $\theta_\mathrm{LG}$ (blue line) and determine the other two angles employing Eqs.~\eqref{eq:Neumann_hor} and~\eqref{eq:Neumann_ver} in long-wave approximation, i.e., Eqs.~\eqref{eq:app:neumann} in Appendix~\ref{app:lw}. The resulting theoretical values are given as dashed lines. At all $U$, the deviation is rather small for $\theta_\mathrm{BG}$ while for $\theta_\mathrm{LG}$ it remains constant at $\approx10\%$ as seen before in the  static case in \autoref{sec:NumerikGG}. By checking the validity of the vertical and horizontal Neumann conditions for all three measured angles, we determine that any deviation is as in the static case mostly due to the vertical Neumann balance. Collecting all terms of Eqs.~\eqref{eq:Neumann_hor} and~\eqref{eq:Neumann_ver} on one side, we find that the expressions for $U=0$ deviate from zero by absolute errors of 0.003 (horizontal condition) and 0.174 (vertical condition).

The continued approximate validity of the Neumann law implies that the Neumann angles are merely rotated with increasing $U$ as known for moving contact lines on an elastic substrate.\cite{KDGP2015nc} The rotation is best seen in the dynamic Young angle that coincides with $\theta_\mathrm{LG}$ (blue line in Fig.~\ref{fig:Braking}~(a)). In the velocity range from \SI{e-5}{} to \SI{e-3}{} it undergoes a 75\% increase accompanied by a 60\% decline in the height of the wetting ridge above the dry brush height (from 25 to 10, see Fig.~\ref{fig:Braking}~(c)). This makes the continued validity of the Neumannn law particularly interesting.

Another remarkable observation is, that at $U\approx\SI{e-2}{}$ the three dynamic angles all pass an extremum, i.e., a further velocity increase results in a smaller rotation of the Neumann angles as compared to their equilibrium values. Indeed, there, the deviation from Neumann's law strongly increases, rendering its application invalid. As next discussed in section~\ref{sec:instability}, the qualitative change in behaviour is closely related to the onset of an instability that occurs at a velocity less than one order of magnitude larger. It gives rise to time-periodic stick-slip motion.

\subsection{Large velocities -- instability mechanism}\label{sec:instability}
\begin{figure}[!htb]
	\centering
	\includegraphics{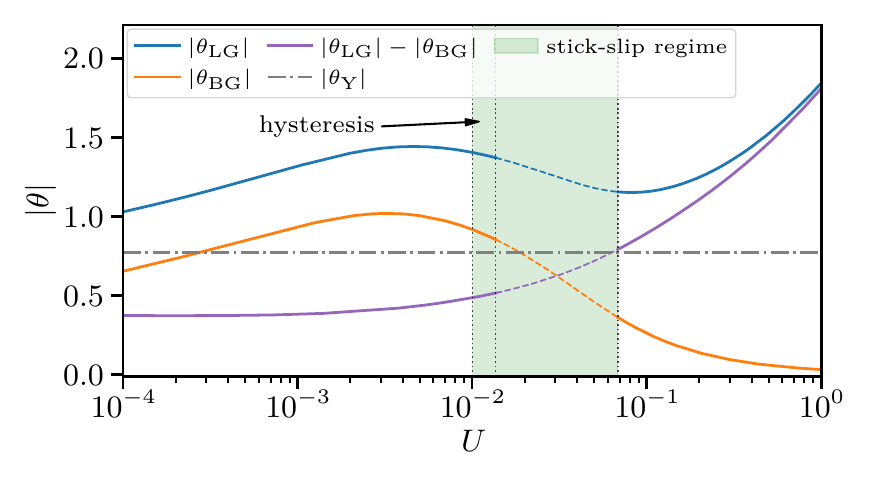}
	\caption{The dynamic angles $|\theta_\mathrm{LG}|$ and $|\theta_\mathrm{BG}|$ as well as their difference are given for steady profiles for a large range of substrate velocities $U$. At small and large $U$, the profiles are linearly stable (solid lines), while they are unstable at intermediate velocities (dashed lines). There, time-periodic stick-slip behaviour occurs (green shading) with a hysteresis range beyond the lower instability threshold (indicated by vertical dotted lines). The remaining parameters are as in Fig.~\ref{fig:Braking}.}
	\label{fig:var_Uhigh}
\end{figure}

Moving to larger velocities in the range \SI{e-2}{} to \SI{e+0}{} we restrict our attention to the angles $\theta_\mathrm{LG}$ and $\theta_\mathrm{BG}$. As the wetting ridge is very low, no well-defined $\theta_\mathrm{BL}$ can be measured. \autoref{fig:var_Uhigh} shows the remaining dynamic angles as a function of $U$. At small and large $U$ the angle $\theta_\mathrm{LG}$ increases with increasing $U$ while it decreases in the intermediate range \SI{5e-3}{}$<U<$\SI{8e-2}{}. In contrast, the angle $\theta_\mathrm{BG}$ first becomes larger and then, for $U>$\SI{3e-3}{}, continuously decreases. Asymptotically it approaches zero as the wetting ridge and the difference in the brush heights inside and outside the liquid both shrink. The latter results from the strong decrease in the advection time scale as compared to the time scale for mass transfer into the brush. This also illustrates that at higher velocities the deviation from the Neumann law must get large.

Evaluating the stability of the stationary states tracked in Fig.~\ref{fig:var_Uhigh} shows that they are linearly stable at small and large $U$ (solid lines), but unstable at intermediate $U$ (dashed lines). In the latter range, time simulations reveal time-periodic stick-slip behaviour. Furthermore, Fig.~\ref{fig:var_Uhigh} already hints at two important conditions for the instability to occur. First, the corresponding $U$-range nearly coincides with the range where $\theta_\mathrm{LG}$ decreases, i.e., where $\partial |\theta_\mathrm{LG}|/\partial U<0$ for the stationary state. In other words, the instability occurs when an increase in velocity results in a decrease of the global dynamic contact angle. This implies that a slowly increasing velocity would favour a decreasing angle, but to decrease the angle the contact line region has to advance even faster -- corresponding to a destabilising feedback loop.

\begin{figure}[!htb]
	\centering
	\includegraphics[width=.45\textwidth]{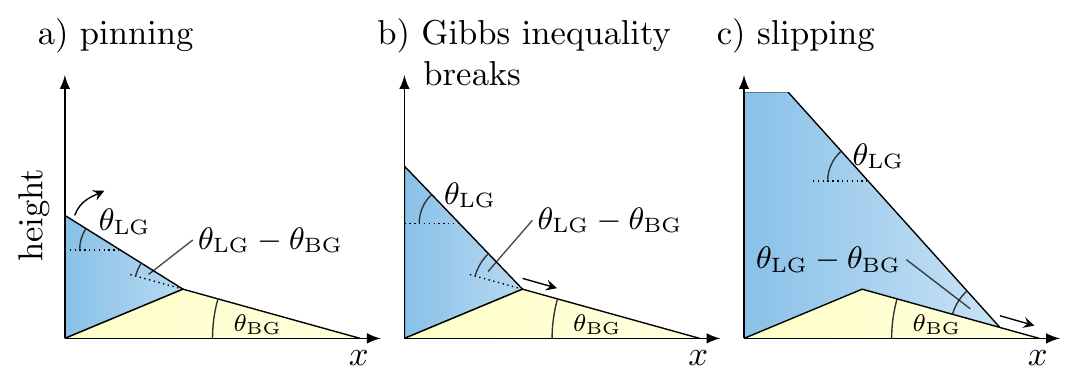}
	\caption{Sketch illustrating the Gibbs condition and the resulting depinning. In panel (a) the contact line is pinned at the ridge as the effective angle between liquid-gas interface and the substrate-gas interface in front of the contact line is lower than the Young angle, i.e., $|\theta_\mathrm{LG}|-|\theta_\mathrm{BG}| < |\theta_\mathrm{Y}|$. The liquid-gas interface steepens until equality (panel (b)). Any further steepening lets the contact line slip off the ridge (panel~(c)).}
	\label{fig:Gibbs}
\end{figure}

Second, the instability occurs at an order of magnitude of $U$, where the difference $|\theta_\mathrm{LG}|-|\theta_\mathrm{BG}|$ starts to strongly increase as the Neumann law does not hold any more. Notably, the upper critical $U$ where the unstable range ends, occurs where the difference $|\theta_\mathrm{LG}|-|\theta_\mathrm{BG}|$, i.e., the effective liquid contact angle at the advancing side of the ridge, exceeds the equilibrium contact angle $|\theta_\mathrm{Y}|$. This is a consequence of the Gibbs condition\cite{Quer2008armr,GASK2018prl} which states that a contact line depins from a heterogeneity of a rigid solid substrate when the contact angle with respect to the local substrate slope exceeds the equilibrium contact angle, see \autoref{fig:Gibbs}. While for $|\theta_\mathrm{LG}|-|\theta_\mathrm{BG}|<|\theta_\mathrm{Y}|$ the pinning at the wetting ridge governs the contact line dynamics (or at least part of it during the stick-slip dynamics), above this critical angle the contact line is unable to pin at all and instead constantly slips. Consequently, the dynamics is purely governed by the moving contact line while the ridge, or swelling gradient, simply follows behind.

\section{Stick-slip dynamics}\label{sec:stickslip}
Having established the existence of a  velocity range where unstable stationary profiles give rise to stick-slip motion, we next analyse in some detail this time-periodic state of the contact line region and its dependency on various parameters.

\subsection{Single stick-slip cycle}

\begin{figure}[!htb]
	\centering
	\includegraphics[width=.5\textwidth]{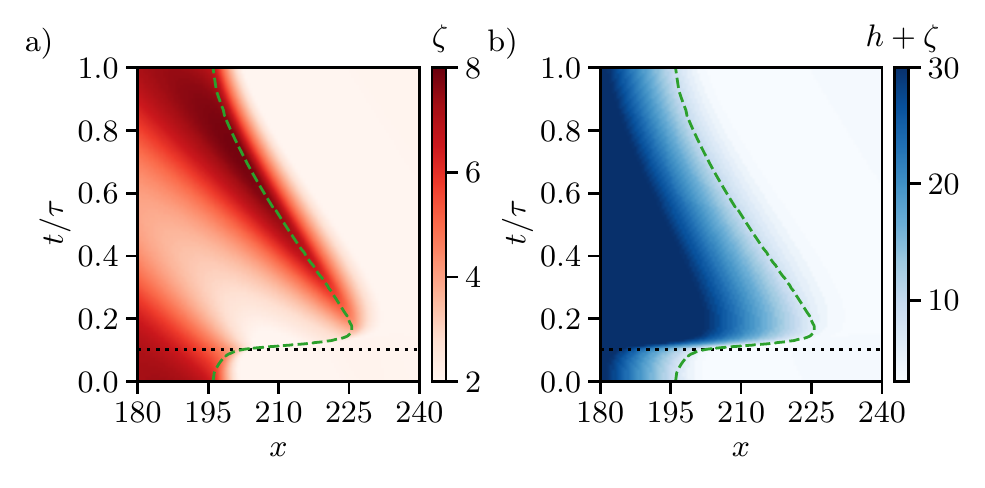}
	\caption{Shown are space-time plots of (a) the brush profile $\zeta$ and (b) the liquid height profile $h+\zeta$ in the contact line region during a single typical stick-slip cycle at  $U=0.014$. The green dashed line illustrates the approximate position of the contact line over time. The black dotted line indicates when the Gibbs condition is fulfilled and the ensuing depinning results in a slipping motion. The remaining parameters are as in Fig.~\ref{fig:Braking}. The period of the cycle is $\tau=4800$. A video of the dynamics displayed here is available under \url{https://zenodo.org/record/7886530}.}
	\label{fig:spacetime}
\end{figure}

\begin{figure*}[!htb]
	\centering
	\includegraphics{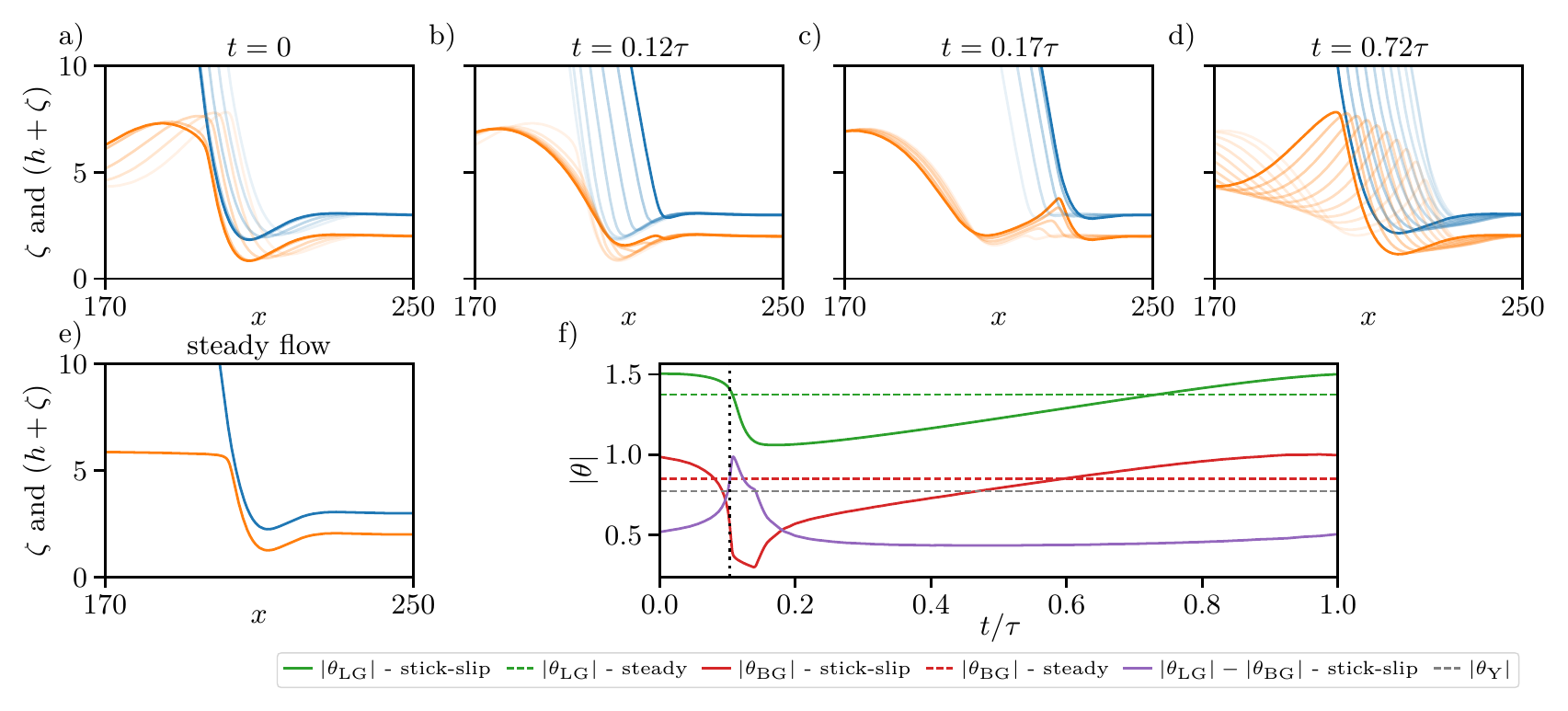}
	\caption{The top row (a-d) displays time sequences of snapshots from four different stages of the single cycle of stick-slip motion in Fig.~\ref{fig:spacetime}. Shown are profiles of the liquid-gas (blue) and brush-liquid (orange) interface in the contact line region. The final snapshot from each phase is shown in darker colours than the preceding ones. For a discussion see main text. The bottom row shows a linearly stable stationary profile at identical parameters (e), and  the temporal change in contact angles (f) over one period $\tau$ for the top row dynamics (solid lines) compared to the values for the stationary state (dashed lines). The vertical dotted line indicates when the Gibbs condition is fulfilled.}\label{fig:stickslip}
\end{figure*}

\begin{figure*}[!htb]
	\centering
	\includegraphics{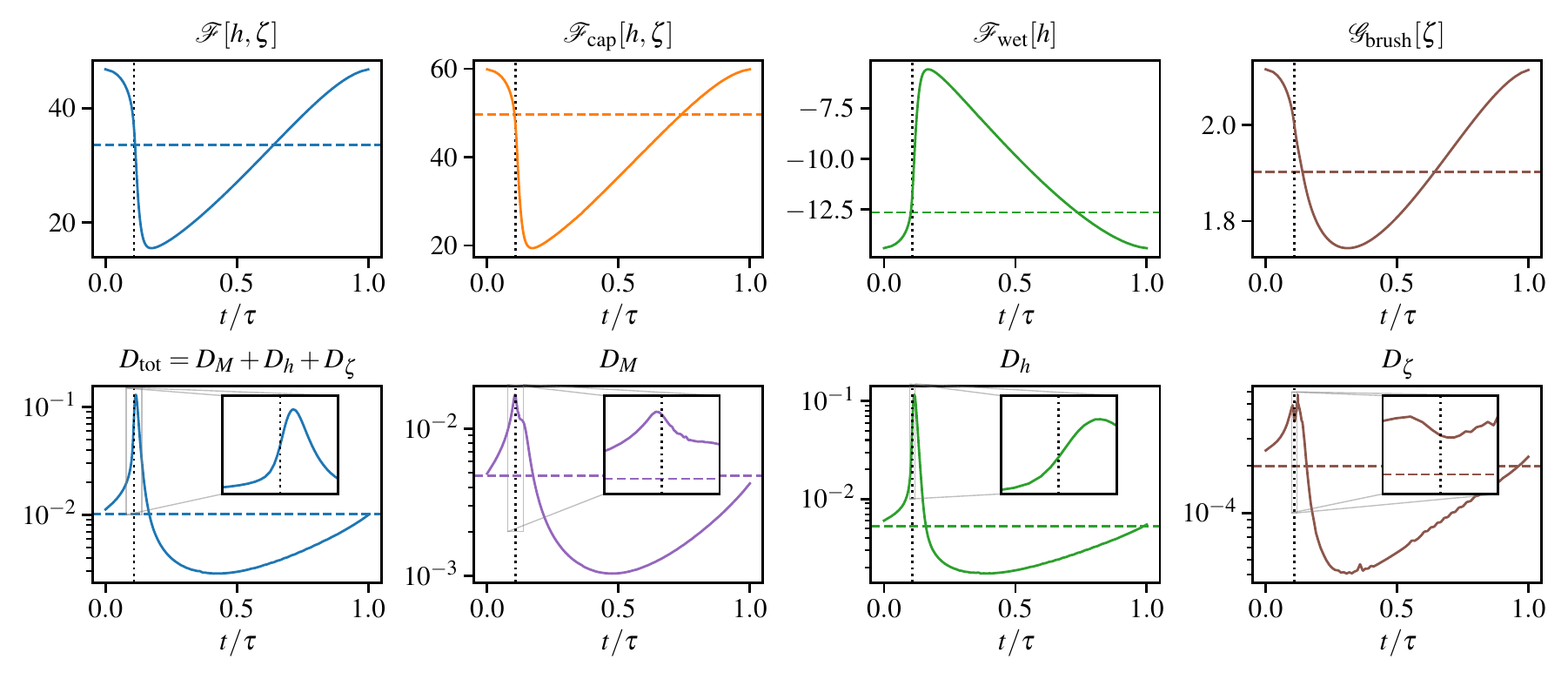}
	\caption{The top and bottom row present contributions to energy and dissipation, respectively, 
	for the single cycle of stick-slip motion shown in Figs.~\ref{fig:spacetime} and~\ref{fig:stickslip} (solid lines). In particular, the total free energy $\mathcal{F}$ and its constituents $\mathcal{F}_\mathrm{cap}$, $\mathcal{F}_\mathrm{wet}$ and $\mathcal{G}_\mathrm{brush}$ defined through Eq.~(\ref{eq:free_energy}) are given as differences w.r.t.\ to their values at equilibrium ($U=0$). The total energy dissipation $D_\mathrm{tot}$ is compared to contributions from the three dissipation channels $D_h$, $D_\zeta$, and $D_M$ defined through Eq.~(\ref{eq:Dissipation}) in section \ref{sec:dissipation_derivation}. The horizontal dashed lines give the corresponding values for the linearly stable stationary state that exists at identical parameters. All values are obtained by integrating over the spatial domain $x\in [0,420]$. In each panel the vertical dotted line marks the time when the Gibbs condition is fulfilled. The insets magnify the vicinity of this instant. }
	\label{fig:Energiedissipation}
\end{figure*}

A single typical stick-slip cycle is displayed in different representations in Figs.~\ref{fig:spacetime} and~\ref{fig:stickslip}. The former gives space-time plots of the interface profiles while the latter presents detailed views of sequences of individual profiles close to crucial moments within the cycle as well as measurements of dynamic angles during the cycle. The accompanying \autoref{fig:Energiedissipation} presents an analysis of changes during the same cycle in overall free energy $F[h,\zeta]$ and its constituents as well as in total dissipation and the individual dissipation channels as defined in Appendix~\ref{sec:dissipation_derivation}. Note that for Figs.~\ref{fig:spacetime}, \ref{fig:stickslip} and \ref{fig:Energiedissipation} we set the beginning of a cycle ($t=0$) to be the time where the contact angle $|\theta_{\mathrm{LG}}|$ reaches its maximum value.{}

We start with a discussion of the dynamics of the interfaces, where we split one cycle of the periodic motion into phases of qualitatively different behaviour, that is characterised  in the following{}: During the phase where the contact line is pinned to the fully developed wetting ridge ($0.17\leq t/\tau\leq 0.72$), the ridge slowly recedes towards the left carrying the pinned contact line along, which is illustrated by the brighter lines in Fig.~\ref{fig:stickslip}~(d) displaying the profile at previous times in this phase of the cycle{}.  However, in parallel, the contact line slowly creeps onto the right flank of the ridge (Fig.~\ref{fig:stickslip}~(a)), i.e., it surfs the ridge (as described for elastic substrates by~\citet{KDGP2015nc,GASK2018prl}). Because the surfing contact line separates from the peak of the wetting ridge, the upwards traction at the peak decreases and the ridge becomes rounder. In this phase $|\theta_\mathrm{LG}|$ remains nearly constant while $|\theta_\mathrm{BG}|$ decreases more and more. This occurs, in particular, in the surfing phase when the difference $|\theta_\mathrm{LG}|-|\theta_\mathrm{BG}|$ sharply increases (Fig.~\ref{fig:stickslip}~(f)). In consequence, at $t/\tau\approx 0.10$ the difference $|\theta_\mathrm{LG}|-|\theta_\mathrm{BG}|$ passes the equilibrium angle thereby fulfilling the Gibbs condition. As a result, the contact line depins and suddenly moves to the right leaving the ridge behind that then very slowly shrinks (Fig.~\ref{fig:stickslip}~(b)). In this short slipping phase $|\theta_\mathrm{LG}|$ sharply decreases while $|\theta_\mathrm{BG}|$ remains nearly constant but is irrelevant as it belongs to the ridge that is already detached from the contact line. When the contact line slows down a new wetting ridge starts to emerge and pins the contact line (Fig.~\ref{fig:stickslip} (c-d)). In this long phase $|\theta_\mathrm{LG}|$ slowly increases as does $|\theta_\mathrm{BG}|$, now related to the new ridge (Fig.~\ref{fig:stickslip} (f)).  Nevertheless, $|\theta_\mathrm{LG}|-|\theta_\mathrm{BG}|$ stays almost constant in this phase. Then the cycle starts again.

Focusing next on \autoref{fig:Energiedissipation} we note that the dominant contribution to the overall energy $\mathcal{F}$ is in all phases the capillary energy  $\mathcal{F}_\mathrm{cap}$. This is even more pronounced when an entire droplet is considered instead of a  domain limited to the contact line region. When the new ridge grows with the contact line sticking to it, the increasing $|\theta_\mathrm{LG}|$ results in an increase in capillary energy. In parallel, the wetting energy  $\mathcal{F}_\mathrm{wet}$ decreases as in an increasing part of the domain the brush is only covered by the energetically favourable adsorption layer. During the fast slipping process,  $\mathcal{F}_\mathrm{cap}$ and with it $\mathcal{F}$ sharply decrease, relaxing to their lowest values within the cycle.  $\mathcal{F}_\mathrm{wet}$ behaves inversely and sharply increases. The contribution of the brush $\mathcal{G}_\mathrm{brush}$ is comparatively small in the considered parameter range, and qualitatively similar to $\mathcal{F}_\mathrm{cap}$.
Dissipation occurs through three channels: viscous dissipation $D_h$ within the liquid, mostly focused in the contact line region, dissipation $D_\zeta$ due to liquid diffusion within the brush, and dissipation $D_M$ due to mass transfer between liquid layer and brush. The corresponding expressions are derived in section \ref{sec:dissipation_derivation}. The total dissipation $D_{\mathrm{tot}}$ is dominated by $D_h$, followed by the about one magnitude weaker $D_M$, and the rather small $D_\zeta$. All contributions show a slow build-up in the stick-phase that develops into a sharp peak in the surf- and slip-phase with the maxima of the individual channels slightly shifted w.r.t.\ each other. The trough to peak contrast is about one magnitude for all channels. The total dissipation peaks after the Gibbs condition is reached, i.e., clearly in the slip-phase. In contrast, $D_M$ peaks earlier confirming that depinning is triggered by a mass transfer-induced deformation of the substrate. The very weak $D_\zeta$ seems to have a double-peak structure with the minimum between the two maxima coinciding with the fulfilment of the Gibbs condition.

Note, finally, that due to hysteresis there exists a stationary state at the same parameter values where the analyzed stick-slip cycle occurs. It is shown in Fig.~\ref{fig:stickslip}~(e) with the corresponding angles, energies and dissipation given as horizontal dashed lines in Fig.~\ref{fig:stickslip}~(f), Fig.~\ref{fig:Energiedissipation}~(top) and Fig.~\ref{fig:Energiedissipation}~(bottom), respectively. On average the stick-slip cycle has a lower total energy and shows also a lower mean dissipation than the stationary state.

After having discussed a single stick-slip cycle in detail for a particular set of parameters, in the following we explore how the behaviour depends on important system parameters. We focus on the driving substrate velocity in section~\ref{sec:para-velocity}, on the mass transfer rate between liquid layer and brush in section~\ref{sec:para-transfer}, and on interface and brush energies in section~\ref{sec:para-energy}.

\subsection{Dependence on substrate velocity}\label{sec:para-velocity}
It is our main interest to systematically establish in which parameter ranges stationary and time-periodic states dominate. To obtain these ranges we employ two numerical techniques: First, we use extensive time simulations of the liquid meniscus dynamics where we increase (decrease) the substrate velocity in small steps and make sure that the integration time of each step is long enough for transient behaviour to die out. This allows us to access stable dynamical states. Second, we employ path continuation to access the stationary state for arbitrary values of the substrate velocity and to obtain its linear stability. We proceed to combine the results of both approaches into one diagram.
As solution measures for the stationary states we use the height of the wetting ridge $\zeta_\mathrm{wr}$, and the dynamic contact angle $|\theta_\mathrm{LG}|$. For the time-periodic stick-slip motion we show the corresponding minimal and maximal values within a cycle. Thereby, for $\zeta_\mathrm{wr}$ we only consider the particular ridge located in the vicinity of the contact line, (practically, we request a liquid film height $h<5$). This ensures that the newly growing ridge is analysed and not the former one that is slowly decaying underneath the liquid. Additionally, we consider the frequency $\nu=1/\tau$ obtained as the inverse of the time period. Note that $\nu=0$ for the stationary states.

\begin{figure}[!htb]
	\centering
	\includegraphics{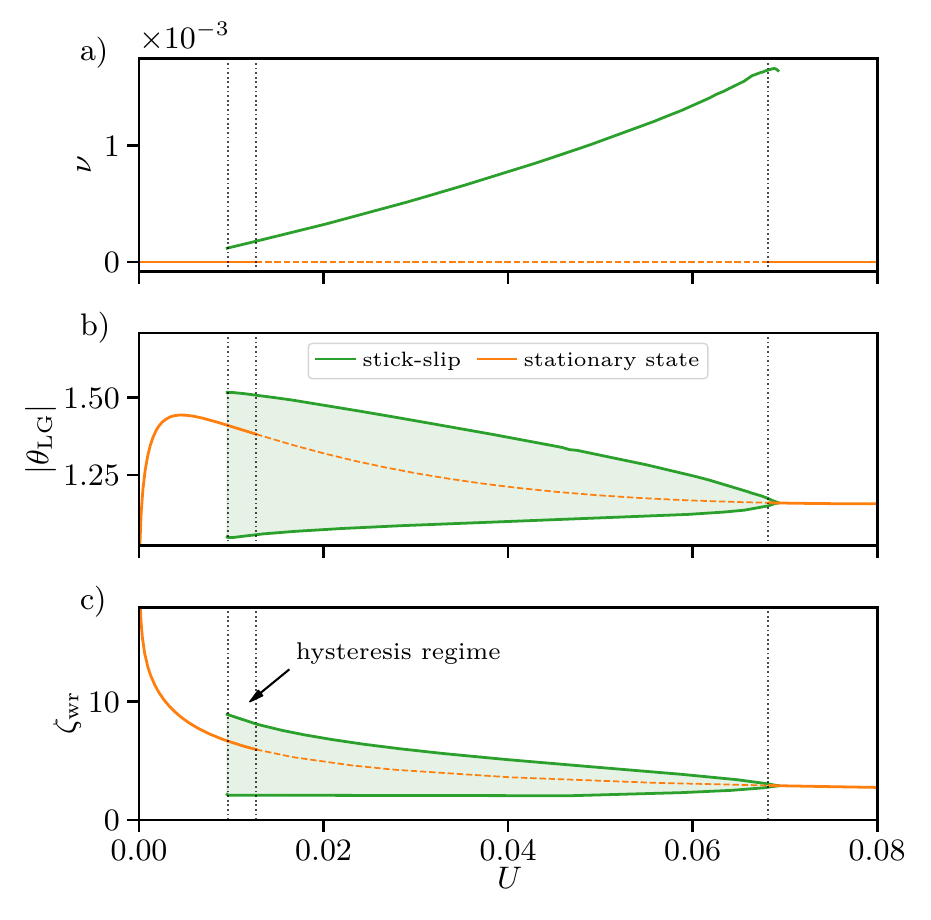}
	\caption{Stationary states and stick-slip cycles are analyzed in dependence of substrate velocity $U$. Shown are (a) frequency $\nu$, (b) the dynamic liquid-gas contact angle $|\theta_\mathrm{LG}|$, and (c) the wetting ridge peak height $\zeta_\mathrm{wr}$. Solid [dashed] orange lines indicate linearly stable [unstable] stationary states, while green lines give the corresponding range for linearly stable stick-slip cycles. Linear stability thresholds and existence ranges are indicated by vertical dotted lines, thereby marking the stick-slip and hysteresis regimes. The remaining parameters are $T=0.02$, $\sigma=0.3$, $\gamma_\mathrm{bl}=0.3$, $\ell=20$, $M=0.1$ and $D=\SI{4e-3}{}$.}
	\label{fig:var_U}
\end{figure}

\autoref{fig:var_U} gives the described quantities in dependence of substrate velocity $U$. Inspecting the figure we see that there exists a regime of stick-slip behaviour in the range of decreasing angles  $|\theta_\mathrm{LG}|$ as already established in section~\ref{sec:instability}. At large velocities, it ends at $U\approx0.068$ in a supercritical Hopf bifurcation: the amplitude of the temporal modulation of all quantities approaches zero while the frequency approaches a finite value. This is further confirmed by the fact that at the same velocity the stationary state changes its linear stability via a pair of complex conjugate eigenvalues that crosses the imaginary axis. The situation is more involved at small velocities where a range of hysteresis exists, i.e., where bistability occurs between stationary state and time-periodic stick-slip behaviour. The latter is first observed at $U\approx0.010$ while the stationary state looses stability in a (seemingly subcritical) Hopf bifurcation at $U\approx0.013$. This indicates that the time simulation stops to pick up the stick-slip motion close to a saddle-node bifurcation of limit cycles. There, the subcritical branch of cycles from the primary Hopf bifurcation of the stationary state at $U\approx0.013$ stabilises and the branch folds back towards larger $U$ becoming the numerically observed stable cycle between $U\approx 0.010$ and $U\approx0.068$.{} At this point the frequency is small but finite and the amplitude of the temporal modulation of $|\theta_\mathrm{LG}|$ and $\zeta_\mathrm{wr}$ reach their largest values. Both amplitudes monotonically decrease with increasing $U$. Note that we find no indication of more complicated behaviour like, e.g., period doubling.

\subsection{Dependence on transfer constant}\label{sec:para-transfer}

\begin{figure}[]
	\centering
	\includegraphics{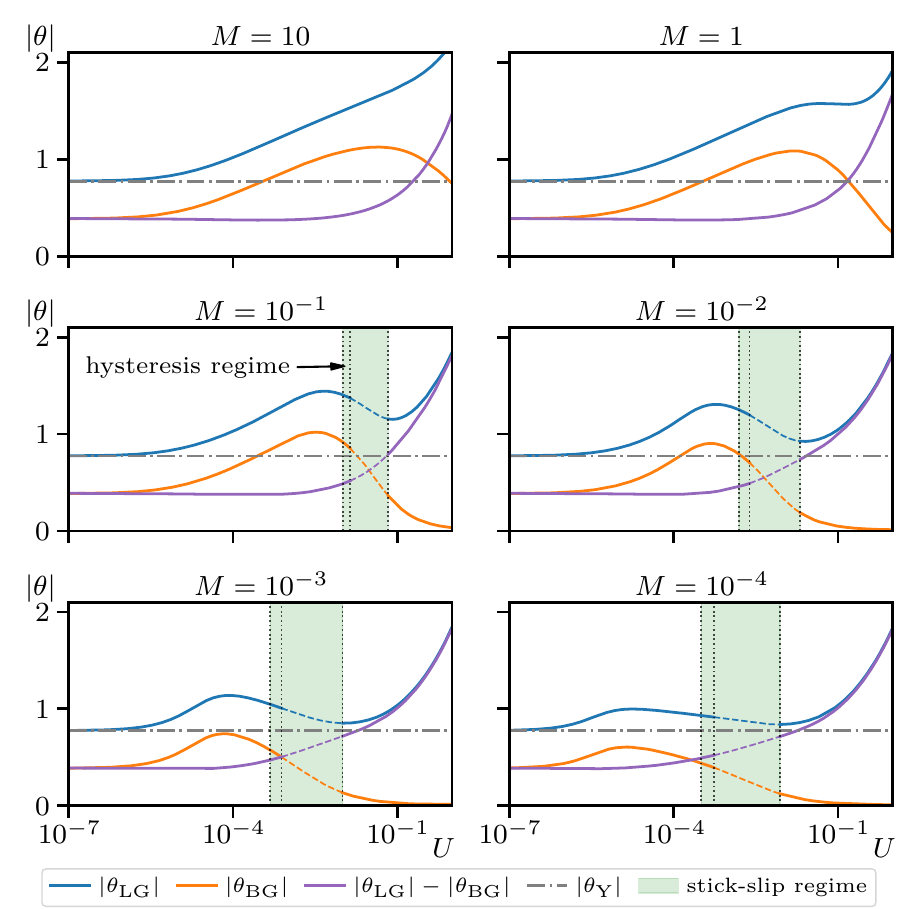}
	\caption{For selected values of the transfer rate constant $M$ (stated above each panel) the 	dependencies of the dynamic angles $|\theta_\mathrm{LG}|$ and $|\theta_\mathrm{BG}|$ as well as of their difference on substrate velocity $U$ are given. As reference the equilibrium contact angle $|\theta_\mathrm{Y}|$ is indicated by a horizontal line. The green shading marks the range of stick-slip motion including the hysteresis regime. The remaining parameters and line styles are as in Fig.~\ref{fig:var_U}.}
	\label{fig:var_M}
\end{figure}

\begin{figure}[]
	\centering
	\includegraphics{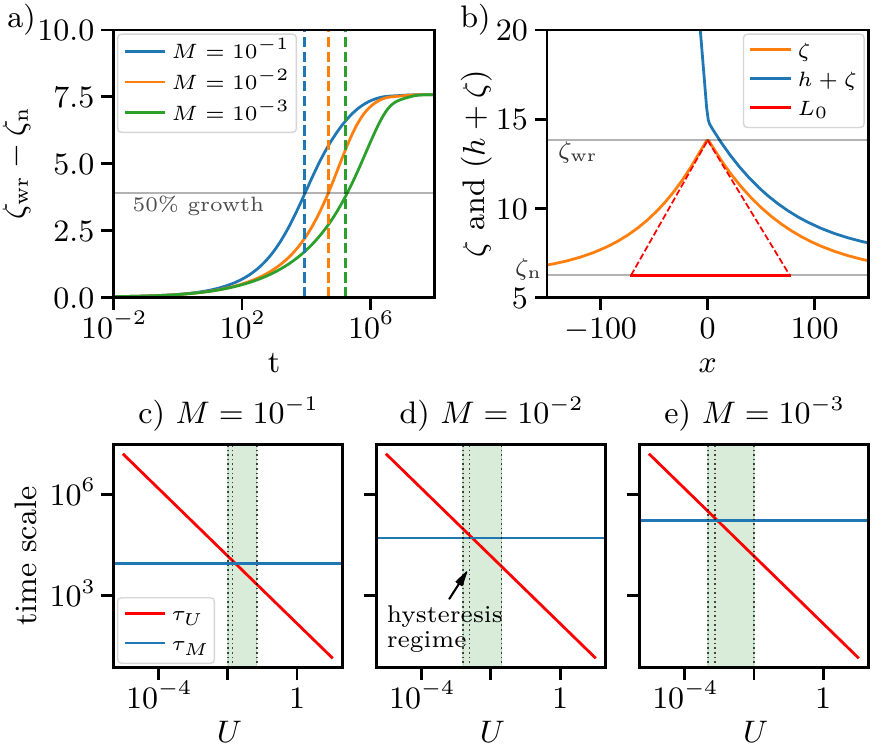}
	\caption{Illustration of the definitions of (a) the characteristic timescale $\tau_M$ (vertical lines) for ridge growth as the time when $\zeta_\mathrm{wr}-\zeta_\mathrm{n}$ reaches half its maximal equilibrium height, and (b) the characteristic equilibrium ridge width $L_0$ defined by a triangular approximation using the equilibrium values of $\theta_\mathrm{BL}$ and $\theta_\mathrm{BG}$ (see \autoref{eq:L0}). Panels (c-e) compare for selected values of the transfer rate $M$ the timescale $\tau_M$ and a characteristic timescale of substrate motion $\tau_U=L_0/U$. Their equality coincides with the lower border of the unstable $U$-range highlighted by the green shading.}\label{fig:scales}
\end{figure}

In section~\ref{sec:para-velocity} we have analysed how the system behaviour depends on substrate velocity for a fixed set of all other parameters. Thereby, we have identified a $U$-range where stick-slip behaviour occurs. Next, we explore how the stick-slip range changes when the transfer rate constant $M$ is varied. Accordingly, \autoref{fig:var_M} shows dependencies of dynamic angles $|\theta_\mathrm{LG}|$ and $|\theta_\mathrm{BG}|$ as well as of their difference $|\theta_\mathrm{LG}|-|\theta_\mathrm{BG}|$ on $U$ for six different values of $M$ between $M=10$ (very fast mass transfer between liquid meniscus and brush) and $M=10^{-4}$ (very slow transfer). The panel at intermediate $M=0.1$ reproduces Figure~\ref{fig:var_Uhigh}. Comparing the panels of Fig.~\ref{fig:var_M} we first note that for fast transfer $|\theta_\mathrm{LG}|$ is a monotonically increasing function of velocity $U$. There is no range of negative inclination of $|\theta_\mathrm{LG}|$ vs. $U$ and no stick-slip motion occurs. Both features only appear if $M$ is sufficiently small, here $M\ge0.1$. Further decreasing $M$ by three orders of magnitude widens the $U$-range of negative inclination (the dip get shallower though) as well as the range of stick-slip motion.

Both (necessary) criteria for the instability uncovered in section~\ref{sec:instability} still hold for the entire range of considered transfer constants $M$: On the one hand, the instability only occurs, when $|\theta_\mathrm{LG}|(U)$ has a negative slope. On the other hand, the maximal $U$ where the stationary state is unstable coincides with the fulfilment of the Gibbs condition, i.e., $|\theta_\mathrm{LG}|-|\theta_\mathrm{BG}|$ crosses $\theta_Y$. However, considering the sequence from $M=0.1$ to $M=10^{-4}$ it also becomes obvious that the negative slope criterion is not sufficient for the instability to occur. While at $M=0.1$ the instability threshold follows not too far behind the maximum of $|\theta_\mathrm{LG}|(U)$, the distance becomes larger for smaller $M$. For instance, for the slowest considered transfer, at $M=10^{-4}$, the unstable $U$-range only corresponds to the second half of the range with negative slope. Therefore we need to refine our criterion for the onset of instability.

To do so, we define a characteristic timescale $\tau_M$ for the growth of a wetting ridge. It is important as during the stick-slip motion periodically a new ridge is created (see \autoref{fig:var_U}). For simplicity, we assume that $U$ has only a minor influence on the growth and consider the relaxation from a flat brush state towards an equilibrium wetting ridge at $U=0$. The corresponding growth of $\zeta_\mathrm{wr}$ is shown for three values of $M$ in Fig.~\ref{fig:scales}~(a). Based on these simulations, we define $\tau_M$ as the time when the wetting ridge reaches half of its equilibrium height. This characteristic time (marked by vertical lines in  Fig.~\ref{fig:scales}~(a)) increases by about one order of magnitude when $M$ is decreased from $0.1$ to $10^{-3}$.

Additionally, we define a typical horizontal length scale $L_0$ characterising a typical ridge width based on the triangular approximation of the equilibrium profile{} illustrated in Fig.~\ref{fig:scales}~(b), or, by formula
\begin{equation}
	L_0=(\zeta_\mathrm{wr}-\zeta_\mathrm{n})\left[\frac{1}{|\theta_\mathrm{BL}|}+\frac{1}{|\theta_\mathrm{BG}|}\right]	\label{eq:L0}
\end{equation}
as obtained in long-wave approximation. This length scale allows us to define a second characteristic time $\tau_U=\frac{L_0}{U}$ that characterises the forced motion of the substrate.
The values for the two defined timescales $\tau_M, \tau_U$ are plotted in dependence of the velocity $U$ in Figs.~\ref{fig:scales}~(c), (d) and (e) for $M=0.1$, $10^{-2}$, and $10^{-3}$, respectively. One immediately notices that the low-velocity border of the region of unstable stationary states very closely coincides with the velocity where the two time scales cross. In other words, there, the stationary state is linearly stable as long as the wetting ridge grows much faster than it is moved by the substrate (note Figs.~\ref{fig:scales}~(c-e) are log-log plots). One may say, the ridge is slaved to the liquid that moves with the substrate. In contrast, when the time scales become similar all influences compete on equal terms and an instability is possible.

As a result of this consideration, we expect that the effect of a variation of the transfer coefficient $M$ is reciprocal to the effect of a variation of the velocity $U$, i.e., when increasing $M$ from small values, we expect the stick-slip motion to start with a large finite frequency and zero amplitude. Then the frequency decreases and  the amplitude increases with further increasing $M$ until the periodic behaviour ceases when the saddle-node bifurcation of limit cyles is passed at large finite $M$.

\subsection{Dependence on energies}\label{sec:para-energy}
\begin{figure*}[!htb]
	\centering
	\includegraphics[width=\textwidth]{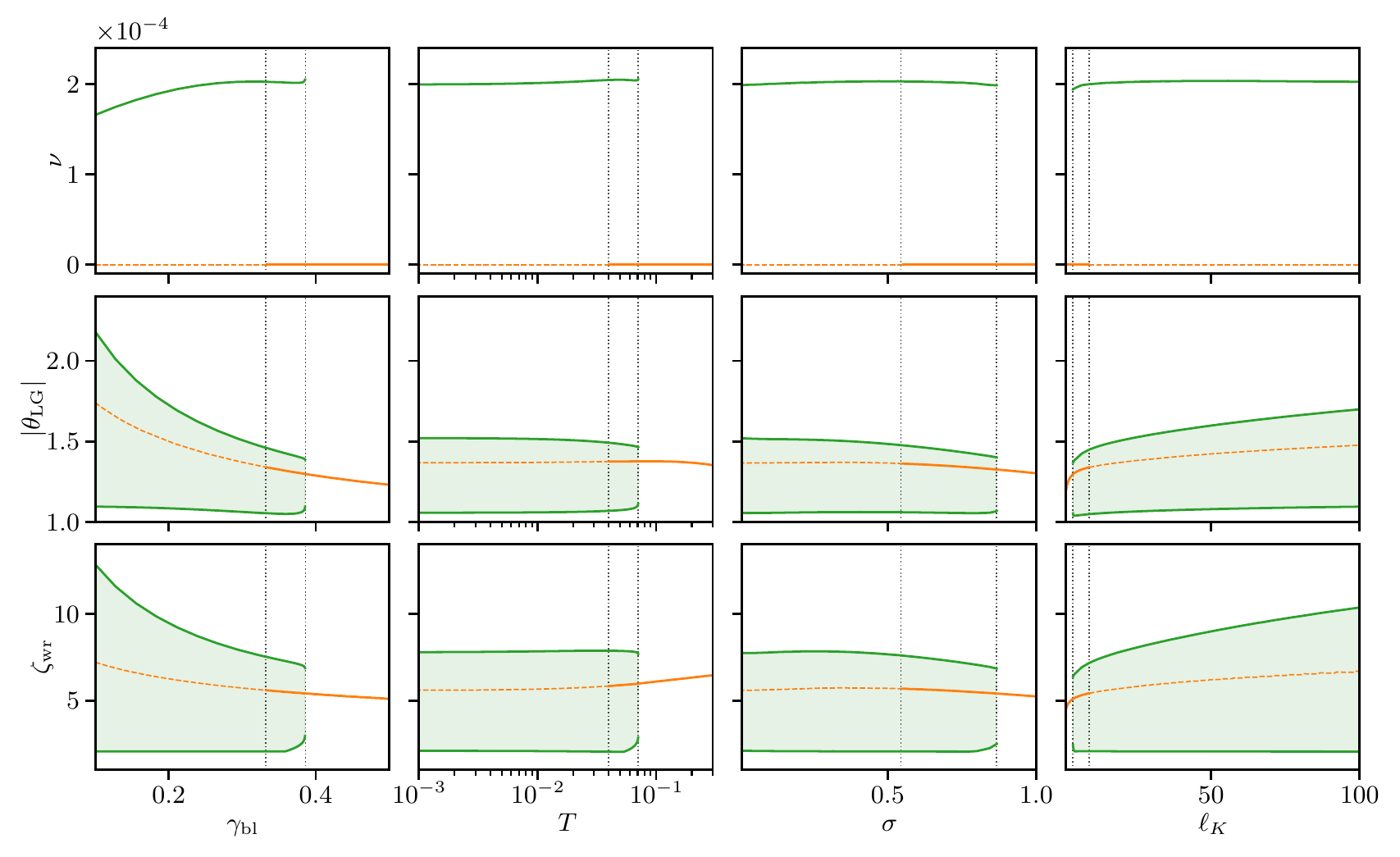}
	\caption{Influence of various parameters on stationary states and stick-slip cycles: (top row) Frequency $\nu$, (center row) contact angle $|\theta_\mathrm{LG}|$, and (bottom row) height of the wetting ridge $\zeta_\mathrm{wr}$ are given as function of (1st column) brush-liquid interface energy $\gamma_\mathrm{bl}$, (2nd column) effective temperature $T$, (3rd column) brush grafting density $\sigma$ and (4th column) brush Kuhn length $\ell_K$. Line styles and respective remaining parameters are as in Fig.~\ref{fig:var_U}.
	}
	\label{fig:dynVariation}
\end{figure*}

Finally, we briefly investigate the influence of a number of further parameters related to capillary and brush energies. An overview is given in Fig.~\ref{fig:dynVariation} where dependencies on brush-liquid interface energy $\gamma_\mathrm{bl}$, effective temperature $T$, brush grafting density $\sigma$ and the Kuhn length $\ell$ of the brush polymers are shown at fixed $U=\SI{0.014}{}$, i.e., at the velocity of our reference case in Fig.~\ref{fig:stickslip}. Both, stationary states and stick-slip cycles are included. Note that each of the chosen parameters allows us to approach the limiting case of a rigid, non-adaptive substrate, i.e., with vanishing wetting ridge. The  respective limits correspond to $\gamma_\mathrm{bl}\to\infty$, $T\to\infty$, $\sigma\to\infty$ and $\ell\to0$. This implies that in each of these cases the stick-slip behaviour will eventually disappear when moving towards the corresponding limit.

Inspecting Fig.~\ref{fig:dynVariation} we note that hysteresis is observed for all parameter dependencies indicating that in each case a subcritical Hopf bifurcation occurs. It is remarkable that the frequency $\nu$ of the stick-slip cycle barely depends on the brush-related parameters and only shows a small decrease with decreasing interface energy $\gamma_\mathrm{bl}$. A similar observation holds for the amplitude of variations in the height of the wetting ridge over a stick-slip cycle. The abrupt stop of stick-slip motion with increasing $\gamma_\mathrm{bl}$, $T$, and $\sigma$ can in all cases be related to the occurrence of a saddle-node bifurcation of cycles, in particular, for $\gamma_\mathrm{bl}$ and $T$ curves become nearly vertical clearly indicating a fold. This is less pronounced for $\sigma$ and $\ell$.

Taken together these observations imply that there is a ``critical softness'' or ``critical adaptivity'' that must be reached to induce a stick-slip behaviour. The stick-slip motion itself only depends relatively weakly on the energetic parameters. This is in stark contrast to the much larger influence found in sections~\ref{sec:para-velocity} and \ref{sec:para-transfer}, respectively, for the dynamic parameters substrate velocity and transfer rate constant. Their direct correspondence to the timescales $\tau_U$ and $\tau_M$ reinforces our finding that their ratio controls the occurrence of stick-slip behaviour.

\section{Conclusion}\label{sec:conclusions}
We have presented a mesoscopic model for the dynamic wetting of a polymer brush-covered substrate. It accounts for the coupled dynamics of a liquid drop/meniscus on the brush and of the imbibed liquid within the brush as well as for the imbibition process itself. Due to its central gradient dynamics structure, the dynamics is driven by a single underlying free energy functional. The model extends the work by \citet{ThHa2020epjt} by introducing a refined energy functional that additionally accounts for the dependencies of the wetting and interface energies on the brush state. Moreover, now the energies may be employed in their full-curvature formulation as well as in their long-wave approximation.

The refined mesoscopic model has first allowed us to employ the full-curvature formulation to consider a static contact line region. There, we have analytically recovered both, the Young law for the global liquid-gas contact angle and the Neumann law that locally relates the contact angle to the opening angle of the occurring static wetting ridge. While our calculations have mostly been performed in the mesoscopic picture, the derivation of macroscale equivalents for both laws has also been sketched. This has  allowed us to establish consistency conditions that relate mesoscale and macroscale quantities in a similar spirit as recently presented for surfactant-covered drops.\cite{TSTJ2018l} In passing, analytic considerations have provided an estimate for the shape and size of the wetting ridge. Due to its nanometric size, a quantification remains a challenge for experiments. However, observations of wetting ridges are reported for Molecular Dynamics simulations. \cite{LeMu2011jcp,MeSB2019m}

Second, we have employed numerical simulations not only to confirm our analytical findings but also to test how they change beyond the static equilibrium case when considering  a more intricate forced wetting scenario. We have therefore investigated the case of an externally driven liquid meniscus moving over the brush substrate at an imposed velocity modelling, e.g., a brush-covered plate that is plunged into a liquid bath at constant relatively small velocity. In other words, the employed geometry resembles a dip coating process that is modelled by adding a simple advection term and adjusting the boundary conditions. This corresponds to an inverted Landau-Levich setting as usually one studies the transfer of liquid onto a plate that is drawn out  of a bath.\cite{SADF2007jfm,MRRQ2011jcis,TWGT2019prf} 
Note that the case of a receding contact line could be equally described using our model, but may develop more intricate behaviour, e.g., due to effects of liquid deposition. Preliminary calculations have not shown stick-slip motion in this case, in agreement with available experimental results \cite{SHNF2021acis}.

For the brush-covered substrate that is immersed into the bath, at low velocities we have observed a deformation of the interfaces in the contact line region. Specifically, the liquid contact angle as well as the inclination angles of the flanks of the wetting ridge all change under the additional stress that forces the ridge to advance along the substrate. However, we have found that at small velocities the three angles continue to satisfy the Neumann law even in this  out-of-equilibrium scenario. In other words, the interfaces close to the contact line region undergo a nearly rigid rotation. This agrees with earlier observations for moving contact lines on viscoelastic substrates (i.e., without mass transfer into the substrate). \cite{KDGP2015nc,AnSn2020arfm} The motion of the wetting ridge we have observed here on the swelling polymer brush has been found to cause a similar dissipative braking of the motion, that counteracts the imposed movement of the liquid meniscus, and results in a steepening of the liquid contact angle. 

At larger velocities, first, the deviation of the relation between the angles from the Neumann relations starts to increase such that the Neumann law does not hold anymore. Then, one enters the velocity range where the dynamic brush-gas and liquid-gas contact angles decrease with increasing plate velocity. Eventually, the growth of the ridge due to mass transfer is not able to keep up with the substrate motion, the stationary state becomes unstable. In the unstable regime, the liquid-gas interface depins from the wetting ridge and advances over the substrate in a rapid slipping motion. The contact line slows down and stops when the contact angle relaxes. Subsequently, the brush starts to form a new ridge via some liquid imbibition near the contact line, thereby pinning the front again. In consequence, the process starts over and a time-periodic stick-slip motion emerges. There exists a hysteresis between stationary state and stick-slip cycles in the vicinity of the stability threshold.
While a few experimental observations of stick-slip behaviour in the forced spreading of drops on polymer brushes exist,\cite{WMYT2010scc,SHNF2021acis} to our knowledge, the present work establishes the first theoretical description and analysis of the mechanism.

Overall, the phases of a stick-slip cycle are rather similar to the ones described for other systems where stick-slip motion occurs, e.g., for evaporation-induced moving contact lines of solutions and suspensions that periodically deposit material\cite{FrAT2012sm} and contact lines moving on viscoelastic substrates.\cite{PBDJ2017sm,KDGP2015nc,AlMo2021ijnme} Moreover, we have identified that the depinning of the contact line from the ridge within a stick-slip cycle coincides with the fulfilment of a Gibbs condition\cite{Quer2008armr} as it is known from the wetting of rough or otherwise topographically heterogeneous substrates. Our data suggest that this condition governs both the onset of slip within a cycle and the size of the stick-slip range of forcing velocities. Note that the relevance of Gibbs' condition is also reported for other out-of-equilibrium settings, namely, related to evaporation.\cite{TDGR2014l}

Similarly, our analysis has shown that the growth rate of the ridge that newly grows during a cycle is critical to the existence of a stick-slip motion. If the time scale of growth (here limited by the rate of imbibition) is much shorter than the time scale of contact line motion over a characteristic length, the wetting ridge is able to move along with the contact line without depinning. In other words, a rapid (de-)swelling of the polymer brush allows the contact line to surf the comoving ridge. If, however, the contact line is forced to advance at a much larger velocity than the swelling rate of the polymers, the wetting ridge is unable to form at all. It is only when the time scales of imbibition and motion are comparable that a periodic ridge formation, pinning and subsequent depinning are observed.

We acknowledge that the time scale of ridge growth is a theoretical measure that could be difficult to access in experiments. Using our findings on the stick-slip phenomenon in applied scenarios may therefore require a related characteristic quantity as a proxy, e.g., the swelling time of a homogeneous brush from vapour. The prevalence of stick-slip motion in numerous wetting problem renders this a very interesting subject for future investigations. We point out that some of our key findings should allow for a straightforward comparison with experimental works and theoretical results employing other methods. Prominently, our model predicts a hysteretic transition between the stationary regime at low velocities and the stick-slip regime. The recent theoretical work by \citet{MoAK2022el} on the forced wetting of soft elastic substrates gives comparable results. Namely, their investigation of the transitions in and out of the stick-slip regime reveals a similar, respectively sub- and supercritical behaviour although hysteresis is not explicitly mentioned.
In a comparison to experimental data models like ours could in the future be used to quantify microscopic properties of a brush, e.g. the speed of wetting ridge formation, from macroscopic features of the stick-slip motion.

We have concluded our investigation with an extensive study of the dependence of the dynamics on various control parameters thereby providing an overview of the stick-slip range in parameter space. This parameter study will help to understand and compare the role of the various parameters in different experimental settings and even in the analysis of stick-slip behaviour in related problems. For example, the presented results show that the periodic stick-slip motion can be suppressed by using a liquid of a low liquid-gas interface energy. Similarly, we predict that brushes of lower grafting density should be more prone to stick-slip behaviour.
Note that the validity of the discussed criteria for stick-slip motion was not accessed in this parameter study as the time-scales would need to be re-evaluated for every parameter configuration. This could be a subject for future studies.

Finally, we stress that the presented gradient dynamics model allows for refinements in numerous ways. While for the sake of efficiency all of our numerical results have been performed using the free energy in its long-wave approximation, we expect only subtle shifts in the results when computed using the full-curvature model. Moreover, an extended model may additionally account for miscibility effects (e.g., by using a Flory-Huggins energy with $\chi \neq 0$), substrate elasticity,\cite{HeST2021sm} or the effects of evaporation into an ambient vapour phase using, e.g., the approach suggested by \citet{HDJT2022apa}.

\section*{Author Contributions}
\textbf{Daniel Greve:} Writing -- Review \& Editing, Writing -- Original Draft, Investigation;
\textbf{Simon Hartmann:} Writing -- Review \& Editing, Writing -- Original Draft, Investigation, Software;
\textbf{Uwe Thiele:} Writing -- Review \& Editing, Supervision, Funding acquisition.

\section*{Conflicts of interest}
There are no conflicts to declare.

\section*{Acknowledgements}
This work was supported by the Deutsche Forschungsgemeinschaft (DFG) within SPP~2171 by Grants No.\ TH781/12-1 and TH781/12-2. We acknowledge fruitful discussions on adaptive viscous and viscoelastic substrates with Jan Diekmann, Christopher Henkel and Jacco Snoeijer, as well discussions on the interaction of liquids and polymer brushes with the group of Sissi de Beer.

\bibliography{literature}

\providecommand*{\mcitethebibliography}{\thebibliography}
\csname @ifundefined\endcsname{endmcitethebibliography}
{\let\endmcitethebibliography\endthebibliography}{}
\begin{mcitethebibliography}{86}
\providecommand*{\natexlab}[1]{#1}
\providecommand*{\mciteSetBstSublistMode}[1]{}
\providecommand*{\mciteSetBstMaxWidthForm}[2]{}
\providecommand*{\mciteBstWouldAddEndPuncttrue}
  {\def\EndOfBibitem{\unskip.}}
\providecommand*{\mciteBstWouldAddEndPunctfalse}
  {\let\EndOfBibitem\relax}
\providecommand*{\mciteSetBstMidEndSepPunct}[3]{}
\providecommand*{\mciteSetBstSublistLabelBeginEnd}[3]{}
\providecommand*{\EndOfBibitem}{}
\mciteSetBstSublistMode{f}
\mciteSetBstMaxWidthForm{subitem}
{(\emph{\alph{mcitesubitemcount}})}
\mciteSetBstSublistLabelBeginEnd{\mcitemaxwidthsubitemform\space}
{\relax}{\relax}

\bibitem[de~Gennes(1985)]{Genn1985rmp}
P.-G. de~Gennes, \emph{Rev. Mod. Phys.}, 1985, \textbf{57}, 827--863\relax
\mciteBstWouldAddEndPuncttrue
\mciteSetBstMidEndSepPunct{\mcitedefaultmidpunct}
{\mcitedefaultendpunct}{\mcitedefaultseppunct}\relax
\EndOfBibitem
\bibitem[Teletzke \emph{et~al.}(1988)Teletzke, Davis, and
  Scriven]{TeDS1988rpap}
G.~F. Teletzke, H.~T. Davis and L.~E. Scriven, \emph{Rev. Phys. Appl. (Paris)},
  1988, \textbf{23}, 989--1007\relax
\mciteBstWouldAddEndPuncttrue
\mciteSetBstMidEndSepPunct{\mcitedefaultmidpunct}
{\mcitedefaultendpunct}{\mcitedefaultseppunct}\relax
\EndOfBibitem
\bibitem[Starov and Velarde(2009)]{StVe2009jpm}
V.~M. Starov and M.~G. Velarde, \emph{J. Phys.-Condens. Matter}, 2009,
  \textbf{21}, 464121\relax
\mciteBstWouldAddEndPuncttrue
\mciteSetBstMidEndSepPunct{\mcitedefaultmidpunct}
{\mcitedefaultendpunct}{\mcitedefaultseppunct}\relax
\EndOfBibitem
\bibitem[Bonn \emph{et~al.}(2009)Bonn, Eggers, Indekeu, Meunier, and
  Rolley]{BEIM2009rmp}
D.~Bonn, J.~Eggers, J.~Indekeu, J.~Meunier and E.~Rolley, \emph{Rev. Mod.
  Phys.}, 2009, \textbf{81}, 739--805\relax
\mciteBstWouldAddEndPuncttrue
\mciteSetBstMidEndSepPunct{\mcitedefaultmidpunct}
{\mcitedefaultendpunct}{\mcitedefaultseppunct}\relax
\EndOfBibitem
\bibitem[Craster and Matar(2009)]{CrMa2009rmp}
R.~V. Craster and O.~K. Matar, \emph{Rev. Mod. Phys.}, 2009, \textbf{81},
  1131--1198\relax
\mciteBstWouldAddEndPuncttrue
\mciteSetBstMidEndSepPunct{\mcitedefaultmidpunct}
{\mcitedefaultendpunct}{\mcitedefaultseppunct}\relax
\EndOfBibitem
\bibitem[Snoeijer and Andreotti(2013)]{SnAn2013arfm}
J.~H. Snoeijer and B.~Andreotti, \emph{Annu. Rev. Fluid Mech.}, 2013,
  \textbf{45}, 269--292\relax
\mciteBstWouldAddEndPuncttrue
\mciteSetBstMidEndSepPunct{\mcitedefaultmidpunct}
{\mcitedefaultendpunct}{\mcitedefaultseppunct}\relax
\EndOfBibitem
\bibitem[Engelnkemper \emph{et~al.}(2016)Engelnkemper, Wilczek, Gurevich, and
  Thiele]{EWGT2016prf}
S.~Engelnkemper, M.~Wilczek, S.~V. Gurevich and U.~Thiele, \emph{Phys. Rev.
  Fluids}, 2016, \textbf{1}, 073901\relax
\mciteBstWouldAddEndPuncttrue
\mciteSetBstMidEndSepPunct{\mcitedefaultmidpunct}
{\mcitedefaultendpunct}{\mcitedefaultseppunct}\relax
\EndOfBibitem
\bibitem[Bico \emph{et~al.}(2018)Bico, Reyssat, and Roman]{BiRR2018arfm}
J.~Bico, E.~Reyssat and B.~Roman, \emph{Annu. Rev. Fluid Mech.}, 2018,
  \textbf{50}, 629--659\relax
\mciteBstWouldAddEndPuncttrue
\mciteSetBstMidEndSepPunct{\mcitedefaultmidpunct}
{\mcitedefaultendpunct}{\mcitedefaultseppunct}\relax
\EndOfBibitem
\bibitem[Butt \emph{et~al.}(2018)Butt, Berger, Steffen, Vollmer, and
  Weber]{BBSV2018l}
H.-J. Butt, R.~Berger, W.~Steffen, D.~Vollmer and S.~A.~L. Weber,
  \emph{Langmuir}, 2018, \textbf{34}, 11292--11304\relax
\mciteBstWouldAddEndPuncttrue
\mciteSetBstMidEndSepPunct{\mcitedefaultmidpunct}
{\mcitedefaultendpunct}{\mcitedefaultseppunct}\relax
\EndOfBibitem
\bibitem[Andreotti and Snoeijer(2020)]{AnSn2020arfm}
B.~Andreotti and J.~H. Snoeijer, \emph{Annu. Rev. Fluid Mech.}, 2020,
  \textbf{52}, 285--308\relax
\mciteBstWouldAddEndPuncttrue
\mciteSetBstMidEndSepPunct{\mcitedefaultmidpunct}
{\mcitedefaultendpunct}{\mcitedefaultseppunct}\relax
\EndOfBibitem
\bibitem[Young(1805)]{Youn1805ptrs}
T.~Young, \emph{Phil. Trans. R. Soc.}, 1805, \textbf{95}, 65--87\relax
\mciteBstWouldAddEndPuncttrue
\mciteSetBstMidEndSepPunct{\mcitedefaultmidpunct}
{\mcitedefaultendpunct}{\mcitedefaultseppunct}\relax
\EndOfBibitem
\bibitem[de~Gennes \emph{et~al.}(2004)de~Gennes, Brochard-Wyart,
  Qu{\'e}r{\'e},\emph{et~al.}]{DeGennesBrochard2004}
P.-G. de~Gennes, F.~Brochard-Wyart, D.~Qu{\'e}r{\'e} \emph{et~al.},
  \emph{Capillarity and Wetting Phenomena: Drops, Bubbles, Pearls, Waves},
  Springer, New York, 2004\relax
\mciteBstWouldAddEndPuncttrue
\mciteSetBstMidEndSepPunct{\mcitedefaultmidpunct}
{\mcitedefaultendpunct}{\mcitedefaultseppunct}\relax
\EndOfBibitem
\bibitem[Marchand \emph{et~al.}(2012)Marchand, Das, Snoeijer, and
  Andreotti]{MDSA2012prl}
A.~Marchand, S.~Das, J.~H. Snoeijer and B.~Andreotti, \emph{Physical review
  letters}, 2012, \textbf{109}, 236101\relax
\mciteBstWouldAddEndPuncttrue
\mciteSetBstMidEndSepPunct{\mcitedefaultmidpunct}
{\mcitedefaultendpunct}{\mcitedefaultseppunct}\relax
\EndOfBibitem
\bibitem[Pandey \emph{et~al.}(2020)Pandey, Andreotti, Karpitschka, van Zwieten,
  van Brummelen, and Snoeijer]{PAKZ2020prx}
A.~Pandey, B.~Andreotti, S.~Karpitschka, G.~J. van Zwieten, E.~H. van Brummelen
  and J.~H. Snoeijer, \emph{Phys. Rev. X}, 2020, \textbf{10}, 031067\relax
\mciteBstWouldAddEndPuncttrue
\mciteSetBstMidEndSepPunct{\mcitedefaultmidpunct}
{\mcitedefaultendpunct}{\mcitedefaultseppunct}\relax
\EndOfBibitem
\bibitem[MacDowell and M{\"u}ller(2005)]{MaMu2005jpm}
L.~G. MacDowell and M.~M{\"u}ller, \emph{J. Phys.: Condens. Matter}, 2005,
  \textbf{17}, 3523--3528\relax
\mciteBstWouldAddEndPuncttrue
\mciteSetBstMidEndSepPunct{\mcitedefaultmidpunct}
{\mcitedefaultendpunct}{\mcitedefaultseppunct}\relax
\EndOfBibitem
\bibitem[L{\'e}onforte and M{\"u}ller(2011)]{LeMu2011jcp}
F.~L{\'e}onforte and M.~M{\"u}ller, \emph{J. Chem. Phys.}, 2011, \textbf{135},
  214703\relax
\mciteBstWouldAddEndPuncttrue
\mciteSetBstMidEndSepPunct{\mcitedefaultmidpunct}
{\mcitedefaultendpunct}{\mcitedefaultseppunct}\relax
\EndOfBibitem
\bibitem[Mensink \emph{et~al.}(2019)Mensink, Snoeijer, and de~Beer]{MeSB2019m}
L.~I.~S. Mensink, J.~H. Snoeijer and S.~de~Beer, \emph{Macromolecules}, 2019,
  \textbf{52}, 2015--2020\relax
\mciteBstWouldAddEndPuncttrue
\mciteSetBstMidEndSepPunct{\mcitedefaultmidpunct}
{\mcitedefaultendpunct}{\mcitedefaultseppunct}\relax
\EndOfBibitem
\bibitem[Etha \emph{et~al.}(2021)Etha, Desai, Sachar, and Das]{EDSD2021m}
S.~A. Etha, P.~R. Desai, H.~S. Sachar and S.~Das, \emph{Macromolecules}, 2021,
  \textbf{54}, 584--596\relax
\mciteBstWouldAddEndPuncttrue
\mciteSetBstMidEndSepPunct{\mcitedefaultmidpunct}
{\mcitedefaultendpunct}{\mcitedefaultseppunct}\relax
\EndOfBibitem
\bibitem[M{\"u}ller-Buschbaum \emph{et~al.}(2011)M{\"u}ller-Buschbaum, Magerl,
  Hengstler, Moulin, Korstgens, Diethert, Perlich, Roth, Burghammer, Riekel,
  Gross, Varnik, Uhlmann, Stamm, Feldkamp, and Schroer]{MMHM2011jpm}
P.~M{\"u}ller-Buschbaum, D.~Magerl, R.~Hengstler, J.~F. Moulin, V.~Korstgens,
  A.~Diethert, J.~Perlich, S.~V. Roth, M.~Burghammer, C.~Riekel, M.~Gross,
  F.~Varnik, P.~Uhlmann, M.~Stamm, J.~M. Feldkamp and C.~G. Schroer, \emph{J.
  Phys.: Condens. Matter}, 2011, \textbf{23}, 184111\relax
\mciteBstWouldAddEndPuncttrue
\mciteSetBstMidEndSepPunct{\mcitedefaultmidpunct}
{\mcitedefaultendpunct}{\mcitedefaultseppunct}\relax
\EndOfBibitem
\bibitem[Thiele and Hartmann(2020)]{ThHa2020epjt}
U.~Thiele and S.~Hartmann, \emph{Eur. Phys. J.-Spec. Top.}, 2020, \textbf{229},
  1819–1832\relax
\mciteBstWouldAddEndPuncttrue
\mciteSetBstMidEndSepPunct{\mcitedefaultmidpunct}
{\mcitedefaultendpunct}{\mcitedefaultseppunct}\relax
\EndOfBibitem
\bibitem[Wong \emph{et~al.}(2020)Wong, Hauer, Naga, Kaltbeitzel, Baumli,
  Berger, D'Acunzi, Vollmer, and Butt]{WHNK2020l}
W.~S.~Y. Wong, L.~Hauer, A.~Naga, A.~Kaltbeitzel, P.~Baumli, R.~Berger,
  M.~D'Acunzi, D.~Vollmer and H.~J. Butt, \emph{Langmuir}, 2020, \textbf{36},
  7236--7245\relax
\mciteBstWouldAddEndPuncttrue
\mciteSetBstMidEndSepPunct{\mcitedefaultmidpunct}
{\mcitedefaultendpunct}{\mcitedefaultseppunct}\relax
\EndOfBibitem
\bibitem[Li \emph{et~al.}(2021)Li, Silge, Saal, Kircher, Koynov, Berger, and
  Butt]{LSSK2021l}
X.~Li, S.~Silge, A.~Saal, G.~Kircher, K.~Koynov, R.~Berger and H.-J. Butt,
  \emph{Langmuir}, 2021, \textbf{37}, 1571--1577\relax
\mciteBstWouldAddEndPuncttrue
\mciteSetBstMidEndSepPunct{\mcitedefaultmidpunct}
{\mcitedefaultendpunct}{\mcitedefaultseppunct}\relax
\EndOfBibitem
\bibitem[Henkel \emph{et~al.}(2021)Henkel, Snoeijer, and Thiele]{HeST2021sm}
C.~Henkel, J.~H. Snoeijer and U.~Thiele, \emph{Soft Matter}, 2021, \textbf{17},
  10359--10375\relax
\mciteBstWouldAddEndPuncttrue
\mciteSetBstMidEndSepPunct{\mcitedefaultmidpunct}
{\mcitedefaultendpunct}{\mcitedefaultseppunct}\relax
\EndOfBibitem
\bibitem[Wan \emph{et~al.}(2010)Wan, Meng, Yang, Tian, and Xu]{WMYT2010scc}
L.~Wan, X.~Meng, Y.~Yang, J.~Tian and Z.~Xu, \emph{Sci. China Chem.}, 2010,
  \textbf{53}, 183--189\relax
\mciteBstWouldAddEndPuncttrue
\mciteSetBstMidEndSepPunct{\mcitedefaultmidpunct}
{\mcitedefaultendpunct}{\mcitedefaultseppunct}\relax
\EndOfBibitem
\bibitem[Schubotz \emph{et~al.}(2021)Schubotz, Honnigfort, Nazari, Fery,
  Sommer, Uhlmann, Braunschweig, and Auernhammer]{SHNF2021acis}
S.~Schubotz, C.~Honnigfort, S.~Nazari, A.~Fery, J.-U. Sommer, P.~Uhlmann,
  B.~Braunschweig and G.~K. Auernhammer, \emph{Adv. Colloid Interface Sci.},
  2021, \textbf{294}, 102442\relax
\mciteBstWouldAddEndPuncttrue
\mciteSetBstMidEndSepPunct{\mcitedefaultmidpunct}
{\mcitedefaultendpunct}{\mcitedefaultseppunct}\relax
\EndOfBibitem
\bibitem[Hinduja \emph{et~al.}(2022)Hinduja, Laroche, Shumaly, Wang, Vollmer,
  Butt, and Berger]{HLSW2022l}
C.~Hinduja, A.~Laroche, S.~Shumaly, Y.~Wang, D.~Vollmer, H.-J. Butt and
  R.~Berger, \emph{Langmuir}, 2022, \textbf{38}, 14635--14643\relax
\mciteBstWouldAddEndPuncttrue
\mciteSetBstMidEndSepPunct{\mcitedefaultmidpunct}
{\mcitedefaultendpunct}{\mcitedefaultseppunct}\relax
\EndOfBibitem
\bibitem[Kajiya \emph{et~al.}(2013)Kajiya, Daerr, Narita, Royon, Lequeux, and
  Limat]{KDNR2013sm}
T.~Kajiya, A.~Daerr, T.~Narita, L.~Royon, F.~Lequeux and L.~Limat, \emph{Soft
  Matter}, 2013, \textbf{9}, 454--461\relax
\mciteBstWouldAddEndPuncttrue
\mciteSetBstMidEndSepPunct{\mcitedefaultmidpunct}
{\mcitedefaultendpunct}{\mcitedefaultseppunct}\relax
\EndOfBibitem
\bibitem[Kajiya \emph{et~al.}(2014)Kajiya, Brunet, Royon, Daerr, Receveur, and
  Limat]{KBRD2014sm}
T.~Kajiya, P.~Brunet, L.~Royon, A.~Daerr, M.~Receveur and L.~Limat, \emph{Soft
  Matter}, 2014, \textbf{10}, 8888--8895\relax
\mciteBstWouldAddEndPuncttrue
\mciteSetBstMidEndSepPunct{\mcitedefaultmidpunct}
{\mcitedefaultendpunct}{\mcitedefaultseppunct}\relax
\EndOfBibitem
\bibitem[Karpitschka \emph{et~al.}(2015)Karpitschka, Das, van Gorcum, Perrin,
  Andreotti, and Snoeijer]{KDGP2015nc}
S.~Karpitschka, S.~Das, M.~van Gorcum, H.~Perrin, B.~Andreotti and J.~H.
  Snoeijer, \emph{Nat. Commun.}, 2015, \textbf{6}, 7891\relax
\mciteBstWouldAddEndPuncttrue
\mciteSetBstMidEndSepPunct{\mcitedefaultmidpunct}
{\mcitedefaultendpunct}{\mcitedefaultseppunct}\relax
\EndOfBibitem
\bibitem[Park \emph{et~al.}(2017)Park, Bostwick, De~Andrade, and
  Je]{PBDJ2017sm}
S.~J. Park, J.~B. Bostwick, V.~De~Andrade and J.~H. Je, \emph{Soft Matter},
  2017, \textbf{13}, 8331--8336\relax
\mciteBstWouldAddEndPuncttrue
\mciteSetBstMidEndSepPunct{\mcitedefaultmidpunct}
{\mcitedefaultendpunct}{\mcitedefaultseppunct}\relax
\EndOfBibitem
\bibitem[van Gorcum \emph{et~al.}(2018)van Gorcum, Andreotti, Snoeijer, and
  Karpitschka]{GASK2018prl}
M.~van Gorcum, B.~Andreotti, J.~H. Snoeijer and S.~Karpitschka, \emph{Phys.
  Rev. Lett.}, 2018, \textbf{121}, 208003\relax
\mciteBstWouldAddEndPuncttrue
\mciteSetBstMidEndSepPunct{\mcitedefaultmidpunct}
{\mcitedefaultendpunct}{\mcitedefaultseppunct}\relax
\EndOfBibitem
\bibitem[Mokbel \emph{et~al.}(2022)Mokbel, Aland, and Karpitschka]{MoAK2022el}
D.~Mokbel, S.~Aland and S.~Karpitschka, \emph{Europhys. Lett.}, 2022,
  \textbf{139}, 33002\relax
\mciteBstWouldAddEndPuncttrue
\mciteSetBstMidEndSepPunct{\mcitedefaultmidpunct}
{\mcitedefaultendpunct}{\mcitedefaultseppunct}\relax
\EndOfBibitem
\bibitem[Han and Lin(2012)]{HaLi2012acie}
W.~Han and Z.~Lin, \emph{Angew. Chem. Int. Ed.}, 2012, \textbf{51},
  1534--1546\relax
\mciteBstWouldAddEndPuncttrue
\mciteSetBstMidEndSepPunct{\mcitedefaultmidpunct}
{\mcitedefaultendpunct}{\mcitedefaultseppunct}\relax
\EndOfBibitem
\bibitem[Thiele(2014)]{Thie2014acis}
U.~Thiele, \emph{Adv. Colloid Interface Sci.}, 2014, \textbf{206},
  399--413\relax
\mciteBstWouldAddEndPuncttrue
\mciteSetBstMidEndSepPunct{\mcitedefaultmidpunct}
{\mcitedefaultendpunct}{\mcitedefaultseppunct}\relax
\EndOfBibitem
\bibitem[Cubaud and Fermigier(2001)]{CuFe2001el}
T.~Cubaud and M.~Fermigier, \emph{Europhys. Lett.}, 2001, \textbf{55},
  239--245\relax
\mciteBstWouldAddEndPuncttrue
\mciteSetBstMidEndSepPunct{\mcitedefaultmidpunct}
{\mcitedefaultendpunct}{\mcitedefaultseppunct}\relax
\EndOfBibitem
\bibitem[Thiele and Knobloch(2006)]{ThKn2006njp}
U.~Thiele and E.~Knobloch, \emph{New J. Phys.}, 2006, \textbf{8}, 313\relax
\mciteBstWouldAddEndPuncttrue
\mciteSetBstMidEndSepPunct{\mcitedefaultmidpunct}
{\mcitedefaultendpunct}{\mcitedefaultseppunct}\relax
\EndOfBibitem
\bibitem[Zhang and Mi(2009)]{ZhMi2009l}
X.~Y. Zhang and Y.~L. Mi, \emph{Langmuir}, 2009, \textbf{25}, 3212--3218\relax
\mciteBstWouldAddEndPuncttrue
\mciteSetBstMidEndSepPunct{\mcitedefaultmidpunct}
{\mcitedefaultendpunct}{\mcitedefaultseppunct}\relax
\EndOfBibitem
\bibitem[Beltrame \emph{et~al.}(2011)Beltrame, Knobloch, H{\"a}nggi, and
  Thiele]{BKHT2011pre}
P.~Beltrame, E.~Knobloch, P.~H{\"a}nggi and U.~Thiele, \emph{Phys. Rev. E},
  2011, \textbf{83}, 016305\relax
\mciteBstWouldAddEndPuncttrue
\mciteSetBstMidEndSepPunct{\mcitedefaultmidpunct}
{\mcitedefaultendpunct}{\mcitedefaultseppunct}\relax
\EndOfBibitem
\bibitem[Savva and Kalliadasis(2013)]{SaKa2013jfm}
N.~Savva and S.~Kalliadasis, \emph{J. Fluid Mech.}, 2013, \textbf{725},
  462--491\relax
\mciteBstWouldAddEndPuncttrue
\mciteSetBstMidEndSepPunct{\mcitedefaultmidpunct}
{\mcitedefaultendpunct}{\mcitedefaultseppunct}\relax
\EndOfBibitem
\bibitem[Varagnolo \emph{et~al.}(2013)Varagnolo, Ferraro, Fantinel, Pierno,
  Mistura, Amati, Biferale, and Sbragaglia]{VFFP2013prl}
S.~Varagnolo, D.~Ferraro, P.~Fantinel, M.~Pierno, G.~Mistura, G.~Amati,
  L.~Biferale and M.~Sbragaglia, \emph{Phys. Rev. Lett.}, 2013, \textbf{111},
  066101\relax
\mciteBstWouldAddEndPuncttrue
\mciteSetBstMidEndSepPunct{\mcitedefaultmidpunct}
{\mcitedefaultendpunct}{\mcitedefaultseppunct}\relax
\EndOfBibitem
\bibitem[Varagnolo \emph{et~al.}(2014)Varagnolo, Schiocchet, Ferraro, Pierno,
  Mistura, Sbragaglia, Gupta, and Amati]{VSFP2014l}
S.~Varagnolo, V.~Schiocchet, D.~Ferraro, M.~Pierno, G.~Mistura, M.~Sbragaglia,
  A.~Gupta and G.~Amati, \emph{Langmuir}, 2014, \textbf{30}, 2401--2409\relax
\mciteBstWouldAddEndPuncttrue
\mciteSetBstMidEndSepPunct{\mcitedefaultmidpunct}
{\mcitedefaultendpunct}{\mcitedefaultseppunct}\relax
\EndOfBibitem
\bibitem[Sbragaglia \emph{et~al.}(2014)Sbragaglia, Biferale, Amati, Varagnolo,
  Ferraro, Mistura, and Pierno]{SBAV2014pre}
M.~Sbragaglia, L.~Biferale, G.~Amati, S.~Varagnolo, D.~Ferraro, G.~Mistura and
  M.~Pierno, \emph{Phys. Rev. E}, 2014, \textbf{89}, 012406\relax
\mciteBstWouldAddEndPuncttrue
\mciteSetBstMidEndSepPunct{\mcitedefaultmidpunct}
{\mcitedefaultendpunct}{\mcitedefaultseppunct}\relax
\EndOfBibitem
\bibitem[Sch{\"a}ffer and Wong(1998)]{ScWo1998prl}
E.~Sch{\"a}ffer and P.~Z. Wong, \emph{Phys. Rev. Lett.}, 1998, \textbf{80},
  3069--3072\relax
\mciteBstWouldAddEndPuncttrue
\mciteSetBstMidEndSepPunct{\mcitedefaultmidpunct}
{\mcitedefaultendpunct}{\mcitedefaultseppunct}\relax
\EndOfBibitem
\bibitem[Cubaud and Fermigier(2004)]{CuFe2004jcis}
T.~Cubaud and A.~Fermigier, \emph{J. Colloid Interface Sci.}, 2004,
  \textbf{269}, 171--177\relax
\mciteBstWouldAddEndPuncttrue
\mciteSetBstMidEndSepPunct{\mcitedefaultmidpunct}
{\mcitedefaultendpunct}{\mcitedefaultseppunct}\relax
\EndOfBibitem
\bibitem[Tavana \emph{et~al.}(2006)Tavana, Yang, Yip, Appelhans, Zschoche,
  Grundke, Hair, and Neumann]{TYYA2006l}
H.~Tavana, G.~C. Yang, C.~M. Yip, D.~Appelhans, S.~Zschoche, K.~Grundke, M.~L.
  Hair and A.~W. Neumann, \emph{Langmuir}, 2006, \textbf{22}, 628--636\relax
\mciteBstWouldAddEndPuncttrue
\mciteSetBstMidEndSepPunct{\mcitedefaultmidpunct}
{\mcitedefaultendpunct}{\mcitedefaultseppunct}\relax
\EndOfBibitem
\bibitem[Savva \emph{et~al.}(2010)Savva, Kalliadasis, and
  Pavliotis]{SaKP2010prl}
N.~Savva, S.~Kalliadasis and G.~A. Pavliotis, \emph{Phys. Rev. Lett.}, 2010,
  \textbf{104}, 084501\relax
\mciteBstWouldAddEndPuncttrue
\mciteSetBstMidEndSepPunct{\mcitedefaultmidpunct}
{\mcitedefaultendpunct}{\mcitedefaultseppunct}\relax
\EndOfBibitem
\bibitem[Bodiguel \emph{et~al.}(2010)Bodiguel, Doumenc, and
  Guerrier]{BoDG2010l}
H.~Bodiguel, F.~Doumenc and B.~Guerrier, \emph{Langmuir}, 2010, \textbf{26},
  10758--10763\relax
\mciteBstWouldAddEndPuncttrue
\mciteSetBstMidEndSepPunct{\mcitedefaultmidpunct}
{\mcitedefaultendpunct}{\mcitedefaultseppunct}\relax
\EndOfBibitem
\bibitem[Larson(2014)]{Lars2014aj}
R.~G. Larson, \emph{Aiche J.}, 2014, \textbf{60}, 1538--1571\relax
\mciteBstWouldAddEndPuncttrue
\mciteSetBstMidEndSepPunct{\mcitedefaultmidpunct}
{\mcitedefaultendpunct}{\mcitedefaultseppunct}\relax
\EndOfBibitem
\bibitem[Jabal \emph{et~al.}(2018)Jabal, Egbaria, Zigelman, Thiele, and
  Manor]{JEZT2018l}
M.~A. Jabal, A.~Egbaria, A.~Zigelman, U.~Thiele and O.~Manor, \emph{Langmuir},
  2018, \textbf{34}, 11784--11794\relax
\mciteBstWouldAddEndPuncttrue
\mciteSetBstMidEndSepPunct{\mcitedefaultmidpunct}
{\mcitedefaultendpunct}{\mcitedefaultseppunct}\relax
\EndOfBibitem
\bibitem[Spratte \emph{et~al.}(1994)Spratte, Chi, and Riegler]{SpCR1994el}
K.~Spratte, L.~F. Chi and H.~Riegler, \emph{Europhys. Lett.}, 1994,
  \textbf{25}, 211--217\relax
\mciteBstWouldAddEndPuncttrue
\mciteSetBstMidEndSepPunct{\mcitedefaultmidpunct}
{\mcitedefaultendpunct}{\mcitedefaultseppunct}\relax
\EndOfBibitem
\bibitem[Li \emph{et~al.}(2012)Li, K{\"o}pf, Gurevich, Friedrich, and
  Chi]{LKGF2012s}
L.~Q. Li, M.~H. K{\"o}pf, S.~V. Gurevich, R.~Friedrich and L.~F. Chi,
  \emph{Small}, 2012, \textbf{8}, 488--503\relax
\mciteBstWouldAddEndPuncttrue
\mciteSetBstMidEndSepPunct{\mcitedefaultmidpunct}
{\mcitedefaultendpunct}{\mcitedefaultseppunct}\relax
\EndOfBibitem
\bibitem[Deegan(2000)]{Deeg2000pre}
R.~D. Deegan, \emph{Phys. Rev. E}, 2000, \textbf{61}, 475--485\relax
\mciteBstWouldAddEndPuncttrue
\mciteSetBstMidEndSepPunct{\mcitedefaultmidpunct}
{\mcitedefaultendpunct}{\mcitedefaultseppunct}\relax
\EndOfBibitem
\bibitem[Zhang \emph{et~al.}(2017)Zhang, Guy, Lasheras, and del
  Alamo]{ZGLA2017jpdp}
S.~Zhang, R.~D. Guy, J.~C. Lasheras and J.~C. del Alamo, \emph{J. Phys. D-Appl.
  Phys.}, 2017, \textbf{50}, 204004\relax
\mciteBstWouldAddEndPuncttrue
\mciteSetBstMidEndSepPunct{\mcitedefaultmidpunct}
{\mcitedefaultendpunct}{\mcitedefaultseppunct}\relax
\EndOfBibitem
\bibitem[Ron \emph{et~al.}(2020)Ron, Monzo, Gauthier, Voituriez, and
  Gov]{RMGV2020prr}
J.~E. Ron, P.~Monzo, N.~C. Gauthier, R.~Voituriez and N.~S. Gov, \emph{Phys.
  Rev. Research}, 2020, \textbf{2}, 033237\relax
\mciteBstWouldAddEndPuncttrue
\mciteSetBstMidEndSepPunct{\mcitedefaultmidpunct}
{\mcitedefaultendpunct}{\mcitedefaultseppunct}\relax
\EndOfBibitem
\bibitem[Fra{\v{s}}tia \emph{et~al.}(2012)Fra{\v{s}}tia, Archer, and
  Thiele]{FrAT2012sm}
L.~Fra{\v{s}}tia, A.~J. Archer and U.~Thiele, \emph{Soft Matter}, 2012,
  \textbf{8}, 11363--11386\relax
\mciteBstWouldAddEndPuncttrue
\mciteSetBstMidEndSepPunct{\mcitedefaultmidpunct}
{\mcitedefaultendpunct}{\mcitedefaultseppunct}\relax
\EndOfBibitem
\bibitem[K{\"o}pf and Thiele(2014)]{KoTh2014n}
M.~H. K{\"o}pf and U.~Thiele, \emph{Nonlinearity}, 2014, \textbf{27},
  2711--2734\relax
\mciteBstWouldAddEndPuncttrue
\mciteSetBstMidEndSepPunct{\mcitedefaultmidpunct}
{\mcitedefaultendpunct}{\mcitedefaultseppunct}\relax
\EndOfBibitem
\bibitem[Tewes \emph{et~al.}(2019)Tewes, Wilczek, Gurevich, and
  Thiele]{TWGT2019prf}
W.~Tewes, M.~Wilczek, S.~V. Gurevich and U.~Thiele, \emph{Phys. Rev. Fluids},
  2019, \textbf{4}, 123903\relax
\mciteBstWouldAddEndPuncttrue
\mciteSetBstMidEndSepPunct{\mcitedefaultmidpunct}
{\mcitedefaultendpunct}{\mcitedefaultseppunct}\relax
\EndOfBibitem
\bibitem[Mitas \emph{et~al.}(2021)Mitas, Manor, and Thiele]{MiMT2021prf}
K.~D.~J. Mitas, O.~Manor and U.~Thiele, \emph{Phys. Rev. Fluids}, 2021,
  \textbf{6}, 094002\relax
\mciteBstWouldAddEndPuncttrue
\mciteSetBstMidEndSepPunct{\mcitedefaultmidpunct}
{\mcitedefaultendpunct}{\mcitedefaultseppunct}\relax
\EndOfBibitem
\bibitem[Thiele(2018)]{Thie2018csa}
U.~Thiele, \emph{Colloid Surf. A}, 2018, \textbf{553}, 487--495\relax
\mciteBstWouldAddEndPuncttrue
\mciteSetBstMidEndSepPunct{\mcitedefaultmidpunct}
{\mcitedefaultendpunct}{\mcitedefaultseppunct}\relax
\EndOfBibitem
\bibitem[Mitlin(1993)]{Mitl1993jcis}
V.~S. Mitlin, \emph{J. Colloid Interface Sci.}, 1993, \textbf{156},
  491--497\relax
\mciteBstWouldAddEndPuncttrue
\mciteSetBstMidEndSepPunct{\mcitedefaultmidpunct}
{\mcitedefaultendpunct}{\mcitedefaultseppunct}\relax
\EndOfBibitem
\bibitem[Thiele(2010)]{Thie2010jpcm}
U.~Thiele, \emph{J. Phys.: Condens. Matter}, 2010, \textbf{22}, 084019\relax
\mciteBstWouldAddEndPuncttrue
\mciteSetBstMidEndSepPunct{\mcitedefaultmidpunct}
{\mcitedefaultendpunct}{\mcitedefaultseppunct}\relax
\EndOfBibitem
\bibitem[Pototsky \emph{et~al.}(2005)Pototsky, Bestehorn, Merkt, and
  Thiele]{PBMT2005jcp}
A.~Pototsky, M.~Bestehorn, D.~Merkt and U.~Thiele, \emph{J. Chem. Phys.}, 2005,
  \textbf{122}, 224711\relax
\mciteBstWouldAddEndPuncttrue
\mciteSetBstMidEndSepPunct{\mcitedefaultmidpunct}
{\mcitedefaultendpunct}{\mcitedefaultseppunct}\relax
\EndOfBibitem
\bibitem[Bommer \emph{et~al.}(2013)Bommer, Cartellier, Jachalski, Peschka,
  Seemann, and Wagner]{BCJP2013epje}
S.~Bommer, F.~Cartellier, S.~Jachalski, D.~Peschka, R.~Seemann and B.~Wagner,
  \emph{Eur. Phys. J. E}, 2013, \textbf{36}, 87\relax
\mciteBstWouldAddEndPuncttrue
\mciteSetBstMidEndSepPunct{\mcitedefaultmidpunct}
{\mcitedefaultendpunct}{\mcitedefaultseppunct}\relax
\EndOfBibitem
\bibitem[Thiele \emph{et~al.}(2016)Thiele, Archer, and Pismen]{ThAP2016prf}
U.~Thiele, A.~J. Archer and L.~M. Pismen, \emph{Phys. Rev. Fluids}, 2016,
  \textbf{1}, 083903\relax
\mciteBstWouldAddEndPuncttrue
\mciteSetBstMidEndSepPunct{\mcitedefaultmidpunct}
{\mcitedefaultendpunct}{\mcitedefaultseppunct}\relax
\EndOfBibitem
\bibitem[Thiele \emph{et~al.}(2013)Thiele, Todorova, and Lopez]{ThTL2013prl}
U.~Thiele, D.~V. Todorova and H.~Lopez, \emph{Phys. Rev. Lett.}, 2013,
  \textbf{111}, 117801\relax
\mciteBstWouldAddEndPuncttrue
\mciteSetBstMidEndSepPunct{\mcitedefaultmidpunct}
{\mcitedefaultendpunct}{\mcitedefaultseppunct}\relax
\EndOfBibitem
\bibitem[Hartmann \emph{et~al.}(2022)Hartmann, Diddens, Jalaal, and
  Thiele]{HDJT2022apa}
S.~Hartmann, C.~Diddens, M.~Jalaal and U.~Thiele, \emph{arXiv preprint
  arXiv:2206.14595}, 2022\relax
\mciteBstWouldAddEndPuncttrue
\mciteSetBstMidEndSepPunct{\mcitedefaultmidpunct}
{\mcitedefaultendpunct}{\mcitedefaultseppunct}\relax
\EndOfBibitem
\bibitem[Trinschek \emph{et~al.}(2018)Trinschek, John, and Thiele]{TrJT2018sm}
S.~Trinschek, K.~John and U.~Thiele, \emph{Soft Matter}, 2018, \textbf{14},
  4464--4476\relax
\mciteBstWouldAddEndPuncttrue
\mciteSetBstMidEndSepPunct{\mcitedefaultmidpunct}
{\mcitedefaultendpunct}{\mcitedefaultseppunct}\relax
\EndOfBibitem
\bibitem[Stegemerten \emph{et~al.}(2022)Stegemerten, John, and
  Thiele]{StJT2022sm}
F.~Stegemerten, K.~John and U.~Thiele, \emph{Soft Matter}, 2022, \textbf{18},
  5823--5832\relax
\mciteBstWouldAddEndPuncttrue
\mciteSetBstMidEndSepPunct{\mcitedefaultmidpunct}
{\mcitedefaultendpunct}{\mcitedefaultseppunct}\relax
\EndOfBibitem
\bibitem[Oron \emph{et~al.}(1997)Oron, Davis, and Bankoff]{OrDB1997rmp}
A.~Oron, S.~H. Davis and S.~G. Bankoff, \emph{Rev. Mod. Phys.}, 1997,
  \textbf{69}, 931--980\relax
\mciteBstWouldAddEndPuncttrue
\mciteSetBstMidEndSepPunct{\mcitedefaultmidpunct}
{\mcitedefaultendpunct}{\mcitedefaultseppunct}\relax
\EndOfBibitem
\bibitem[Thiele \emph{et~al.}(2018)Thiele, Snoeijer, Trinschek, and
  John]{TSTJ2018l}
U.~Thiele, J.~H. Snoeijer, S.~Trinschek and K.~John, \emph{Langmuir}, 2018,
  \textbf{34}, 7210--7221\relax
\mciteBstWouldAddEndPuncttrue
\mciteSetBstMidEndSepPunct{\mcitedefaultmidpunct}
{\mcitedefaultendpunct}{\mcitedefaultseppunct}\relax
\EndOfBibitem
\bibitem[Pismen(2001)]{Pism2001pre}
L.~M. Pismen, \emph{Phys. Rev. E}, 2001, \textbf{64}, 021603\relax
\mciteBstWouldAddEndPuncttrue
\mciteSetBstMidEndSepPunct{\mcitedefaultmidpunct}
{\mcitedefaultendpunct}{\mcitedefaultseppunct}\relax
\EndOfBibitem
\bibitem[de~Gennes(1991)]{Genn1991crasi}
P.~G. de~Gennes, \emph{C. R. Acad. Sci. II}, 1991, \textbf{313},
  1117--1122\relax
\mciteBstWouldAddEndPuncttrue
\mciteSetBstMidEndSepPunct{\mcitedefaultmidpunct}
{\mcitedefaultendpunct}{\mcitedefaultseppunct}\relax
\EndOfBibitem
\bibitem[Alexander(1977)]{Alex1977jp}
S.~Alexander, \emph{J Phys-Paris}, 1977, \textbf{38}, 983--987\relax
\mciteBstWouldAddEndPuncttrue
\mciteSetBstMidEndSepPunct{\mcitedefaultmidpunct}
{\mcitedefaultendpunct}{\mcitedefaultseppunct}\relax
\EndOfBibitem
\bibitem[Sommer(2017)]{Somm2017m}
J.~Sommer, \emph{Macromolecules}, 2017, \textbf{50}, 2219--2228\relax
\mciteBstWouldAddEndPuncttrue
\mciteSetBstMidEndSepPunct{\mcitedefaultmidpunct}
{\mcitedefaultendpunct}{\mcitedefaultseppunct}\relax
\EndOfBibitem
\bibitem[Flory(1953)]{Flor1953}
P.~J. Flory, \emph{Principles of Polymer Chemistry}, Cornell University Press,
  Ithaca, 1953\relax
\mciteBstWouldAddEndPuncttrue
\mciteSetBstMidEndSepPunct{\mcitedefaultmidpunct}
{\mcitedefaultendpunct}{\mcitedefaultseppunct}\relax
\EndOfBibitem
\bibitem[Gauglitz and Radke(1988)]{GaRa1988ces}
P.~A. Gauglitz and C.~J. Radke, \emph{Chem. Eng. Sci.}, 1988, \textbf{43},
  1457--1465\relax
\mciteBstWouldAddEndPuncttrue
\mciteSetBstMidEndSepPunct{\mcitedefaultmidpunct}
{\mcitedefaultendpunct}{\mcitedefaultseppunct}\relax
\EndOfBibitem
\bibitem[Snoeijer(2006)]{Snoe2006pf}
J.~H. Snoeijer, \emph{Phys. Fluids}, 2006, \textbf{18}, 021701\relax
\mciteBstWouldAddEndPuncttrue
\mciteSetBstMidEndSepPunct{\mcitedefaultmidpunct}
{\mcitedefaultendpunct}{\mcitedefaultseppunct}\relax
\EndOfBibitem
\bibitem[von Borries~Lopes \emph{et~al.}(2018)von Borries~Lopes, Thiele, and
  Hazel]{BoTH2018jfm}
A.~von Borries~Lopes, U.~Thiele and A.~L. Hazel, \emph{J. Fluid Mech.}, 2018,
  \textbf{835}, 540--574\relax
\mciteBstWouldAddEndPuncttrue
\mciteSetBstMidEndSepPunct{\mcitedefaultmidpunct}
{\mcitedefaultendpunct}{\mcitedefaultseppunct}\relax
\EndOfBibitem
\bibitem[Tretyakov \emph{et~al.}(2013)Tretyakov, M\"{u}ller, Todorova, and
  Thiele]{TMTT2013jcp}
N.~Tretyakov, M.~M\"{u}ller, D.~Todorova and U.~Thiele, \emph{J. Chem. Phys.},
  2013, \textbf{138}, 064905\relax
\mciteBstWouldAddEndPuncttrue
\mciteSetBstMidEndSepPunct{\mcitedefaultmidpunct}
{\mcitedefaultendpunct}{\mcitedefaultseppunct}\relax
\EndOfBibitem
\bibitem[Heil and Hazel(2006)]{HeHa2006}
M.~Heil and A.~L. Hazel, in \emph{Fluid-Structure Interaction: Modelling,
  Simulation, Optimisation}, ed. H.-J. Bungartz and M.~Sch{\"a}fer, Springer,
  Berlin, Heidelberg, 2006, pp. 19--49\relax
\mciteBstWouldAddEndPuncttrue
\mciteSetBstMidEndSepPunct{\mcitedefaultmidpunct}
{\mcitedefaultendpunct}{\mcitedefaultseppunct}\relax
\EndOfBibitem
\bibitem[Quere(2008)]{Quer2008armr}
D.~Quere, \emph{Ann. Rev. Mater. Res.}, 2008, \textbf{38}, 71--99\relax
\mciteBstWouldAddEndPuncttrue
\mciteSetBstMidEndSepPunct{\mcitedefaultmidpunct}
{\mcitedefaultendpunct}{\mcitedefaultseppunct}\relax
\EndOfBibitem
\bibitem[Snoeijer \emph{et~al.}(2007)Snoeijer, Andreotti, Delon, and
  Fermigier]{SADF2007jfm}
J.~H. Snoeijer, B.~Andreotti, G.~Delon and M.~Fermigier, \emph{J. Fluid Mech.},
  2007, \textbf{579}, 63--83\relax
\mciteBstWouldAddEndPuncttrue
\mciteSetBstMidEndSepPunct{\mcitedefaultmidpunct}
{\mcitedefaultendpunct}{\mcitedefaultseppunct}\relax
\EndOfBibitem
\bibitem[Maleki \emph{et~al.}(2011)Maleki, Reyssat, Restagno, Qu{\'e}r{\'e},
  and Clanet]{MRRQ2011jcis}
M.~Maleki, M.~Reyssat, F.~Restagno, D.~Qu{\'e}r{\'e} and C.~Clanet, \emph{J.
  Colloid Interface Sci.}, 2011, \textbf{354}, 359--363\relax
\mciteBstWouldAddEndPuncttrue
\mciteSetBstMidEndSepPunct{\mcitedefaultmidpunct}
{\mcitedefaultendpunct}{\mcitedefaultseppunct}\relax
\EndOfBibitem
\bibitem[Aland and Mokbel(2021)]{AlMo2021ijnme}
S.~Aland and D.~Mokbel, \emph{Int. J. Numer. Methods Eng.}, 2021, \textbf{122},
  903--918\relax
\mciteBstWouldAddEndPuncttrue
\mciteSetBstMidEndSepPunct{\mcitedefaultmidpunct}
{\mcitedefaultendpunct}{\mcitedefaultseppunct}\relax
\EndOfBibitem
\bibitem[Tsoumpas \emph{et~al.}(2014)Tsoumpas, Dehaeck, Galvagno, Rednikov,
  Ottevaere, Thiele, and Colinet]{TDGR2014l}
Y.~Tsoumpas, S.~Dehaeck, M.~Galvagno, A.~Rednikov, H.~Ottevaere, U.~Thiele and
  P.~Colinet, \emph{Langmuir}, 2014, \textbf{30}, 11847--11852\relax
\mciteBstWouldAddEndPuncttrue
\mciteSetBstMidEndSepPunct{\mcitedefaultmidpunct}
{\mcitedefaultendpunct}{\mcitedefaultseppunct}\relax
\EndOfBibitem
\bibitem[Lide(2004)]{Lide2004}
D.~R. Lide, \emph{{CRC} handbook of chemistry and physics}, CRC Press, Boca
  Raton, 85th edn, 2004\relax
\mciteBstWouldAddEndPuncttrue
\mciteSetBstMidEndSepPunct{\mcitedefaultmidpunct}
{\mcitedefaultendpunct}{\mcitedefaultseppunct}\relax
\EndOfBibitem
\end{mcitethebibliography}
\bibliographystyle{rsc}

\appendix
\section{Appendix}\label{sec:appendix}

\subsection{Dimensionless formulation and long-wave expansion of the model equations}\label{sec:nondim_and_longwave}
Here, we give the details of the steps we have performed to prepare the model for the numerical analysis in order to reduce the computational effort, i.e., the nondimensionalisation and rescaling of the variables and parameters, and the long-wave expansion of the interface terms. We then proceed to give details on the numerical implementation, i.e.\ the employed techniques, software, and the initial and boundary conditions.

\subsubsection{Full model equations}\label{app:full_model}
For reference, we start with a brief summary of the complete model. As per Eq.~\eqref{eq:gradient_model} the dimensional model reads: 
\begin{equation}
    \begin{aligned}
        \partial_t h     &= \nabla \cdot \left[\frac{h^3}{3\eta}\,\nabla\frac{\delta\mathcal{F}}{\delta h}\right] \ \ - \ M\left[\frac{\delta\mathcal{F}}{\delta h}-\frac{\delta\mathcal{F}}{\delta\zeta}\right] \ + \ U\nabla h\\
        \partial_t \zeta &=\nabla \cdot \left[D\zeta\,\nabla\frac{\delta\mathcal{F}}{\delta \zeta}\right] \ - \ M\left[\frac{\delta\mathcal{F}}{\delta \zeta}-\frac{\delta\mathcal{F}}{\delta h}\right] \ + \ U\nabla \zeta,
    \end{aligned}\label{eq:full_model1}
\end{equation}
The variations of the energy functional $\mathcal{F}$ are [Eqs.~\eqref{eq:dFdh} \& \eqref{eq:dFdz}]
\begin{align}
    \frac{\delta \mathcal{F}}{\delta h}     =& -\gamma\frac{\nabla^2(h+\zeta)}{\xi_{h+\zeta}^{3}} + \xi_\zeta\partial_h f_\mathrm{wet}\label{eq:full_model2}\\
    \frac{\delta \mathcal{F}}{\delta \zeta} =& -\gamma\frac{\nabla^2(h+\zeta)}{\xi_{h+\zeta}^{3}} - \nabla \cdot \left[(\gamma_\mathrm{bl} + f_\mathrm{wet})\frac{\nabla\zeta}{\xi_\zeta}\right]\nonumber\\
    &+ \xi_\zeta \partial_\zeta (\gamma_\mathrm{bl} + f_\mathrm{wet}) + \partial_\zeta g_\mathrm{brush}.\label{eq:full_model3}
\end{align}
The derivative of the wetting potential (Eq.~\eqref{eq:wetting_potential}) gives the Derjaguin (disjoining) pressure that we further rewrite using Young's law~\eqref{eq:young} into
\begin{equation}
    \Pi(h, \zeta) = -\partial_h f_\mathrm{wet} = \frac{B(\zeta)}{h^6}-\frac{A(\zeta)}{h^3} = \frac{5}{3}\gamma h_p^2 \theta_Y^2 \left(\frac{h_p^3}{h^6} - \frac{1}{h^3}\right). \label{eq:disjoining_pressure}
\end{equation}
The derivative of the brush potential evaluates to
\begin{equation}
    \partial_\zeta g_\mathrm{brush}(\zeta) = \frac{k_B T}{\ell_K^3}\left[\sigma^2/c + c + \log(1-c)\right]\label{eq:d_g_brush}
\end{equation}
with the polymer volume fraction
\begin{equation}
    c = \frac{H_\mathrm{dry}}{H_\mathrm{dry} + \zeta}
\end{equation}
and the dry brush height $H_\mathrm{dry} = \sigma N \ell_K$.

In a first simplification, we consider one spatial dimension, i.e. $\mathbf{x} \to x$.
Moreover, following~\citet{ThHa2020epjt} we neglect the dependency of the wetting potential and brush-liquid interface energy on the brush state, namely, $f_\mathrm{wet}(h, \zeta) = f_\mathrm{wet}(h)$ and ${\gamma_\mathrm{bl}=\text{const}}$.

\subsubsection{Scaling and long-wave expansion}
\label{app:lw}
A dimensionless formulation reduces the model's complexity by eliminating some parameters and by converting the magnitude of all variables to scales that minimise numerical errors. Thus, we introduce a height scale $h_0=h_p$, a lateral length scale $x_0=\sqrt{\gamma/A_0}h_p^2$ and a time scale $t_0=3\eta \gamma h_p^5/A_0^2$ and transform to the dimensionless variables (marked with a tilde) by using
\begin{align}
    h = h_p \tilde h,\quad
    & &\zeta = h_p \tilde \zeta,\quad
    & &x = x_0 \tilde x,\quad
    & &t = t_0 \tilde t.
\end{align}
Here, the parameter $A_0$ denotes the Hamaker constant for a dry brush, see Eq.~\eqref{eq:brush_dependencies}.

For shallow drop, film and brush profiles, we can assume that all occurring height scales are much smaller than the lateral length scales. We hence introduce a ``smallness'' parameter $\epsilon=h_0/x_0$ and expand the model up to second order in $\epsilon$. This is known as the long-wave approximation and is the standard approach for deriving simplified equation for the interface dynamics from the Stokes equation.\cite{OrDB1997rmp} In the long-wave limit, the metric factors then expand to
\begin{equation}
    \begin{aligned}
        &\xi_\zeta=\sqrt{1+\partial_x \zeta^2}\approx 1+\frac{\epsilon^2}{2}(\partial_{\tilde x}\tilde \zeta)^2\\
        \text{and}\quad &\xi_{h+\zeta}=\sqrt{1+(\partial_x h + \partial_x \zeta)^2}\approx 1+\frac{\epsilon^2}{2}[\partial_{\tilde x} (\tilde h+\tilde \zeta)]^2.
    \end{aligned}
\end{equation}

The rescaled variations of the free energy then become
\begin{equation}
    \begin{aligned}
        \frac{\delta \tilde{\mathcal{F}}}{\delta \tilde h}     &= -\partial_{\tilde x}^2(\tilde h + \tilde \zeta) + \frac{1}{h^6} - \frac{1}{h^3}\\
        \frac{\delta \tilde{\mathcal{F}}}{\delta \tilde \zeta} &= -\partial_{\tilde x}^2(\tilde h + \tilde \zeta) - \tilde \gamma_\mathrm{bl} \partial_{\tilde x}^2 \tilde \zeta + \tilde T \left[\sigma^2/c + c + \log(1-c)\right],
    \end{aligned}\label{eq:dFdh_dFdz_dimless}
\end{equation}
where we have introduced the dimensionless interface energy $\tilde \gamma_\mathrm{bl} = \gamma_\mathrm{bl} / \gamma$ and the dimensionless temperature $\tilde T = (h_p^3 k_B T)/(A_0 \ell_K^3)$. The functional variations have units of a pressure and transform as $\delta \mathcal{F}/\delta h = A_0 / h_p^3\,\delta \tilde{\mathcal{F}}/\delta \tilde h$.
Note that the metric factor in front of the wetting potential becomes one because the wetting potential is intrinsically of order~$\epsilon^2$, as also seen when approximating the mesoscopic Young's law~\eqref{eq:young}.

The dimensionless part per volume polymer concentration is
\begin{equation}
    c = \frac{\sigma \tilde l}{\sigma \tilde l + \tilde \zeta}
\end{equation}
where $\tilde l=N \ell_k / h_p$ is the dimensionless polymer length.

In the final step, we apply the scales to the dynamical equations, effectively eliminating the viscosity $\eta$:
\begin{equation}
    \begin{aligned}
        \partial_{\tilde t} \tilde h     &= \partial_{\tilde x} \cdot \left[\tilde h^3\,\partial_{\tilde x}\frac{\delta \tilde{\mathcal{F}}}{\delta \tilde h}\right] \ \ - \ \tilde M\left[\frac{\delta \tilde{\mathcal{F}}}{\delta \tilde h}-\frac{\delta \tilde{\mathcal{F}}}{\delta \tilde \zeta}\right] \ + \ \tilde U\partial_{\tilde x} \tilde h\\
        \partial_{\tilde t} \tilde \zeta &=\partial_{\tilde x} \cdot \left[\tilde D\tilde \zeta\,\partial_{\tilde x}\frac{\delta \tilde{\mathcal{F}}}{\delta \tilde \zeta}\right] \ - \ \tilde M\left[\frac{\delta \tilde{\mathcal{F}}}{\delta \tilde \zeta}-\frac{\delta\mathcal{F}}{\delta h}\right] \ + \ \tilde U\partial_{\tilde x} \tilde \zeta.
    \end{aligned}\label{eq:model_dimless}
\end{equation}
The remaining model parameters 
\begin{equation}
    \begin{aligned}
        & &\tilde \gamma_\mathrm{bl} = \gamma_\mathrm{bl} / \gamma,\quad
        & &\tilde l=N \ell_k / h_p,\quad
        & &\tilde T = \frac{h_p^3 k_B T}{A_0 \ell_K^3},\\
        & &\tilde M = \frac{3 M \gamma h_p \eta}{A_0},\quad
        & &\tilde U = \frac{3 U h_p^3 \eta \sqrt{\gamma}}{\sqrt{A_0}^3},\quad
        & &\tilde D = \frac{3 \eta D}{h_p^2}.
    \end{aligned}
\end{equation}
are dimensionless combinations of the original dimensional parameters. Recovering the dimensionless wetting potential,
\begin{align}
    \tilde f_\mathrm{wet}(h) = -\frac{1}{2h^2}+\frac{1}{5h^5},\label{eq:f_wet_dimless}
\end{align}
and then following the derivation in Sec.~\ref{sec:young}, we find an approximate expression for the equilibrium contact angle as
\begin{equation}
    \theta_{\mathrm{Y}}=-\sqrt{-2f_\mathrm{wet}(h_p)}.\label{eq:young_CA_approx}
\end{equation}
Further, in the long-wave approximation the Neumann law from Section~\ref{sec:neumann} becomes
\begin{align}\label{eq:app:neumann}
    \theta_\mathrm{BL} &=\theta_\mathrm{LG}+\sqrt{\frac{1+\Tilde{\gamma}_\mathrm{bl}}{\Tilde{\gamma}_\mathrm{bl}}}\sqrt{-2 f_\mathrm{wet}(h_\mathrm{p})}\\
    \theta_\mathrm{BG} &=\theta_\mathrm{LG}+\sqrt{\frac{\Tilde{\gamma}_\mathrm{bl}}{1+\Tilde{\gamma}_\mathrm{bl}}}\sqrt{-2 f_\mathrm{wet}(h_\mathrm{p})}.
\end{align}

\subsubsection{Exemplary dimensional scaling}\label{app:real_scales}

As an example, we demonstrate how to scale back our results to dimensional quantities for comparison, e.g., with experimental data. We use a sample liquid with the properties of water at room temperature, i.e., with the viscosity $\eta=\SI{8.9e-4}{Pa\,s}$ and the liquid-gas interface tension $\gamma=\SI{72.8e-3}{N/m}$, cf. \citet{Lide2004}. The height scale is given by the precursor layer height $h_0=h_p=\SI{100}{nm}$. Assuming a low equilibrium contact angle of \SI{5}{\degree}, i.e., $\theta_\mathrm{Y}=\tan(\SI{5}{\degree})$, Eq.~\eqref{eq:young_CA_approx} implies a Hamaker constant of the order $A_0=\SI{9.3e-18}{J}$. In consequence, the length scale is $x_0\approx\SI{885}{nm}$. While the droplets in this work are usually considered to be infinitely large, the smallest droplet in Fig.~\ref{fig:steady_profiles} has a height of $\SI{20}{\micro m}$ and a volume of $\SI{0.01}{\micro l}$.

The time scale is $t_0=\SI{22.5}{\micro s}$. The velocities are transformed with the scale $U_0 = x_0/t_0$, e.g., the stick-slip motion in the space-time plot in Fig.~\ref{fig:spacetime} is caused by a brush moving at a velocity of $U=\SI{0.55}{mm/s}$ and has a period length of $\tau\,t_0=\SI{0.1}{s}$.
The velocity corresponds to a capillary number of $\mathrm{Ca}=\eta U / \gamma \approx \num{7e-6}$.

For a brush of the relative grafting density $\sigma=0.3$ the employed dimensionless height parameter $\tilde l=20$ corresponds to a dry brush thickness of $H_\mathrm{dry}=\sigma N \ell_K=\SI{600}{nm}$. The dimensionless parameter $\tilde T=0.02$ encodes the energy scale of the brush and together with the other parameters determines the Kuhn length $\ell_K=\SI{28}{nm}$. Consequentially, the absolute grafting density is $\sigma/\ell_K^2=\SI{383}{\micro m^{-2}}$.

Given the above scaling the dimensionless diffusion coefficient $D=\SI{4e-3}{}$ corresponds to a value of \SI{2e-6}{m^2/s} in dimensional terms, if additionally multiplied with the liquid energy scale $\rho_\mathrm{liq} k_BT$, where $\rho_\mathrm{liq}=\SI{3.33e28}{m^{-3}}$ is the liquid particle density.
The transfer coefficient $M$ can be expressed in units of liquid volume per area per pressure (or particle velocity per pressure), where, e.g., $M=1$ corresponds to a value of $\SI{4.8e-7}{m/(Pa\,s)}$.

\subsection{Comparison of dynamic behaviour between full and simplified models}\label{app:full_dynamic_comparison}

\begin{figure}[!htb]
	\centering
	\includegraphics[width=0.5\textwidth]{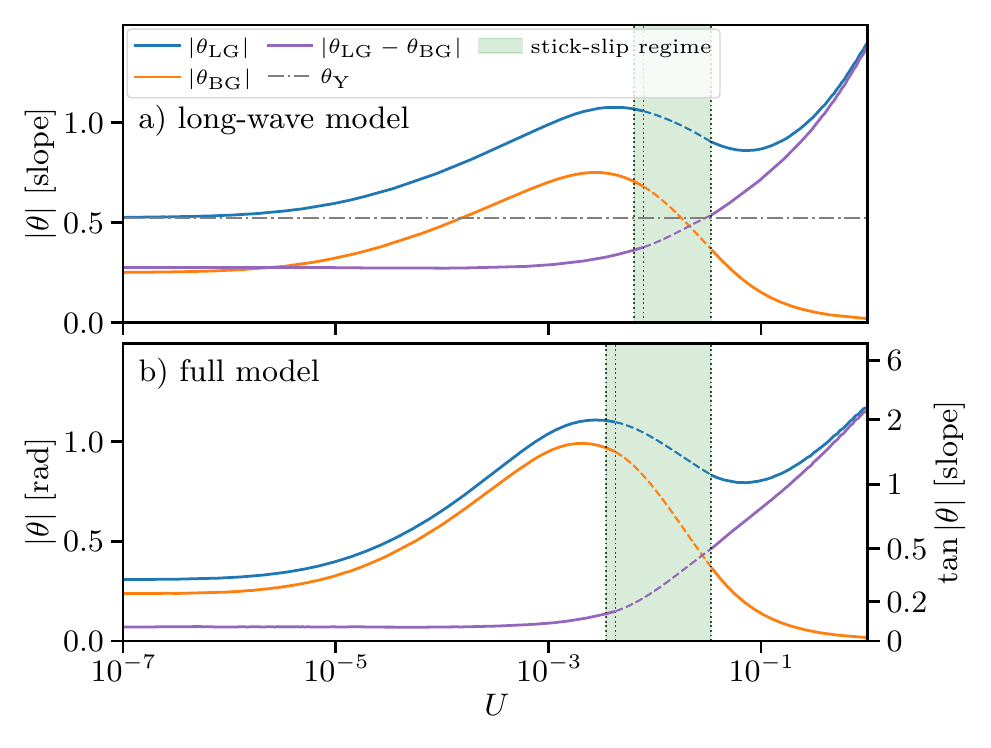}
	\caption{Dynamic contact angles $|\theta_\mathrm{LG}|$ and $|\theta_\mathrm{BG}|$ for a large range of substrate velocities $U$ similar to Fig.~\ref{fig:var_Uhigh}, but here we compare simulations of (a) the simplified model used throughout sections~4 and~5, i.e., the long-wave model without brush state-dependent interface energies [Eqs.~\eqref{eq:dFdh_dFdz_dimless}--\eqref{eq:f_wet_dimless}], and (b) the full model [see Appendix \ref{app:full_model}], i.e., the full-curvature formulation with the brush state-dependent capillarity and wettability [Eqs.~\eqref{eq:wetting_potential}~\&~\eqref{eq:brush_dependencies}]. The stick-slip regime is only slightly shifted between the two models. The parameters are identical to the ones used in Fig.~\ref{fig:var_Uhigh} but the strength of the wetting potential, i.e., the Hamaker constant $A$, was reduced by \SI{25}{\percent}, which lowers the contact angles and eases the numerical effort in particular in the full-curvature computations. Note that in the full-curvature formulation the angles are not approximated by the interface slopes. The conversion via $\tan \theta$ is provided as a second $y$-axis.
 The equilibrium Young angle $\theta_\mathrm{Y}$ is not shown for case (b), as it depends on the dynamic brush state.}
	\label{fig:full_model_comparison}
\end{figure}

In order to reduce the computational complexity all dynamic forced wetting simulations in the main part were conducted using the simplified version of the gradient dynamics model, i.e., by dropping the brush-state dependencies of the interface and wetting energies and employing a long-wave approximation, see Appendix~\ref{app:lw}.
We suggest, that the localised effects acting at the wetting ridge, which are responsible for the stick-slip dynamics, are only weakly affected by this simplification and appear similarly in the full model. Here, with `full model' we refer to the model as presented in section 2 (summarised in Appendix \ref{app:full_model}), in particular, accounting for the correct interface curvature via the metric factors $\xi_{h+\zeta}$ and $\xi_\zeta$ and including the brush state-dependencies of the interface and wetting energies [Eqs.~\eqref{eq:wetting_potential}~\&~\eqref{eq:brush_dependencies}]. To support this statement, in Fig.~\ref{fig:full_model_comparison} we present contact angle measurements in a forced wetting scenario for a wide range of velocities similar to the one presented in Fig.~\ref{fig:var_Uhigh} but for both the full and the simplified model with otherwise identical parameters.

Notably, the behaviour is qualitatively identical and also quantitatively very similar. The stick-slip regime is only slightly shifted between the two models and hysteresis behaviour is equally observed. Due to both full-curvature and adaption effects the contact angles in the full model are slightly different and cover a wider range of interface slopes. Note that the contact angles are converted to radians via the arc tangent of the interface slopes in the full-curvature model.

\subsection{Numerical implementation}\label{sec:numerics}
For the numerical simulations of the model equations~\eqref{eq:dFdh_dFdz_dimless}~and~\eqref{eq:model_dimless} we employ the finite element method implemented in the C++ library \emph{oomph-lib}\cite{HeHa2006} and use the implicit BDF scheme of second order for time stepping. Spatial adaption of the mesh and temporal adaption of the time step allows us to resolve the multi-scale character of the problem, in particular for the stick-slip dynamics. The typical number of used spatial grid points is of the order of 1000.

To investigate the forced motion of a large droplet over a polymer brush in the reference frame of the droplet, we need to adopt sensible boundary conditions. Both partial differential equations~\eqref{eq:model_dimless} are of fourth order in space and thus we require eight individual boundary conditions as specified in the following equations \eqref{eq:BC1}--\eqref{eq:BC3}.

During the forced wetting we have a dry brush entering the comoving frame and a swollen brush leaving it, i.e.\ some amount of liquid is transported through the boundaries and out of the system. As for the analysis of periodic motion it is helpful to conserve the liquid mass in the system, we counter this effect by resupplying the exact amount liquid lost through the boundaries back into the droplet. We express the boundary fluxes via the brush as $j_\zeta$ and the boundary fluxes transported via the liquid layer as $j_h$, where the flux of a variable $\phi$ is defined by
\begin{equation}
    \partial_t \phi = -\partial_x j_\phi + M [\ldots].
\end{equation}
The transport fluxes through the domain boundaries that are caused by the moving substrate and adsorption layer are then
\begin{equation}
    \begin{aligned}
        j_\zeta(x=0) = -U\zeta(x=0),\quad
        j_\zeta(x=L)             &= -U\zeta(x=L),\\
        \text{and}\quad j_h(x=L) &= -Uh(x=L).\label{eq:BC1}
    \end{aligned}
\end{equation}
Note that this corresponds to homogeneous Neumann conditions for the pressure $\partial_x\,\delta \mathcal{F}/\delta \phi$ at the corresponding boundaries.

The liquid flux through the left boundary $j_h(x=0)$, namely, the boundary where the drop is situated, is then set to account for the sum of all other fluxes:
\begin{equation}
    j_h(x=0)=j_\zeta(x=L)+j_h(x=L)-j_\zeta(x=0),\label{eq:BC2}
\end{equation}
thereby compensating all potential losses of liquid volume. With these boundary conditions, the total liquid volume ${\int_0^L (h+\zeta)\,\d x}$ is a conserved quantity also in the comoving frame.

While the gradient of the brush height at the boundaries is assumed to vanish (homogeneous Neumann BC), we fix the slope of the liquid-gas interface profile at the macroscopic Young's angle $\theta_Y$. This implies that the curvature of the droplet approaches zero, which resembles the limiting case of an infinitely large droplet:
\begin{equation}
    \begin{aligned}
        \partial_x h(x=0)     &= \theta_\mathrm{Y}\qquad\text{and}\\
        \partial_x \zeta(x=0) &= \partial_x \zeta(x=L)=\partial_x h(x=L) = 0.\label{eq:BC3}
    \end{aligned}
\end{equation}

\begin{figure}[!htb]
    \includegraphics{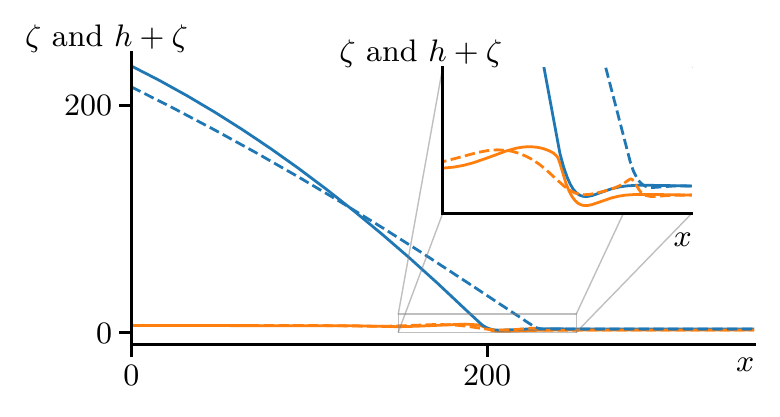}
    \caption{Shown is the numerically realised forced wetting scenario for a stick-slip-motion. The brush-liquid (orange) and liquid-gas interfaces (blue) are displayed at the largest (solid line) and smallest (dashed line) contact angle of a cycle of stick-slip motion. A deformation of the brush only occurs in the vicinity of the contact line.}\label{fig:visstickslip}
\end{figure}

The choice of the macroscopic Young angle as a boundary condition again corresponds to the simulation of a very large droplet and implements the previously discussed equilibrium states as limiting cases at vanishing velocity. The droplet volume is chosen such that the region deformed by the stick-slip motion is far away from the boundaries. This is visualised in \autoref{fig:visstickslip} where the stick-slip dynamics occurs localised in the vicinity of the contact line and far off the boundary.

\subsection{Obtaining dissipation from the gradient dynamics}\label{sec:dissipation_derivation}
Here we discuss how to measure the dissipation due to the various fluxes based on the gradient dynamics structure~\eqref{eq:gradient_model} analogously to the one-field case treated by~\citet{Thie2018csa}. Without loss of generality, we restrict the derivation to the planar case $\mathbf x \to x$.

First, we define the lateral liquid transport fluxes $j_h$ and $j_\zeta$ (within film and brush) including advection and also introduce the vertical transport flux $j_M$ between the two phases as
\begin{equation}
    \begin{aligned}
        j_h &= -\frac{h^3}{3\eta} \partial_x \frac{\delta \mathcal{F}}{\delta h} - Uh,\quad
        j_\zeta = -D \zeta \partial_x \frac{\delta \mathcal{F}}{\delta \zeta} - U\zeta,\\
        j_M &= M \left[ \frac{\delta \mathcal{F}}{\delta h} - \frac{\delta \mathcal{F}}{\delta \zeta} \right].
    \end{aligned}
\end{equation}
Here, a positive sign of $j_M$ corresponds to transport of liquid from the drop into the brush. Next, we rewrite the total time derivative of the free energy in terms of the dynamical equations~\eqref{eq:gradient_model} and the fluxes:
\begin{align}
    \frac{\text{d}\mathcal{F}}{\text{d}t}= &\int\left[\frac{\delta \mathcal{F}}{\delta h}\partial_t h+\frac{\delta \mathcal{F}}{\delta \zeta}\partial_t\zeta\right]\text{d}x \nonumber\\
    =                                      &\int \left[\frac{\delta \mathcal{F}}{\delta h}\left(-\partial_x j_h - j_M\right)+\frac{\delta \mathcal{F}}{\delta \zeta}\left(-\partial_x j_\zeta+j_M\right)\right]\text{d}x\nonumber\\
    \overset{(*)}{=}                       &-\int \frac{h^3}{3\eta}\left(\partial_x\frac{\delta \mathcal{F}}{\delta h}\right)^2\text{d}x-\int D\zeta\left(\partial_x\frac{\delta \mathcal{F}}{\delta\zeta}\right)^2\text{d}x\nonumber\\&-\int M\left(\frac{\delta \mathcal{F}}{\delta h}-\frac{\delta \mathcal{F}}{\delta \zeta}\right)^2\text{d}x+\int(-U)\left(h\partial_x\frac{\delta \mathcal{F}}{\delta h}+\zeta\partial_x\frac{\delta \mathcal{F}}{\delta\zeta}\right)\text{d}x\nonumber\\
    =:                                     &-D_h-D_\zeta-D_M+D_\mathrm{adv}\, \label{eq:Dissipation}
\end{align}
where we have used partial integration with vanishing boundary terms at $(*)$ and identified the integrals with the different channels of energy dissipation:  convective motion within the drop $D_h$, diffusion within the brush $D_\zeta$, liquid transfer between film and brush $D_M$, and caused by imposed advection $D_\mathrm{adv}$. Notably, in the case of $U=0$ this directly implies $\frac{\mathrm{d}\mathcal{F}}{\mathrm{d}t} \leq 0$, i.e. the system approaches an energetic minimum if no driving $U$ is applied.

\subsection{Macroscopic derivation of the Young and Neumann laws}\label{sec:macroscopic}
In Section~\ref{sec:equilibrium} we derive a global Young law and a local Neumann law in the mesoscopic description of a drop on the brush. Here, we present a derivation of the same laws in the macroscopic picture to establish the conditions for the consistency of the two descriptions. This follows in spirit a similar derivation for the case of films covered by an insoluble surfactant presented by \citet{TSTJ2018l}.

On the macroscale, there is a sharp contact line limiting the drop base at radius $x=R$. It divides the brush-covered substrate into a part covered by liquid and a part without liquid on top. Hence, we write for the macroscopic grand potential
\begin{align}
    \mathcal{G}[h, \zeta] = &\int_0^R \left[ \gamma \xi_{h+\zeta} + \gamma_\mathrm{bl} \xi_\zeta
        +  g_\mathrm{brush} - P(h+\zeta) \right] \mathrm{d} x \nonumber\\
    &+ \int_R^\infty \left[\gamma_\mathrm{bg} \xi_\zeta +  g_\mathrm{brush} - P \zeta \right]\, \mathrm{d} x\\
    &+ \lambda_h h(R^-) + \lambda_\zeta (\zeta(R^-)-\zeta(R^+)),\nonumber
\end{align}
i.e.\ the macroscopic analogy to Eq.~\eqref{eq:F_P}.
Here, the Lagrange multiplier \(P\) ensures conservation of liquid volume and the Lagrange multiplier \(\lambda_h \) ensures the condition \(h(R) = 0\). In addition, we need to impose explicitly that the brush height \(\zeta \) is continuous across \(x=R\), which is done by the Lagrange multiplier \(\lambda_\zeta \). In order to distinguish the states on the two sides of the contact line we have defined the limits \(R^{\pm} = \lim_{\varepsilon \to 0} R \pm \varepsilon \).

Employing the same mechanical analogy as in the mesoscopic picture, we interpret the integrand of the grand potential as a Lagrangian \(\mathcal{L}: \ \mathcal{G}=\int \mathcal{L} \ \mathrm{d} x\) and evaluate the Hamiltonian $\mathcal{H}$ with respect to the generalised coordinates $\{h+\zeta,\, \zeta\}$ and their corresponding generalised momenta $\{p_{h+\zeta},\, p_\zeta\}$. In the macroscopic approach, we need to distinguish positions inside and outside of the drop radius and find for the Hamiltonian
\begin{align}
    \mathcal{H}_\mathrm{in}  &= -\frac{\gamma}{\xi_{h+\zeta}}  - \frac{\gamma_\mathrm{bl}}{\xi_\zeta} - g_\mathrm{brush} + P (h+\zeta),\\
    \mathcal{H}_\mathrm{out} &=  -\frac{\gamma_\mathrm{bg}}{\xi_\zeta} - g_\mathrm{brush} + P \zeta.\label{eq:brush-macro-hamiltonian}
\end{align}
Again, it corresponds to an energy density that has to be uniform across the domain. We therefore evaluate \(\mathcal{H}_\mathrm{in}=\mathcal{H}_\mathrm{out}\) far away from the contact line, assuming that \(1/\xi_{h+\zeta}=\cos \theta_\mathrm{LG}=\cos \theta_\mathrm{Y}\) is nearly constant and that the brush has adapted a constant height \(\zeta_{\mathrm{d}}\) inside and \(\zeta_{\mathrm{p}}\) outside of the drop (and $\xi_\zeta=1$). Further, we assume that the drop is very large, hence the Laplace pressure vanishes $P\approx 0$. This gives the ammended Young law as
\begin{equation}
    \gamma \cos \theta_\mathrm{Y} = \gamma_\mathrm{bg}(\zeta_\mathrm{p}) - \gamma_\mathrm{bl}(\zeta_\mathrm{d}) + g_\mathrm{brush}(\zeta_\mathrm{p}) - g_\mathrm{brush}(\zeta_\mathrm{d}). \label{eq:brush-macro-young}
\end{equation}
Note that the slope of the liquid-gas interface is only approximately constant when \(h+\zeta \ll \gamma / P\) but the flat brush is only attained for \(\gamma_\mathrm{bl} / \partial_\zeta g_\mathrm{brush} \ll (R-x) \sim h+\zeta \). Thus, to achieve a form like Young's law requires an intermediate asymptotics \(\gamma_\mathrm{bl} / \partial_\zeta g_\mathrm{brush} \ll h+\zeta \ll \gamma / P \sim R\). This is, however, easily fulfilled for large drops.
The close relation of the macroscopic Young law to its mesoscopic equivalent Eq.~\eqref{eq:young} is particularly transparent if an additional $0=\gamma_\mathrm{bl}(\zeta_\mathrm{p}) - \gamma_\mathrm{bl}(\zeta_\mathrm{p})$ is added to the r.h.s.\ of Eq.~\eqref{eq:brush-macro-young}, see the resulting Eq.~\eqref{eq:young_macro} in the main text.

Similarly, we also recover the two components of the Neumann law by equating the Hamiltonian in the two regions close to the contact line, i.e. for $x\to R^\pm$. There, due to the continuity of $\zeta$, the brush energy approaches the same value and evaluating $\mathcal{H}_\mathrm{in}=\mathcal{H}_\mathrm{out}$ gives
\begin{equation}
    \gamma \cos \theta_\mathrm{LG} + \gamma_\mathrm{bl} \cos \theta_\mathrm{BL} = \gamma_\mathrm{bg} \cos \theta_\mathrm{BG},\label{eq:brush-macro-neumann-horizontal}
\end{equation}
namely, the horizontal Neumann condition in the macroscopic picture. The relation is exactly equivalent to the mesoscopic version (Eq.~\eqref{eq:Neumann_hor}) if the consistency relation \eqref{eq:consistency} is considered.

Assuming a conservation of the generalised momentum ${p_\zeta + p_{h+\zeta}}$ across the contact region $x\in[R^-, R^+]$ results in the vertical Neumann condition
\begin{equation}
    \gamma\sin\theta_\mathrm{LG}+\gamma_\mathrm{bl}\sin\theta_\mathrm{BL} = \gamma_\mathrm{bg}\sin\theta_\mathrm{BG} \label{eq:brush-macro-neumann-vertical}
\end{equation}
as an exact analogy to the mesoscopic equation~\eqref{eq:Neumann_ver}.

\end{document}